\documentclass[letter,twocolumn,pre,showpacs,floatfix]{revtex4}
\usepackage[dvips]{graphicx} 
\usepackage{amssymb} 
\usepackage{amsmath} 
\usepackage{latexsym} 
\usepackage[retainorgcmds]{IEEEtrantools}
\usepackage{subfigure}

\newcommand\be{\begin{equation}}
\newcommand\bea{\begin{eqnarray}}
\newcommand\biea{\begin{IEEEeqnarray}}
\newcommand\eiea{\end{IEEEeqnarray}}
\newcommand\eea{\end{eqnarray}}
\newcommand\ee{\end{equation}}

\newcommand\VE{\varepsilon}

\begin{document}

\title{Mixing-Demixing Phase Diagram for Simple Liquids\\ in Non-Uniform Electric Fields}

\date{\today}
\author{Jennifer Galanis}
\author{Yoav Tsori}
\affiliation{Department of Chemical Engineering, Ben-Gurion University of the Negev,
Beer-Sheva, 84105, Israel}

\begin{abstract}
We deduce the mixing-demixing phase diagram for binary liquid mixtures in an electric
field for various electrode geometries and arbitrary constitutive relation for
the dielectric constant. By focusing on the behavior of the
liquid-liquid interface, we produce simple analytic expressions for the dependence of the
interface location on experimental parameters. We also show that the phase diagram
contains regions where liquid separation cannot occur under \emph{any} applied field. The
analytic expression for the boundary ``electrostatic binodal'' line reveals that the
regions' size and shape depend strongly on the dielectric relation between the liquids.
Moreover, we predict the existence of an ``electrostatic spinodal'' line that identifies
conditions where the liquids are in a metastable state. We finally construct the phase
diagram for closed systems by mapping solutions onto those of an open system via an
effective liquid composition. For closed systems at a fixed temperature and mixture
composition, liquid separation occurs in a finite ``window'' of surface potential (or
charge density). Larger potentials or charge densities counterintuitively destroy the
interface, leading to liquid mixing. These results give valuable guides for experiments by
providing easily testable predictions for how liquids behave in non-uniform electric
fields.

\end{abstract}
%\pacs{64.60.-1, 64.70.Ja, 68.05.-n}
\maketitle

%88888888888888888888888888888888888888888888888888
%  INTRODUCTION
%88888888888888888888888888888888888888888888888888
Phase transitions describe fundamental transformations in substances, where material
properties, such as viscosity, refractive index, etc., often dramatically change. These
changes are not only mediated by intrinsic thermodynamic variables (temperature, pressure,
etc.), but also by external forces (gravitational~\cite{Moldover1979},
magnetic~\cite{Asamitsu1995} and electric~\cite{Landau1957} fields, and shear
flows~\cite{Silberberg1952,Nakatani1996}). Scientific interest in using electric fields to
alter the phase behavior spans over half a century and resulted in theories and
experiments devoted to the application of uniform fields in dielectric liquid
mixtures~\cite{Landau1957,Onuki1995a,Onuki1995b,Stepanow2009,Debye1965,Beaglehole1981,
Early1992, Wirtz1993,Orzechowski1999}. Unfortunately, the liquid-field coupling in
uniform electric fields is very weak since in such cases variations in the field strength occur as a result of variations in the permittivity of the liquid. As a consequence, theories predict that even minuscule changes to the phase diagram require
enormous applied voltages~\cite{Landau1957,Onuki1995a,Debye1965}.

In contrast, recent theoretical and experimental results reveal that nonuniform fields can
effectuate large changes in phase diagrams~\cite{Tsori2004,Marcus2008,Samin2009}.
The externally produced spatial variations in field strength occur even in homogeneous
materials, and lead to liquid rearrangement that can potentially
induce liquid-liquid separation. High-gradient fields readily emerge from a
modest potential or surface charge on misaligned plate capacitors as well as from small
objects with high surface curvature, like nanowires and
colloids~\cite{Tsori2004,Marcus2008,Samin2009}. Thus the relative ease for creating
nonuniform fields underscores the potential to profoundly influence the behavior of
complex liquids.

The challenge of non-uniform fields, however, resides in distinguishing true liquid-liquid 
phase separation from mere concentration gradients. In the more common case of uniform 
fields, the free energy has a double-well form with two coexisting minima. Since the 
system possesses translational invariance, the two liquids can replace each other in space 
without changing the total energy. This does not hold for nonuniform fields where 
translational invariance is broken. Here, the spatial location of the liquids directly 
ties to the free energy, and as a consequence, the total free energy can have a single 
minimum even with two-phase coexistence. 
We point out that not all spatially nonuniform fields display this property, 
as for example in the case of random-field~\cite{Aharony1978} and 
periodic-field~\cite{Vink2012} Ising models.

To overcome the difficulty in determining a transition, we defined phase separation by 
observing a \emph{local} property---the behavior of the interface. 
Using this perspective, we derived analytic expressions for predicting the location of the 
interface from experimental parameters.
We additionally adapted the standard methods used in creating phase diagrams and found the 
electrostatic-equivalent of binodal and spinodal lines as well as critical points. 
The methods presented here can, in principle, apply to any geometry, and we 
explicitly give results for three basic electrode shapes: wedge, cylinder, and sphere. 
Furthermore, these methods can incorporate an arbitrarily complicated dielectric relation 
for the liquid composition, provided that derivatives to the expression exist.

The manuscript is arranged as follows. We describe the theory for liquid mixtures with
electric fields in Sec.~\ref{Sec_theory} and briefly review general properties of phase
diagrams in the absence of external fields in Sec.~\ref{Sec_review}. In Sec.~\ref{Sec_def},
we introduce a useful definition of phase separation in an electric field that
is essential for simplifying theoretical expressions. In
Sec.~\ref{Sect_Interface}, we assume phase separation exists and derive simple expressions
for the location of the liquid-liquid interface. The mixing-demixing regions of the phase
diagram as well as the dividing ``electrostatic binodal'' line are discussed in
Sec.~\ref{Sect_eleBin}, while the theoretical stable-metastable states and dividing
``electrostatic spinodal'' line are presented in Sec.~\ref{Sec_spin}. Finally, we discuss
important differences between open and closed systems in Sec.~\ref{Section_closed}.

%%%%%%%%%%%%%%%%%%%%%%%%%%%%%%%%%%%%%%%%%%%%%%%%
%  THE MODEL
%%%%%%%%%%%%%%%%%%%%%%%%%%%%%%%%%%%%%%%%%%%%%%%%
\section{Theory}\label{Sec_theory}
Using a mean-field approach, we consider a binary mixture of two liquids, $A$ and $B$, in
an electric field $\mathbf{E}$, and write the total free energy $\mathcal{F}$ for a volume
$V$ as 
\begin{equation}
\label{eq_FE_tot}
\mathcal{F} = \int_V \left( \mathcal{F}_{\mathrm m} + \mathcal{F}_{\mathrm e} \right)
\mathrm{d}V~,
 \end{equation}
where $\mathcal{F}_{\mathrm m}$, and $\mathcal{F}_{\mathrm e}$ are the free energy
densities for mixing and electrostatics, respectively.

%========================================
% FREE ENERGY MIXING
%========================================
The liquids, in the absence of an electric field, can mix or demix due to a competition
between entropy and enthalpy, where temperature $T$ adjusts the relative balance. For
concreteness, we use the following Landau free energy of mixing $\mathcal{F}_{\mathrm{m}}
= kTf_{\mathrm{m}}/Nv$ where the expansion is performed around the critical volume
fraction $\phi_c$ 
\begin{equation}
f_{\mathrm{m}} \approx (2-N\chi)\left(\phi-\phi_c\right)^2+
\frac{4}{3}\left(\phi-\phi_c\right)^4 + const. \label{eq_FE_Landau}
\end{equation}
where $k$ is Boltzmann's constant, $\phi$ such that $0<\phi<1$ is the volume fraction of
component $A$, and $\chi\sim 1/T$ is the Flory interaction parameter~\cite{safran_book}.
Without loss of generality, we set $\phi_c=0.5$, and $N\chi = 2T_c/T$, where $T_c$ is the
critical temperature. Simple liquids have $N=1$, while polymers are composed of $N>1$
monomers with volume $v$. Here, we consider the symmetric simple liquid $N = N_A = N_B =
1$. Real interfaces consist of a gradual change in composition. 
In contrast, $f_{\mathrm{m}}$ generates an interface marked by a discontinuity in composition.
We will find, however, that the discontinuity greatly simplifies the analysis to follow.

%========================================
% FREE ENERGY ELECTROSTATICS
%========================================
For electrostatics, the free energy $\mathcal{F}_{\mathrm{e}} = kT f_{\mathrm{e}}/Nv$ is given by
\begin{equation}
\label{eq_FE_es}
\mathcal{F}_{\mathrm{e}} =
\pm\frac{1}{2}\varepsilon_0\varepsilon(\phi)|\nabla\psi|^2
\end{equation}
where $\varepsilon_0$ is the vacuum permittivity, 
and $\psi$ is the electrostatic potential (${\bf E}=-\nabla\psi$). The positive (negative)
sign corresponds to constant charge (potential) boundary conditions. 

The dielectric permittivity at zero frequency $\varepsilon(\phi)$ depends on the relative
liquid-liquid composition. 
For clarity in the discussions, we mainly consider a linear relation, $\varepsilon(\phi) =
(\varepsilon_A-\varepsilon_B)\phi + \varepsilon_B$, where $\varepsilon_A$ and
$\varepsilon_B$ are the dielectric constants for pure liquids $A$ and $B$, respectively.
Excluding the possibility of critical behavior in $\varepsilon(\phi)$ in the immediate 
vicinity of the liquid's critical point $(\phi_c,T_c)$~\cite{Sengers1980,Leys2010}, the
measured $\varepsilon(\phi)$ often approximates a quadratic function for  various liquid
combinations~\cite{Debye1965,Beaglehole1981}. We, therefore,
highlight some significant changes in the results that occur with higher order
$\varepsilon(\phi)$ relations.

%========================================
% EULER LAGRANGE EQUATIONS
%========================================
To determine the equilibrium state in the presence of a field, we minimize $\mathcal{F}$
with respect to $\phi$ and $\psi$ using calculus of variations and obtain the following
Euler-Lagrange equations
\begin{eqnarray}
\frac{\delta \mathcal{F}}{\delta\psi} &=&
\nabla\cdot\left[\varepsilon_0\varepsilon(\phi)\nabla\psi\right] = 0
\label{eq_EL_LandauGinzburg_ele} \\
\frac{\delta \mathcal{F}}{\delta\phi} &=& \mathcal{F}_{\mathrm{m}}^{\prime} -
\frac{\varepsilon_0}{2}\varepsilon^{\prime}(\phi) |\nabla\psi|^2  - \tilde{\mu} = 0
\label{eq_EL_LandauGinzburg}
\end{eqnarray}
where the ``prime'' represents the derivative with respect to $\phi$. The first equation
is Laplace's equation for the potential $\psi$, while the second equation gives the
composition distribution $\phi$. Both $\varepsilon(\phi)$ and $\psi$ couple the two
equations.

The Lagrange multiplier $\tilde{\mu}=kT\mu/Nv$ in eq.~\ref{eq_EL_LandauGinzburg}
differentiates between open and closed systems. For a closed system (canonical ensemble),
$\mu$ is adjusted to satisfy the mass conservation constraint: $\langle \phi \rangle =
\phi_0$, where $\phi_0$ is the average composition. When the system under consideration
is coupled to an infinite reservoir at composition $\phi_0$, $\mu=\mu_0(\phi_0)$ is the
chemical potential of the reservoir. 

\begin{figure}[!tb]
\begin{center}
\subfigure[\label{fig_geom_C}]{\includegraphics[keepaspectratio=true,width=0.232\textwidth
]{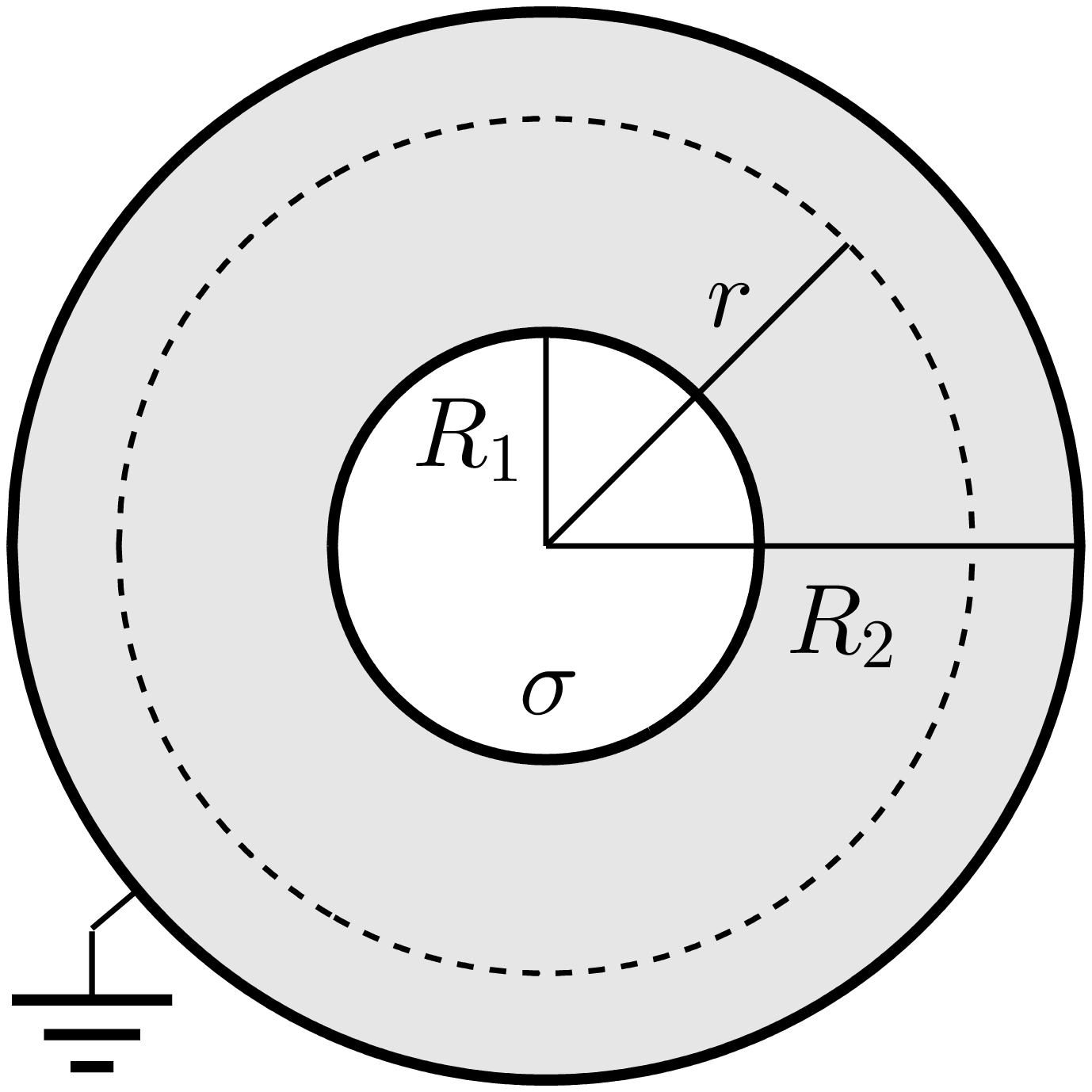}}
\subfigure[\label{fig_geom_W}]{\includegraphics[keepaspectratio=true,width=0.232\textwidth
]{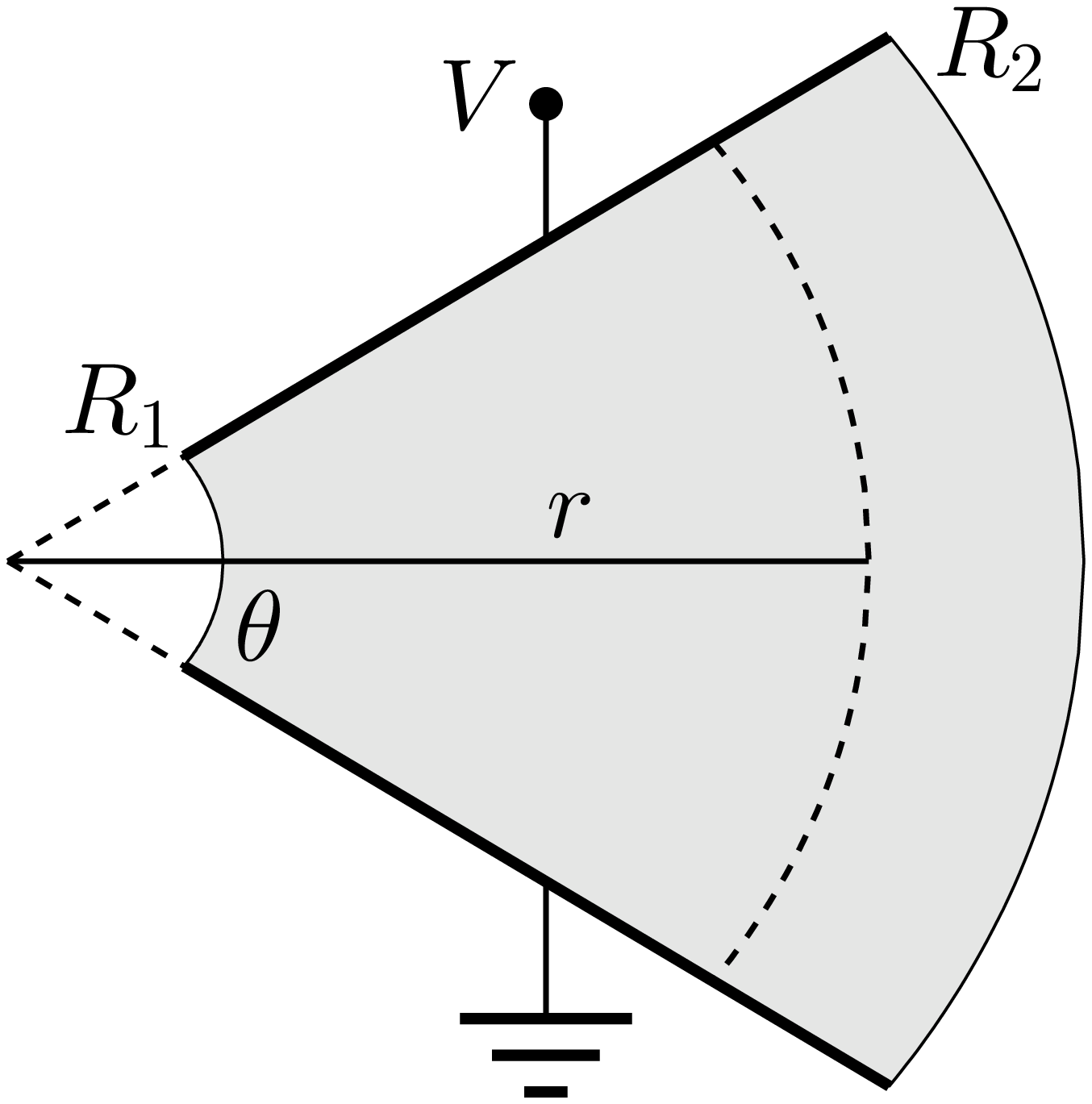}}
\caption{\footnotesize\textsf{Model systems. (a) Cross section through the diameter of
concentric cylinders or spheres with surface charge density $\sigma$. Distance $r$ is
measured
from the center of the cylinder/sphere, and the boundaries are located at $R_1$ and $R_2$.
(b) Cross section of two flat-plate electrodes with an opening angle $\theta$ and
potential difference $V$. Distance $r$ is measured from the ``intersection'' of the two
plates, and the boundaries $R_1$ and $R_2$ mark the ends of the plates. Shading shows the
space occupied by the liquid mixture.}}
\label{fig_Fm}
\end{center}
\end{figure}

%========================================
% GEOMETRIES
%========================================
We conduct detailed investigations of the phase transition with three simple yet
fundamental shapes---cylinder, sphere, and wedge. A closed system with cylindrical
geometry consists of two concentric cylinders with radii $R_1$ and $R_2$, where $R_2
\rightarrow \infty$ produces an open system, Fig.~\ref{fig_geom_C}. We impose cylindrical
symmetry such that $\phi=\phi(r)$ and $\psi=\psi(r)$, where $r$ is the distance from the
inner cylinder's center. Furthermore, the prescribed charge density $\sigma$ on the inner
cylinder allows integration of Gauss's law to obtain an explicit expression for the
electric field. By using a similar construction for spherical geometry
we find that the electric field for both cylindrical and spherical
configurations is $\mathbf{E}(r) = \sigma R_1^n/
(\varepsilon_0\varepsilon(\phi)r^n)\mathbf{\hat{r}}$, where $n=1$ and $2$ for cylinders
and spheres, respectively. Combining this result with $\mathbf{E}=-\nabla\psi$ in
eq.~\ref{eq_EL_LandauGinzburg}, we 
obtain a single equation determining the composition profile $\phi(r)$:
\begin{equation}
f_{\mathrm{m}}^{\prime} - \frac{Nv}{2kT\varepsilon_0} \left(\frac{\sigma
R_1^n}{r^n}\right)^2 \frac{\varepsilon^{\prime}(\phi)}{\varepsilon(\phi)^2} - \mu = 0
\label{eq_EL_cylSph}
\end{equation}

The wedge geometry consists of two ``misaligned'' flat plates with an opening angle
$\theta$, Fig.~\ref{fig_geom_W}. Using a constant potential boundary condition, we obtain
an electric field $\mathbf{E}(r) = (V/r\theta) \mathbf{\hat{\vartheta}}$, where $V$ is the
potential difference across the electrodes, $r$ is the distance from the imaginary
intersection of the two plates, and $\vartheta$ is the azimuthal angle. Combining this
result with eq.~\ref{eq_EL_LandauGinzburg}, we obtain
\begin{equation}
f_{\mathrm{m}}^{\prime} - \frac{Nv}{2kT} \left(\frac{V}{r\theta}\right)^2 \varepsilon_0
\varepsilon^{\prime}(\phi) - \mu = 0
\label{eq_EL_wedge}
\end{equation}

%%%%%%%%%%%%%%%%%%%%%%%%%%%%%%%%%%%%%%%%%%%%%%%%%%%%
% FIGURE
%%%%%%%%%%%%%%%%%%%%%%%%%%%%%%%%%%%%%%%%%%%%%%%%%%%%

In this manuscript we mainly present results for cylindrical geometry. This geometry
intrinsically presents a mathematically unsavory dependence of $f_{\mathrm{e}}$ on
 $\phi$ (via $\VE(\phi)$), and therefore creates more complicated
solutions than, for example, in the wedge. Also, the difference in equational
form between the cylinder and sphere does not present new information for discussion. 
The methods
presented here can easily be adapted to both wedge and sphere geometries.

As will become evident, the precise surface charge density (surface potential) necessary
to induce a transition depends on experimental parameters like the size and relative
concentration of the liquid molecules, size of the charged material, temperature, etc.
We consider a wide range of surface charges $\sigma$, from approximately zero up to
$2\times 10^{-3}\,\mathrm{C/m^2}$ (equivalent to $1.25\times 10^{-2}~ e/\mathrm{nm}^2$). 
For comparison, colloidal
particles immersed in the non-polar phase of an inverse-micelle liquid have been measured
to have large surface potentials, with an estimate of $200$ to $900 e$
charges~\cite{Hsu2005}. This amount of charge on a colloid could induce phase separation
in a binary mixture if its composition is close enough to the demixing curve.
The demixed liquid layer surrounding the colloid is predicted to be several
tens to hundreds of nanometers thick, thereby altering the local environment of the
colloid in an otherwise mixed liquid suspension. Of course, having the ability to
externally apply a field, for example via an electrode, can be useful in some applications.

%88888888888888888888888888888888888888888888888888888888888888888
%  RESULTS
%88888888888888888888888888888888888888888888888888888888888888888
%========================================
% REVIEW OF BINODAL CURVE
%========================================
\section{Phase Diagram without an Electric Field}\label{Sec_review}
We briefly discuss some features of the mixing-demixing phase diagram in the absence of electric fields that are 
essential in the derivations below.
A ``double well'' function (for example eq.~\ref{eq_FE_Landau} when $T<T_c$) possesses
two local minima, one local maximum, and two inflection
points located between the maximum and each minimum. 

%%%%%%%%%%%%%%%%%%%%%%%%%%%%%%%%%%%%%%%%%%%%%%%%%%%%%%%%%%%%%%%%%%
% FIGURE
\begin{figure}[!tb]
\begin{center}
\subfigure[\label{fig_Fm_binodal}]{\includegraphics[keepaspectratio=true,
width=0.232\textwidth]{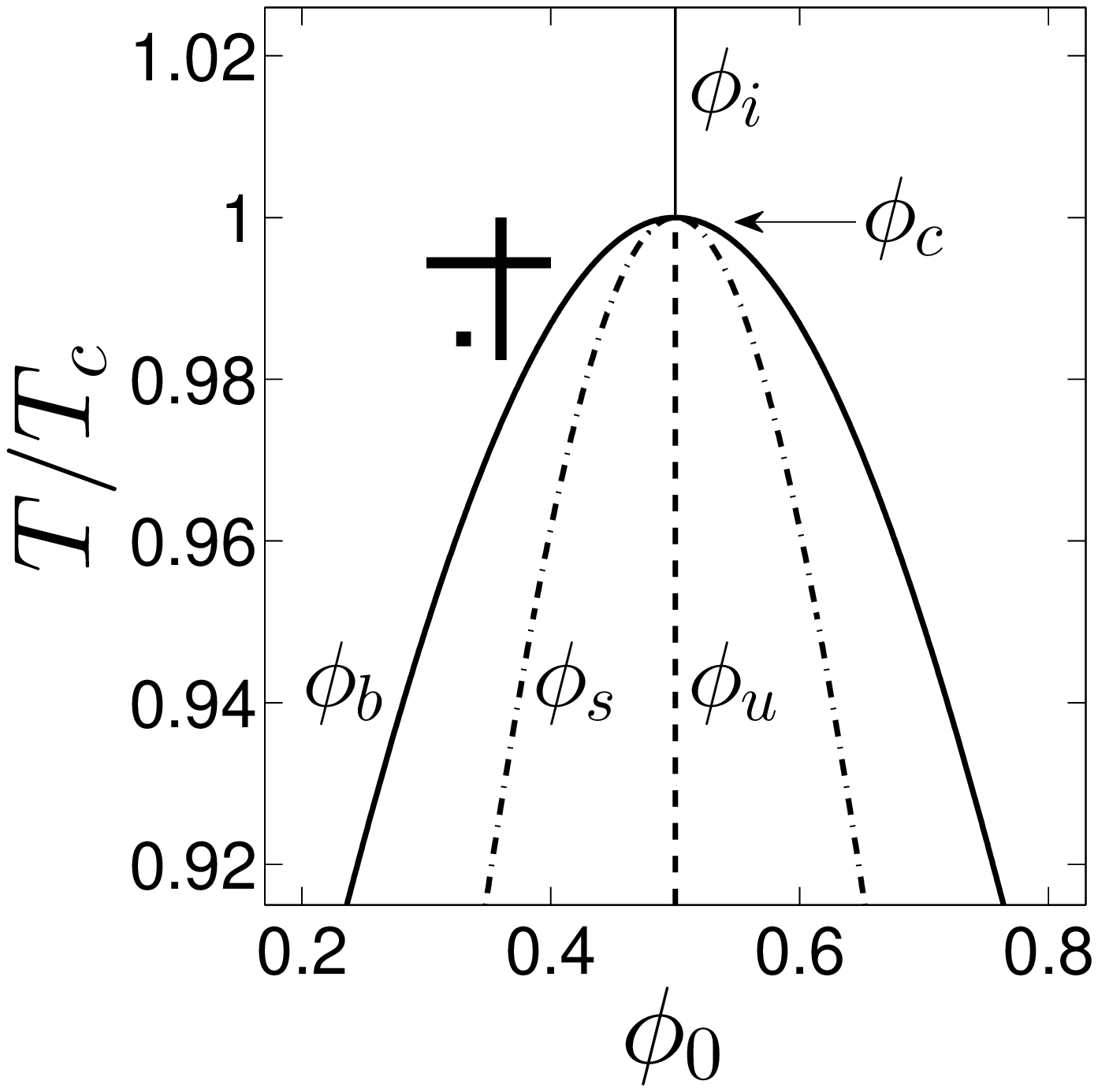}}
\subfigure[\label{fig_Fm_f}]{\includegraphics[keepaspectratio=true,width=0.232
\textwidth]{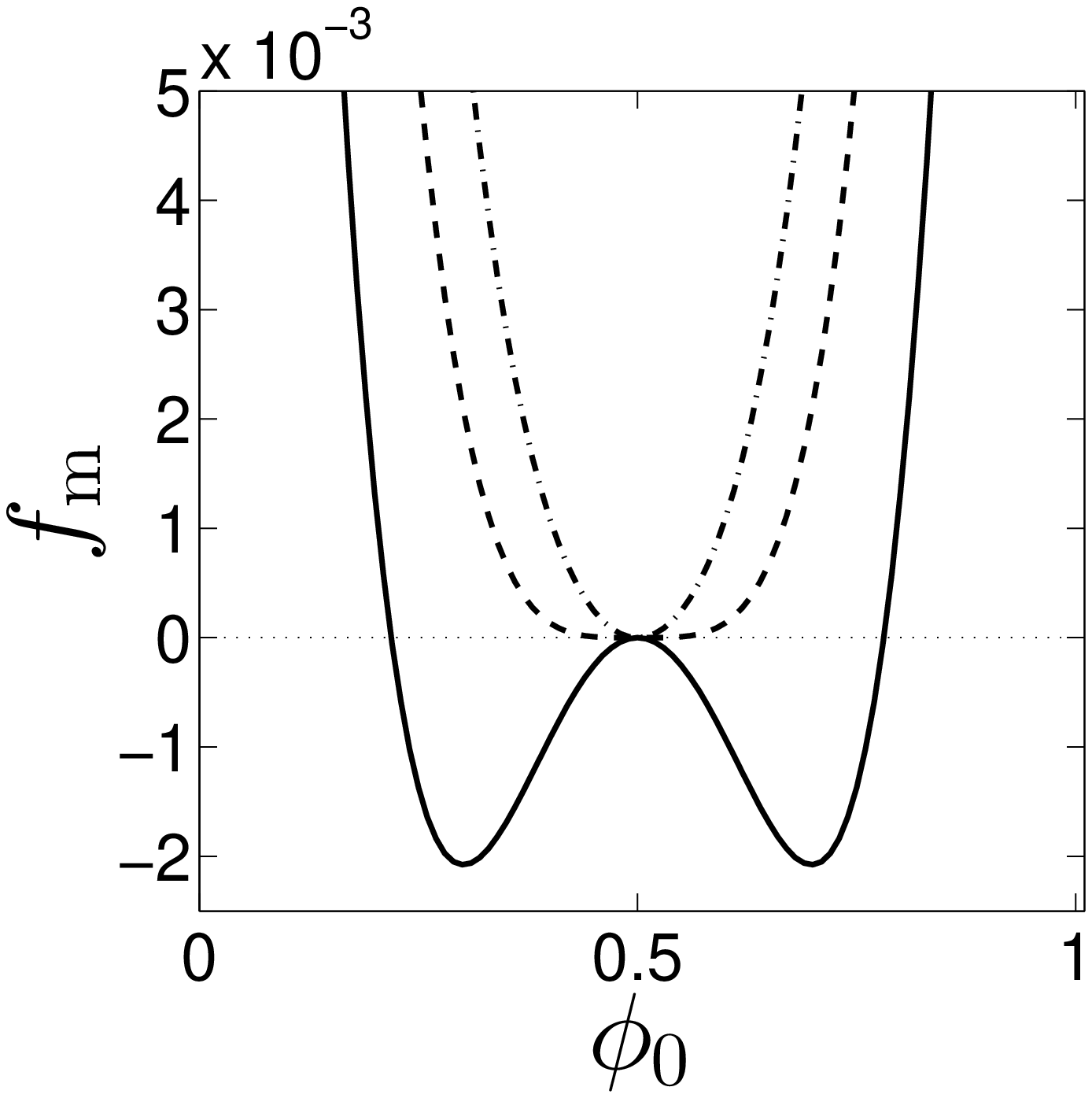}}\\
\subfigure[\label{fig_Fm_dfdp}]{\includegraphics[keepaspectratio=true,
width=0.232\textwidth]{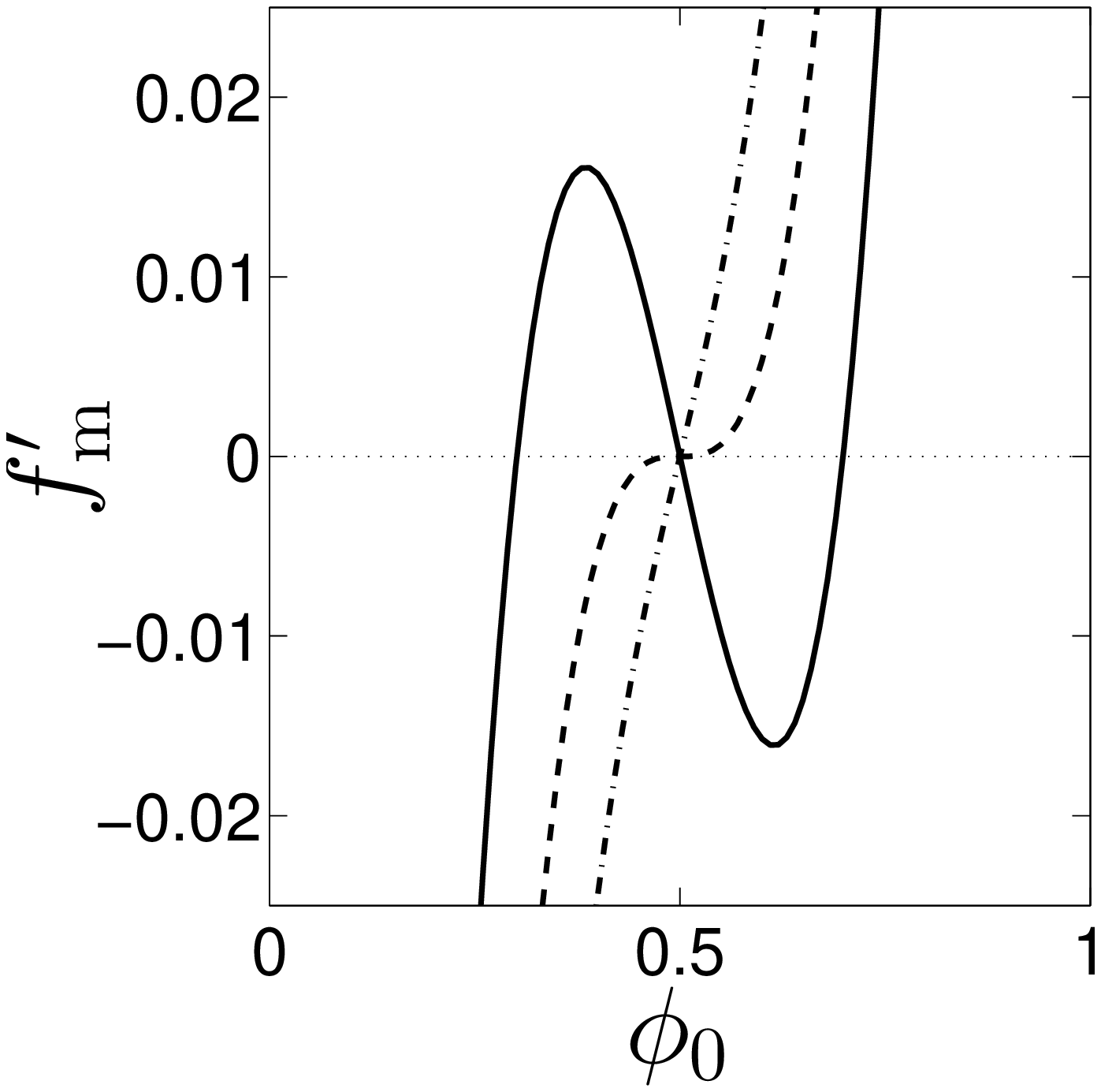}}
\subfigure[\label{fig_Fm_d2fdp2}]{\includegraphics[keepaspectratio=true,
width=0.232\textwidth]{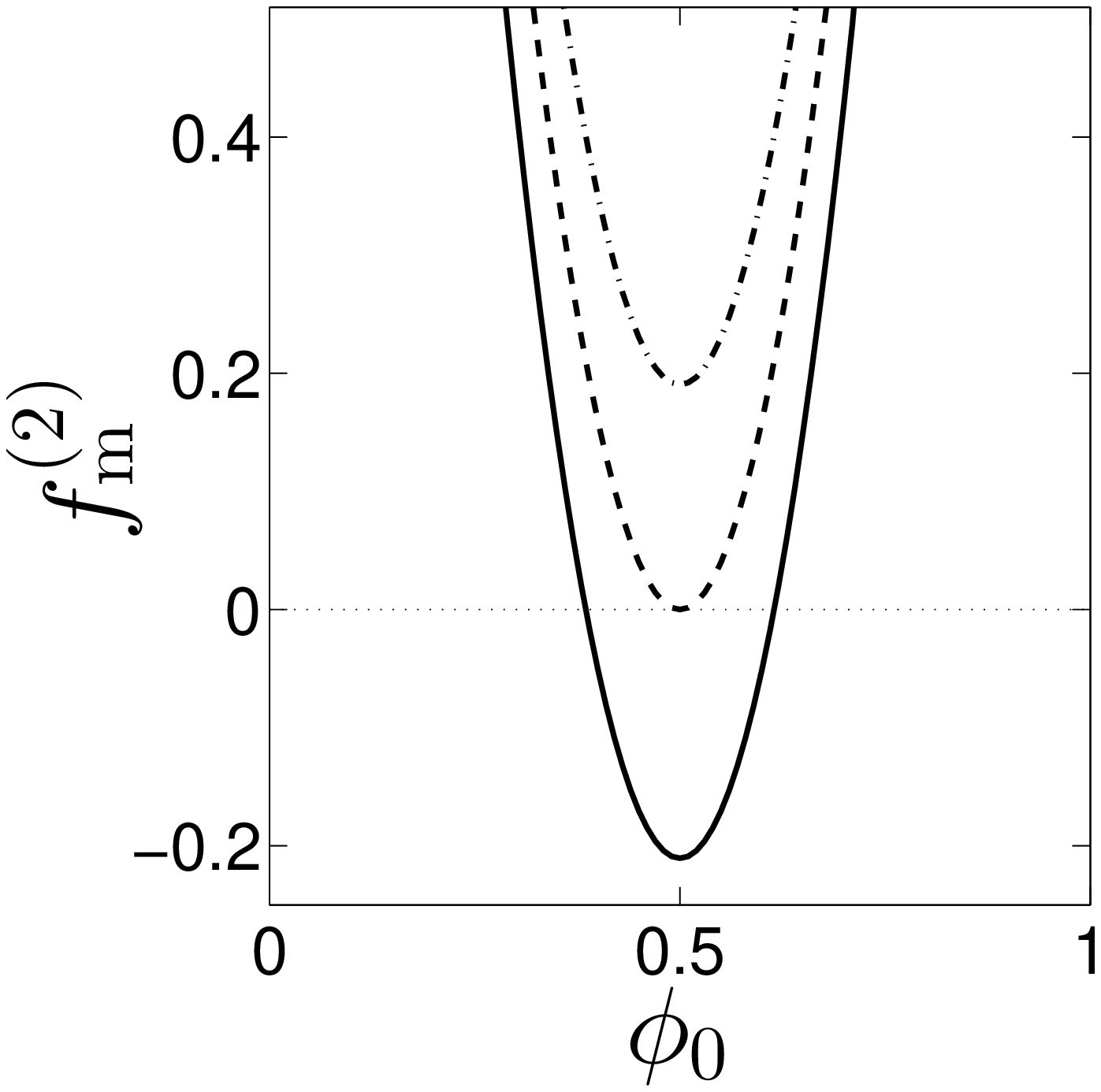}}
 \caption{\footnotesize\textsf{Free energy of mixing $f_{\mathrm{m}}$. (a) Phase diagram
in the $\phi_0-T$ plane showing the zero-field binodal curve ($\phi_b$, thick solid line),
spinodal curve ($\phi_s$, dash-dotted line), unstable solution ($\phi_u$, dashed line),
critical point $\phi_c$, and the minimum of $f_{\mathrm{m}}$ above $T_c$ ($\phi_i$, thin
solid line). Symbol marks the location of data in Fig.~\ref{fig_phi_vs_r_varS}, while
horizontal and vertical bars mark the location of data in Figs.~\ref{fig_phi_vs_r_varP}
and~\ref{fig_phi_vs_r_varT}, respectively. (b-d) Plots of $f_{\mathrm{m}}$ and it's
derivatives with respect to $\phi$ versus $\phi_0$ for $T$ less than (solid line), equal
to (dashed line), and greater than (dash-dotted line) $T_c$. In this and in all
other figures $T_c=298\,\mathrm{K}$, $\VE_A=5$, $\VE_B=3$, $R_1=1\,\mu\mathrm{m}$ and $Nv=1\times
10^{-26}\,\mathrm{m}^3$.
}} 
\end{center}
\end{figure}
%%%%%%%%%%%%%%%%%%%%%%%%%%%%%%%%%%%%%%%%%%%%%%%%%%%%%%%%%%%%%%%%%%

To ascertain the minimum of $f_{\rm m}$ at constant $T$, we find the solution
to $f_{\rm m}^{\prime}=0$, Fig.~\ref{fig_Fm_dfdp}, that also satisfies
$f_{\rm m}^{(2)}>0$, Fig.~\ref{fig_Fm_d2fdp2}, where the derivatives are taken with
respect to $\phi_0$. The two solutions $\phi_b(T) = 1/2 \pm \sqrt{3(T_c-T)/4T)}$ for each
$T<T_c$ create the binodal curve, thick solid line in Fig.~\ref{fig_Fm_binodal}. Fluids
demix if the initial conditions $(\phi_0,T)$ are under the binodal curve, and mix if they
are above this curve. There in fact exists a third solution to
$f_{\mathrm{m}}^{\prime}=0$, Fig.~\ref{fig_Fm_dfdp}---the local maximum at concentration
$\phi_u(T)$, dashed line in Fig.~\ref{fig_Fm_binodal}. Even though this solution is
physically unstable [$f_{\rm m}^{(2)}(\phi_u)<0$, Fig.~\ref{fig_Fm_d2fdp2}], it will be useful in subsequent sections.
If the local minima satisfy $f_{\rm m}^{(2)}>0$ and the local maximum satisfies
$f_{\rm m}^{(2)}<0$, then there must exist inflection points between the extrema that
satisfy $f_{\rm m}^{(2)}=0$, Fig.~\ref{fig_Fm_d2fdp2}. These solutions $\phi_s(T)= 1/2
\pm \sqrt{(T_c-T)/4T}$ for each $T<T_c$ create the spinodal line, dash-dotted line in
Fig.~\ref{fig_Fm_binodal}, and describe liquid behavior dynamically. If the initial point
$(\phi_0,T)$ is located below the spinodal curve, then the liquids demix spontaneously.
If, however, $\phi_0$ exists between $\phi_b$ and $\phi_s$, then the liquid can be
``stuck'' in a local minimum, resulting in a metastable mixed state.

At the critical point $(\phi_c,T_c)$ the shape of $f_{\rm m}$ changes from having a
single to double minima. As $T$ increases to $T_c$, the two minima $\phi_b$, the two
inflection points $\phi_s$, and maximum $\phi_u$ converge and convert into a single
minimum $\phi_c$. To meet these requirements the critical point must
satisfy $f^{\prime}_{\rm m} = f^{(2)}_{\rm m} = f^{(3)}_{\rm m} = 0$ and
$f^{(4)}_{\rm m} > 0$. Figures~\ref{fig_Fm_dfdp} and~\ref{fig_Fm_d2fdp2} display two
of the four requirements. 
Finally, the light solid line in Fig.~\ref{fig_Fm_binodal} shows the single 
solution $\phi_i(T)$ to $f_{\rm m}^{\prime}=0$ above the critical point.

%========================================
% DEFINITION OF PHASE SEPARATION
%========================================
\section{Defining Phase Separation}\label{Sec_def}

\begin{figure}[tb]%
\begin{center}
\subfigure[\label{fig_f}]{\includegraphics[keepaspectratio=true,width=0.232
\textwidth]{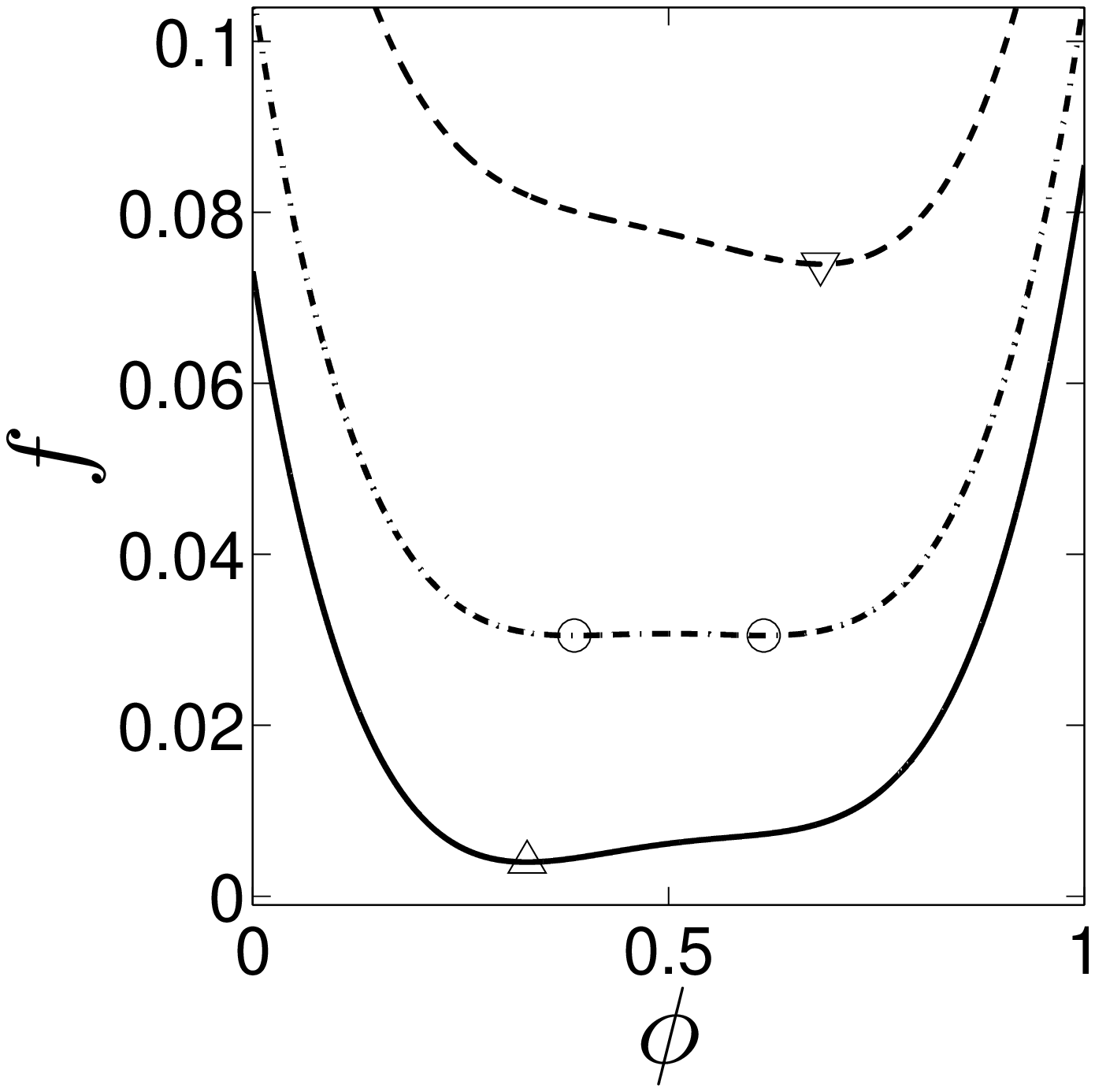}}
\subfigure[\label{fig_phi_r}]{\includegraphics[keepaspectratio=true,width=0.232
\textwidth]{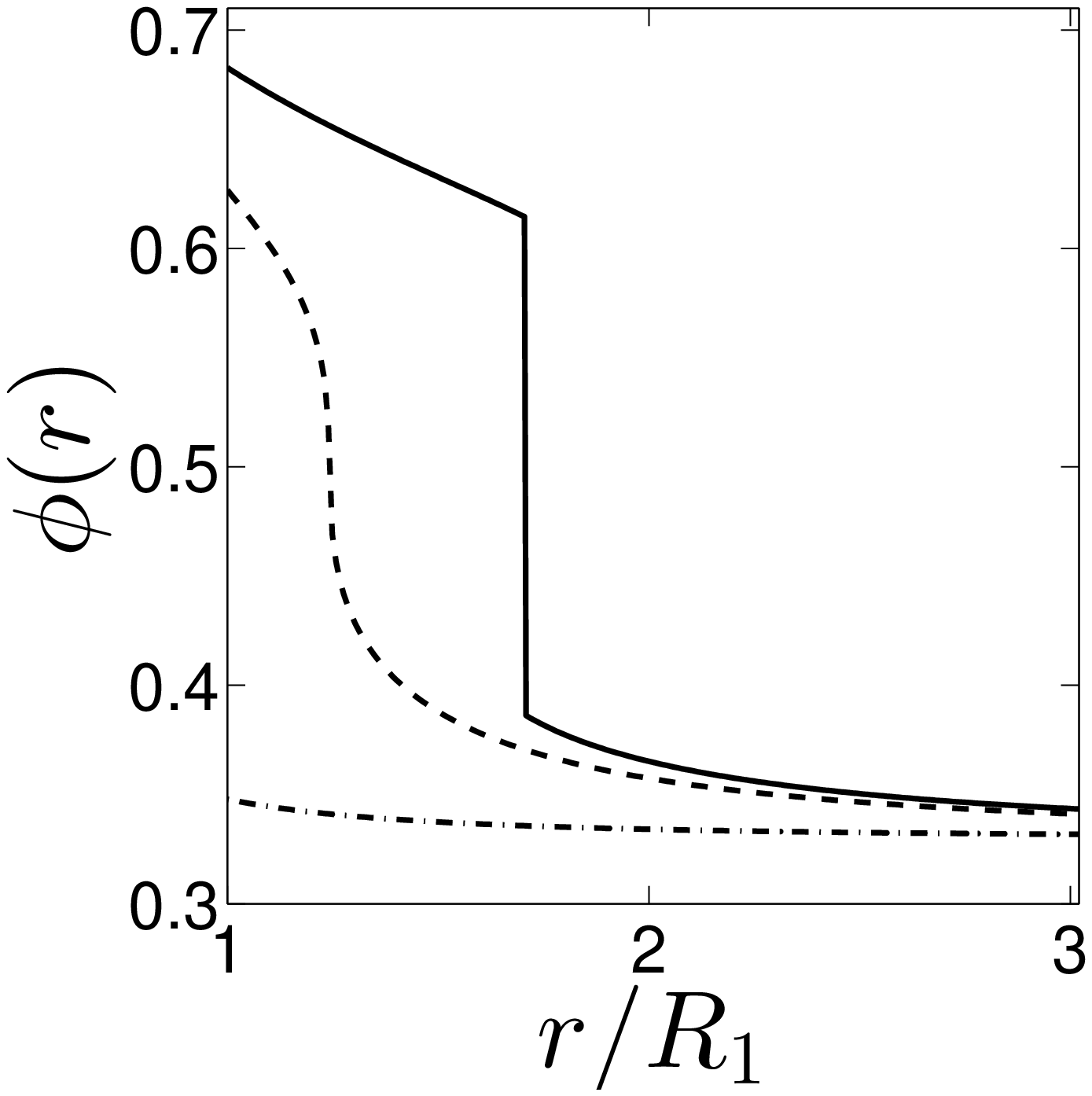}}\\
\subfigure[\label{fig_df}]{\includegraphics[keepaspectratio=true,width=0.232
\textwidth]{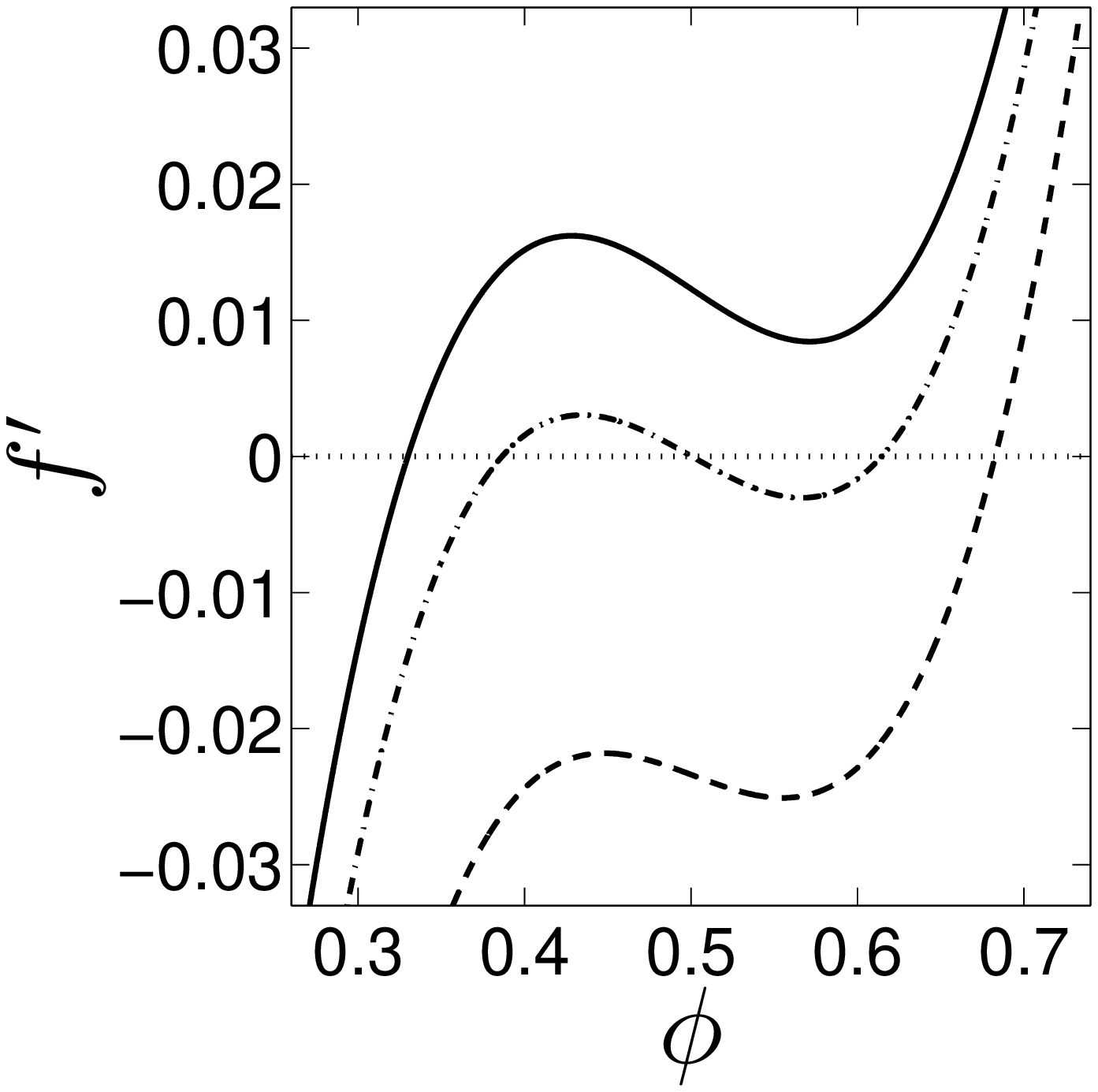}}
\subfigure[\label{fig_d2f}]{\includegraphics[keepaspectratio=true,width=0.232
\textwidth]{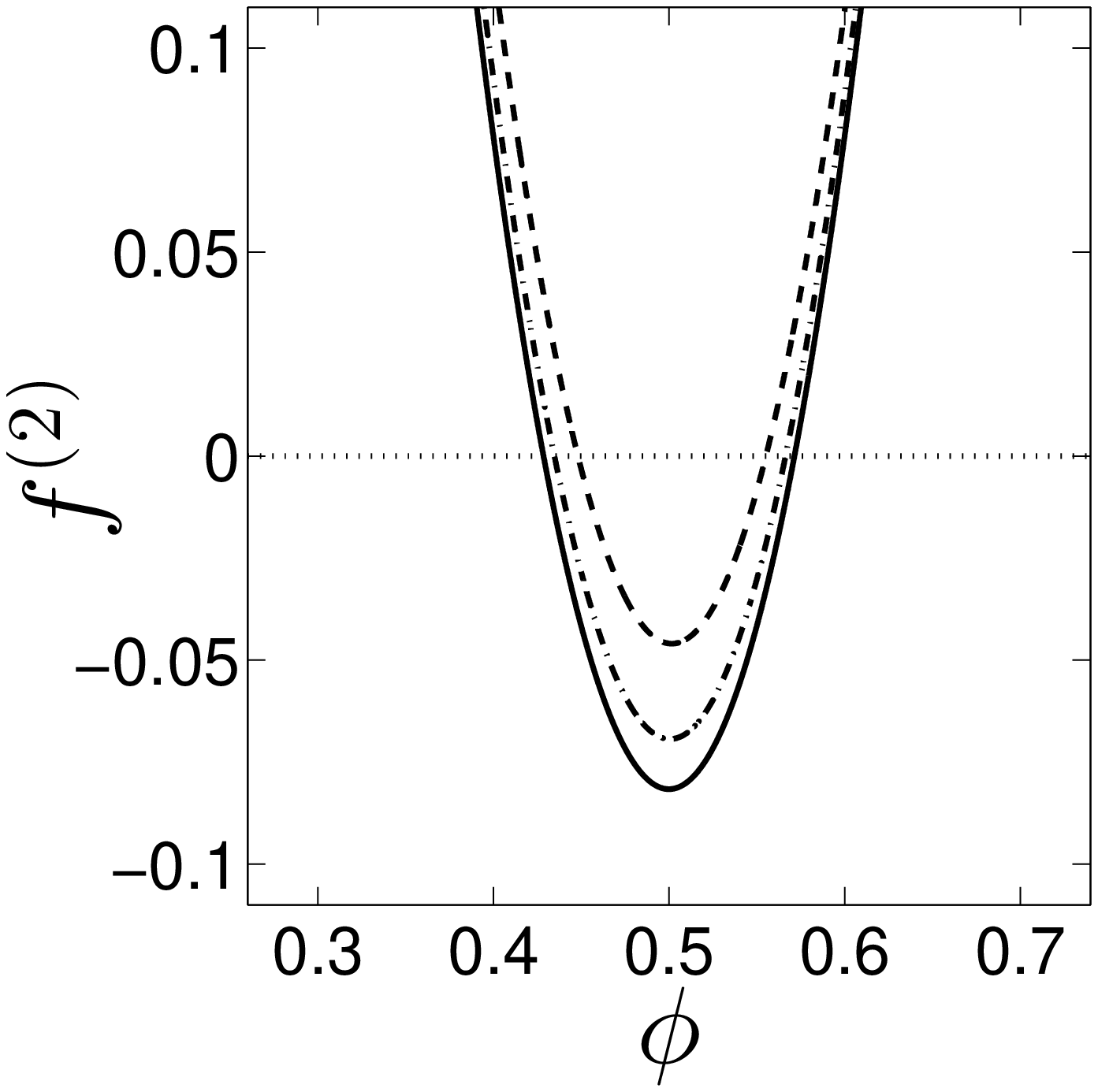}}
\caption{Free energy density $f(\phi,r)$ for an open cylinder system. 
(a,c,d) $f$, $f^{\prime}$, and $f^{(2)}$, respectively, versus $\phi$ at
distance $r=R_1$ (dashed line), $r_i$ (dash-doted line), and a large value (solid line)
for $\phi_0 = 0.33$, $T/T_c = 0.98$, and $\sigma=1.428\times 10^{-3}\,\mathrm{C/m^{2}}$.
Symbols in (a) mark minima for each curve. (b) $\phi(r)$ versus normalized distance $r$.
Solid line is data from (a). Dash-dotted line has same $\phi_0$ and $T$ as (a) but with
smaller $\sigma = 0.540\times 10^{-3}\,\mathrm{C/m^{2}}$. Dashed line has same $\phi_0$
and $\sigma$ as (a) but with larger $T/T_c = 0.995$.}%
\label{F_Fme}%
\end{center}
\end{figure}

Nonuniform electric fields impose a nonuniform ``pull'' on the liquid
mixture, manifesting as an $r$-dependent total free energy density
$f(\phi,r) = f_{\mathrm{m}}(\phi) + f_{\mathrm{e}}(\phi,r) - \phi\mu$. 
The behavior of $f$ can be conceptualized as a competition between mixing and
electrostatic energies. As $r\to\infty$, the electric field is weak, $f_{\mathrm{e}}\to 0$, and
$f\approx f_{\mathrm{m}}-\phi\mu$ governs liquid behavior. The solid line in Fig.~\ref{fig_f}
shows a typical example of $f(\phi,r)$ at a large value of $r$ using $\phi_0=0.33$,
$T/T_c=0.98$, and $\sigma=1.428\times 10^{-3}\,\mathrm{C/m^{2}}$ in an open cylinder
system. The minimum of $f(\phi,r)$, marked by a symbol, gives the value of $\phi(r)$ as
$r\to\infty$, which in this case is $0.33$. At the other distance extreme, $r=R_1$, the
electric field is the strongest, and the dashed line in Fig.~\ref{fig_f} shows the
resulting $f(\phi,r)$. Note the dramatic difference in 
the value of $\phi(r)$ when the value of $r$ is small ($R_1$) versus large.

%%%%%%%%%%%%%%%%%%%%%%%%%%%%%%%%%%%%%%%%%%%%%%%%%%%%%%%%%%%%%%%%
%%%%%%%%%%%%%%%%%%%%%%%%%%%%%%%%%%%%%%%%%%%%%%%%%%%%%%%%%%%%%%%%

By finding the minimum of $f$ for all values of $r$, it is possible to construct the full
concentration profile $\phi(r)$, where the solid line in Fig.~\ref{fig_phi_r} corresponds
to the data from Fig.~\ref{fig_f}. Whether or not a phase transition occurs in the equilibrium solution resides in how
the minimized $f(\phi,r)$ changes as $r$ varies between the two distance extremes.
Specifically, if there exists an $r = r_i$ where $R_1\le r_i \le R_2$ and $f(\phi,r_i)$
contains \emph{two} minima [see dash-dotted line in Fig.~\ref{fig_f}], then $r_i$ is
an interface between the two liquids. Figure~\ref{fig_phi_r} illustrates how the two
minima in $f(\phi,r_i)$ translate into a discontinuity at $\phi(r_i)$, thereby creating a
distinct boundary between the two phases.

A closer inspection of $f(\phi,r)$ at $r=r_i$ reveals important mathematical features
similar to those in $f_{\rm m}(\phi)$ discussed in the previous section. The
similarity is not surprising, since $f_{\mathrm{m}}(\phi)$ is a component of $f(\phi,r)$.
The dash-dotted lines in Figs.~\ref{fig_f},~\ref{fig_df}, and~\ref{fig_d2f} show that
$f(\phi,r_i)$ possesses two local minima we call $\phi_{iH}$ and $\phi_{iL}$, one local
maximum $\phi_u$, and two inflection points we call $\phi_{sH}$ and $\phi_{sL}$. In
addition, $f(\phi,r_i)$ can have critical behavior. We will demonstrate that all these
features at $r_i$ behave analogously to those in Fig.~\ref{fig_Fm_binodal} and show how to
use this information to construct the mixing-demixing phase diagram with an electric
field.

%========================================
% CONCENTRATION PROFILES & CONTROLLING THE INTERFACE
%========================================
\section{Composition Profiles $\phi(r)$ and Location of the
Interface}\label{Sect_Interface}

Not all applied fields induce liquid demixing, and based on our definition of a phase
transition, there are two possible causes. First, $r_i$ exists in ``virtual'' ($r_i<R_1$
or $r_i>R_2$) rather than ``real'' space, dash-dotted line in Fig.~\ref{fig_phi_r}.
Second, $f$ contains a single minimum for \emph{all} $r$, including $r_i$, dashed line in
Fig.~\ref{fig_phi_r}.

\begin{figure}[!tb]%
\begin{center}
\subfigure[\label{fig_phi_vs_r_varS}]{\includegraphics[keepaspectratio=true,width=0.232\textwidth]{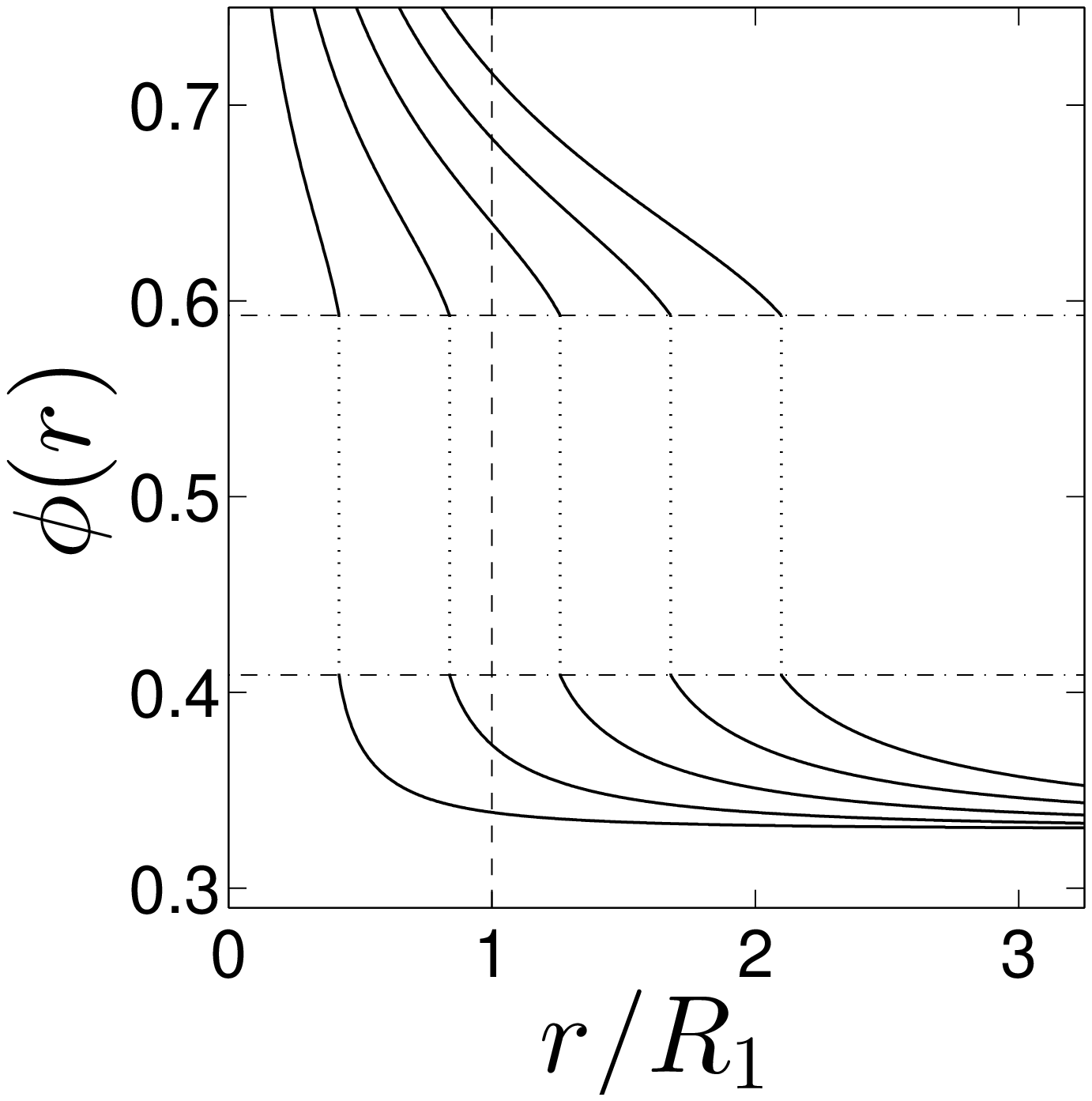}}
\subfigure[\label{fig_phi_vs_r_varS_collapse}]{\includegraphics[keepaspectratio=true,width=0.232\textwidth]{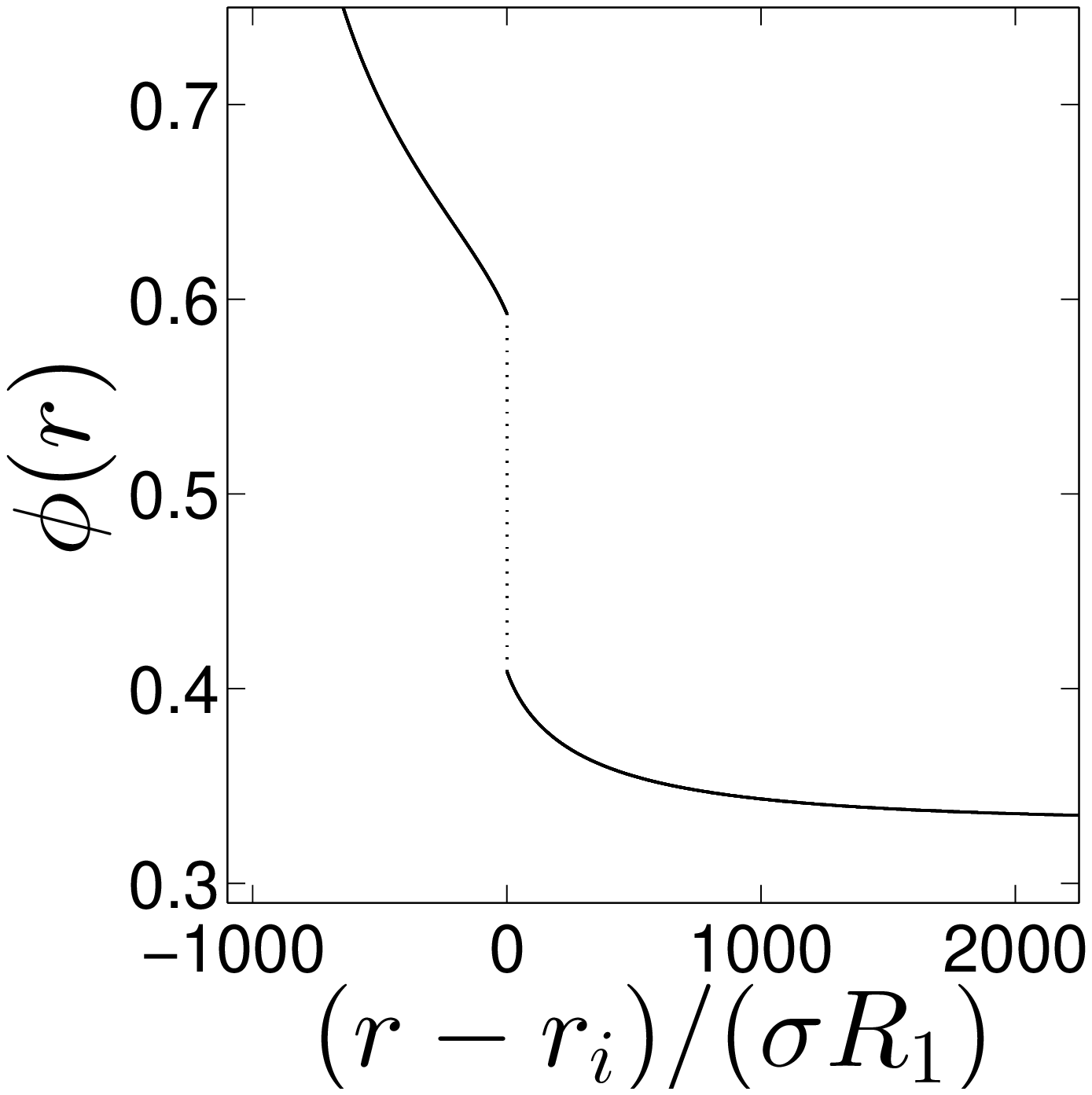}}\\
\subfigure[\label{fig_phi_vs_r_varP}]{\includegraphics[keepaspectratio=true,width=0.232\textwidth]{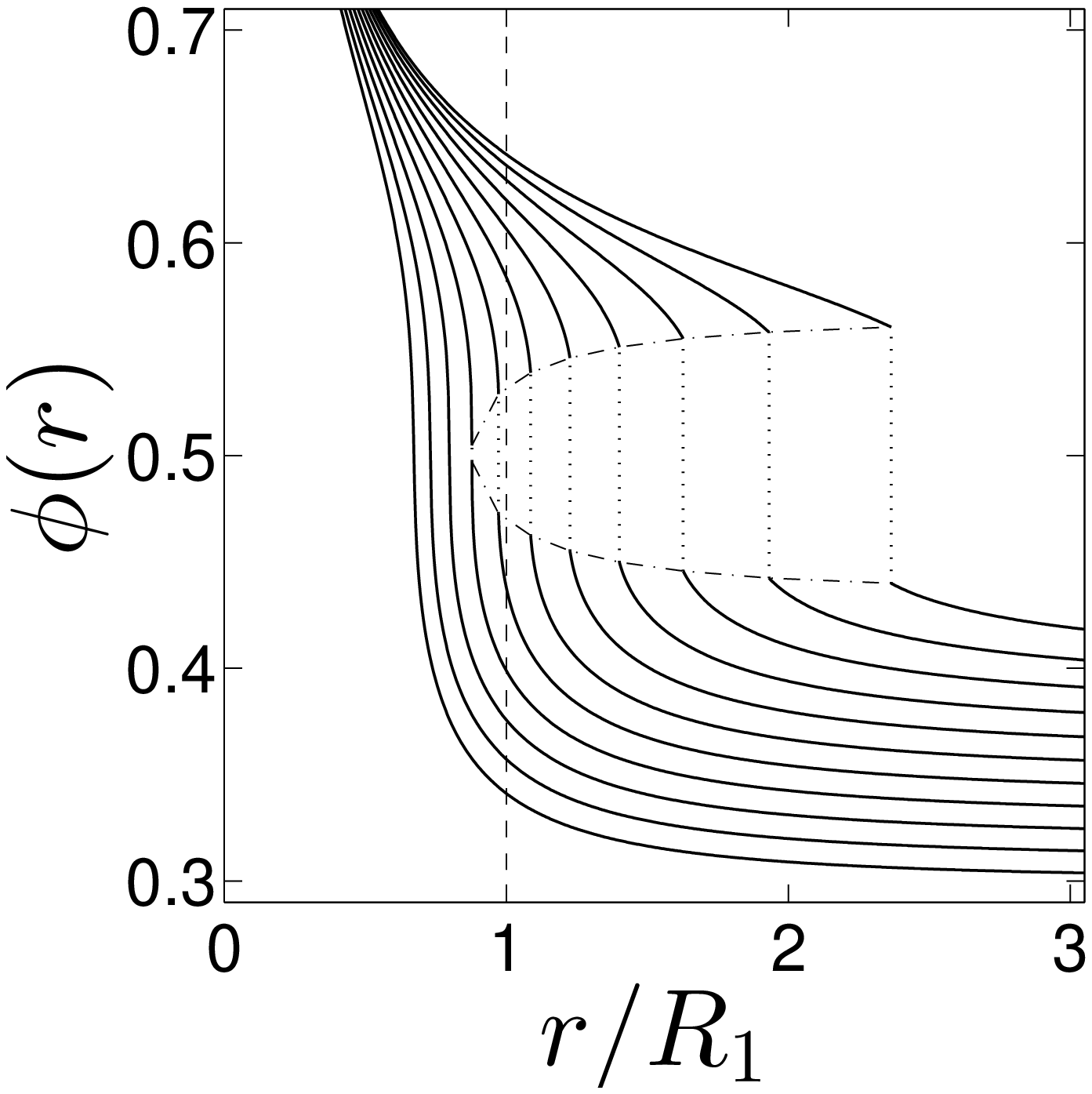}}
\subfigure[\label{fig_phi_vs_r_varT}]{\includegraphics[keepaspectratio=true,width=0.232\textwidth]{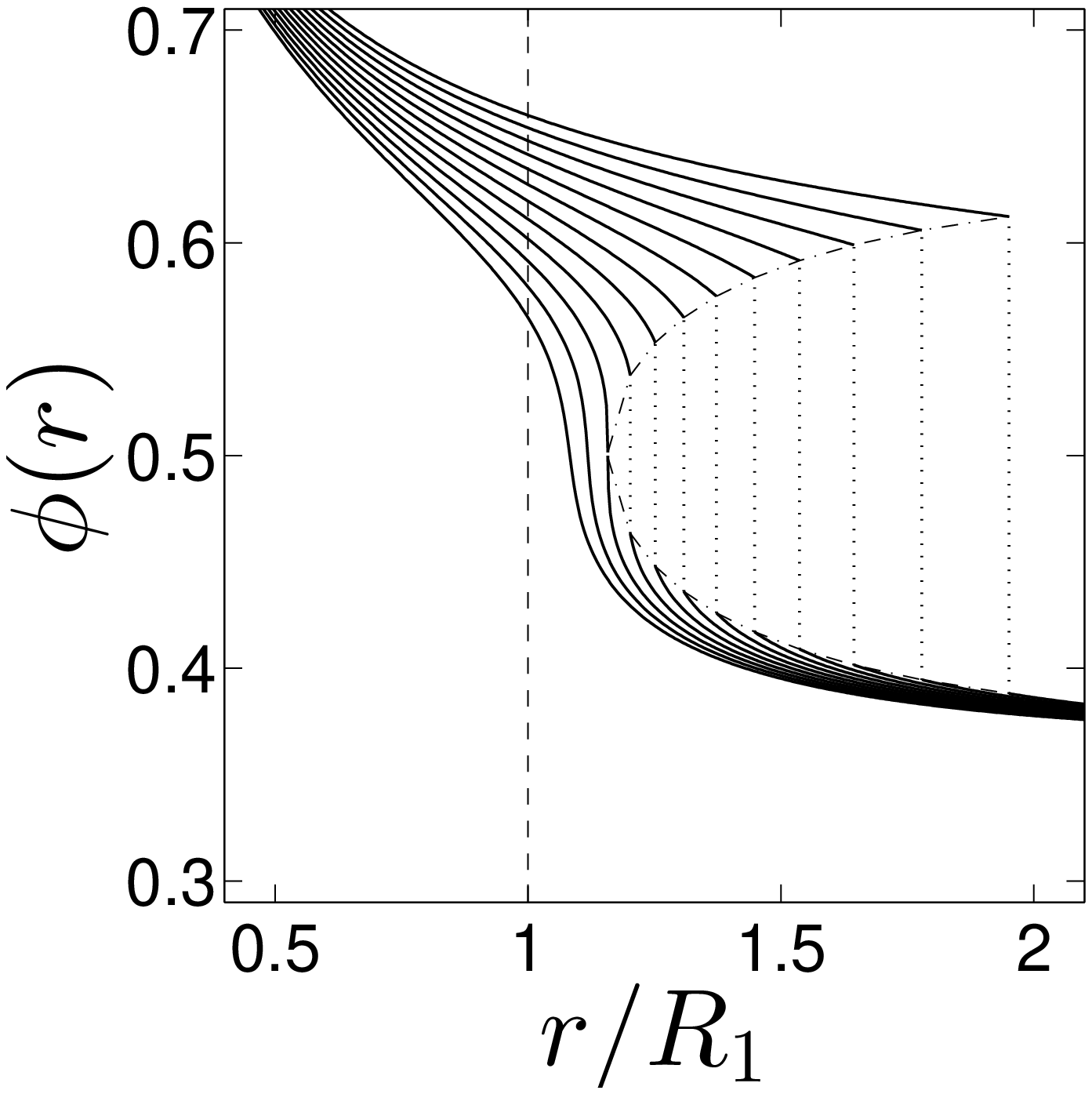}}
\caption{Variation of concentration profiles $\phi(r)$ for an open cylinder system. (a)
$\phi(r)$
 versus normalized distance $r$ for constant $\phi_0=0.33$, $T/T_c=0.985$, and varying
$\sigma=0.4\times 10^{-3}$ to $2.0\times 10^{-3}\,\mathrm{C/m^{2}}$ in $0.4\times 10^{-3}$
increments (lines, left to right). (b) Data in (a) collapses when plotted versus a
rescaled distance $(r-r_i)/(\sigma R_1)$. (c) $\phi(r)$ versus $r$ for constant
$T/T_c\approx 0.994$, $\sigma=1\times 10^{-3}\,\mathrm{C/m^{2}}$, and varying $\phi=0.3$
to $0.4$ in $0.01$ increments (lines, left to right). (d) $\phi(r)$ versus
$r$ for constant $\phi= 0.36$, $\sigma=1\times 10^{-3}\,\mathrm{C/m^{2}}$, and varying
$T/T_c \approx 0.982$ to $1$ in $0.0016$ increments (lines, right to left).}
\label{fig_phi_vs_r_open}%
\end{center}
\end{figure}

We begin with the first cause. 
For a constant $\phi_0$ and $T$, Fig.~\ref{fig_phi_r} shows that certain values of
$\sigma$ induce a transition, whereas others do not. In fact, there exists a transition
$\sigma_t$ that marks the lowest $\sigma$ necessary for liquid-liquid separation.
Figure~\ref{fig_phi_vs_r_varS} also shows how increasing $\sigma$ moves the interface
$r_i$ to larger $r$, using an open cylinder system as an example.
Noting that mathematical solutions exist for all $r$ (including those distances in
non-physical space), the vertical dashed line in 
Fig.~\ref{fig_phi_vs_r_varS} at $r=R_1$ marks the surface of the cylinder. To the right of
this line is real (physical) space, while to the left is the virtual space inside the
electrode (or not between the plates as defined in Fig.~\ref{fig_geom_W} for wedge
geometries). This observation inspires an alternative definition: the surface charge density 
$\sigma_t$ is the
$\sigma$ that places $r_i$ exactly at $R_1$. We stress that 
profiles $\phi(r)$ at constant $\phi_0$ and $T$ in open systems with varying values of 
$\sigma$ all collapse to a single curve when plotted versus a scaled distance
$(r-r_i)/(\sigma R_1)$, Fig.~\ref{fig_phi_vs_r_varS_collapse}.
%%%%%%%%%%%%%%%%%%%%%%%%%%%%%%%%%%%%%%%%%%%%%%%%%%%%%%%%%%%%%%%%%%
%%%%%%%%%%%%%%%%%%%%%%%%%%%%%%%%%%%%%%%%%%%%%%%%%%%%%%%%%%%%%%%%%%

Varying $\phi_0$ (holding $T$ and $\sigma$ constant) and $T$ (holding $\phi_0$ and
$\sigma$ constant) reveals the second cause for no phase separation, illustrated in
Fig.~\ref{fig_phi_vs_r_varP} and~\ref{fig_phi_vs_r_varT}, respectively. In these cases,
both the interface location $r_i$ \emph{and} the size of the discontinuity change.
Importantly, the discontinuity can even vanish, as the high and low concentrations
$\phi_{iH}$ and $\phi_{iL}$ at the interface merge to the same value at certain $\phi_0$
or $T$. Notice the remarkable similarity between the behavior of the discontinuity at
$r_i$, dash-dotted lines in Figs.~\ref{fig_phi_vs_r_varP} and~\ref{fig_phi_vs_r_varT}, and
the binodal curve, Fig.~\ref{fig_Fm_binodal}.

\begin{figure}[!tb]%
\begin{center}
\subfigure[\label{fig_sigma_vs_ri}]{\includegraphics[keepaspectratio=true,width=0.232\textwidth]{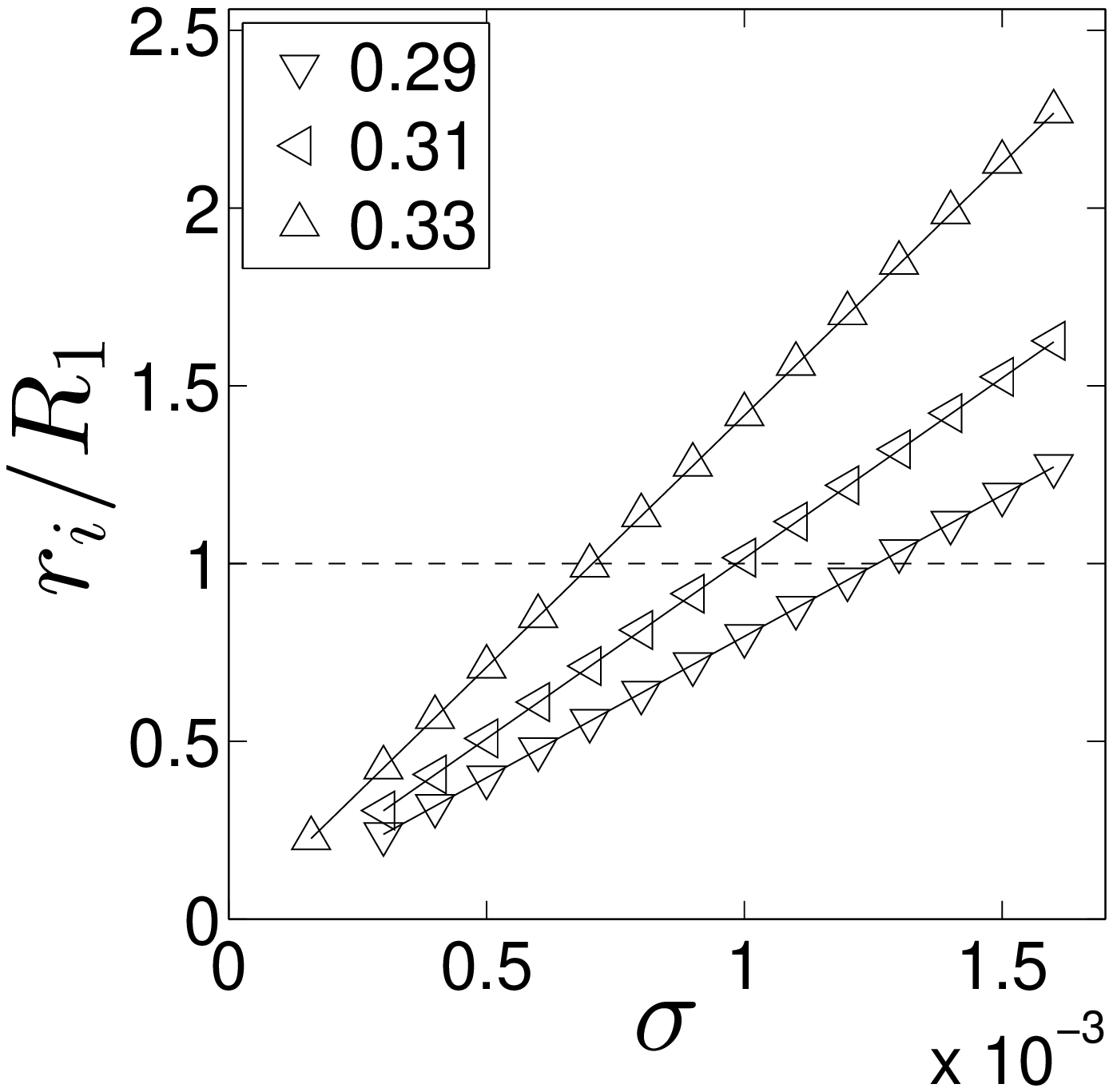}}
\subfigure[\label{fig_phi0_vs_ri}]{\includegraphics[keepaspectratio=true,width=0.232\textwidth]{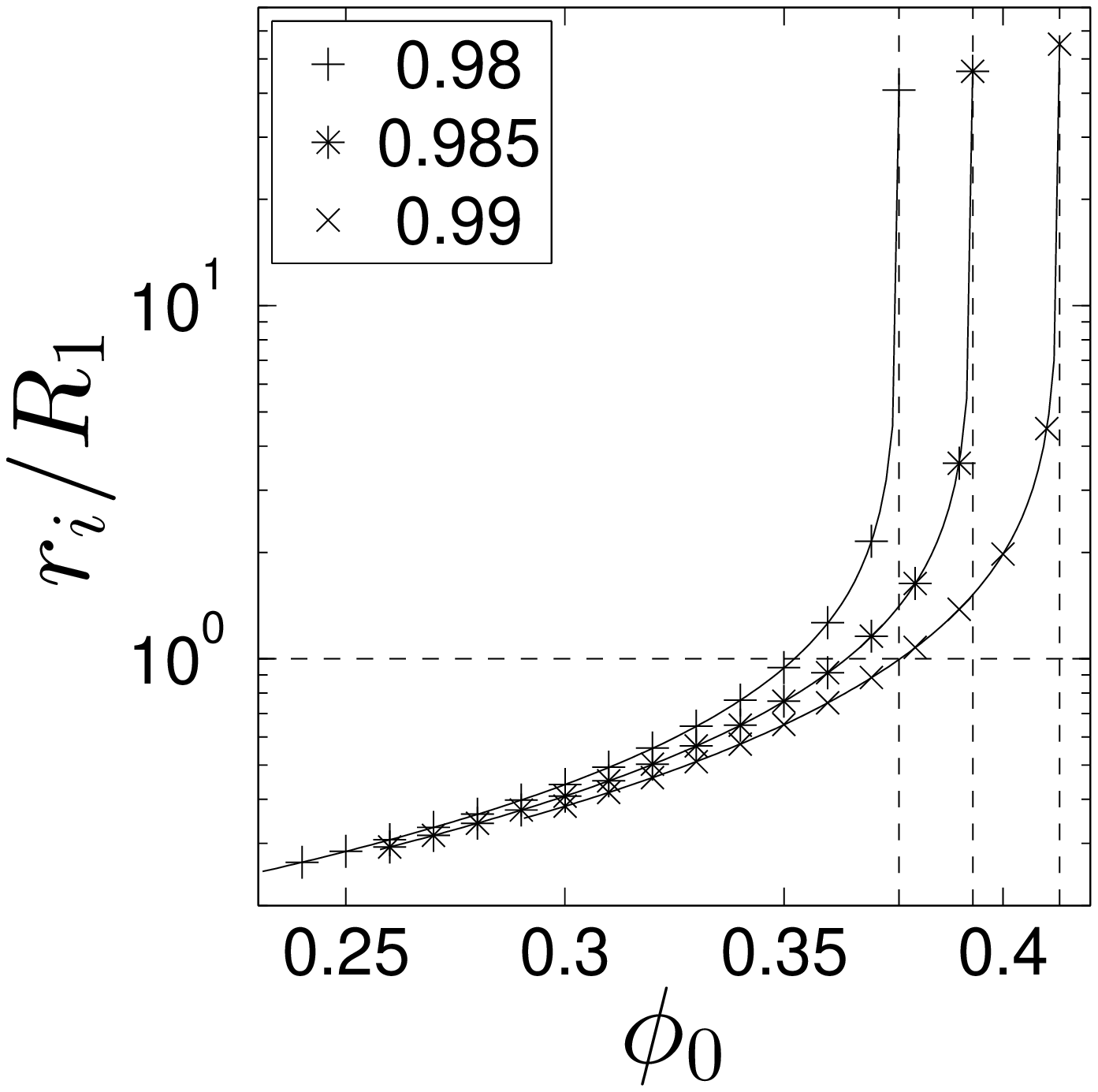}}\\
\subfigure[\label{fig_T_vs_ri}]{\includegraphics[keepaspectratio=true,width=0.232\textwidth]{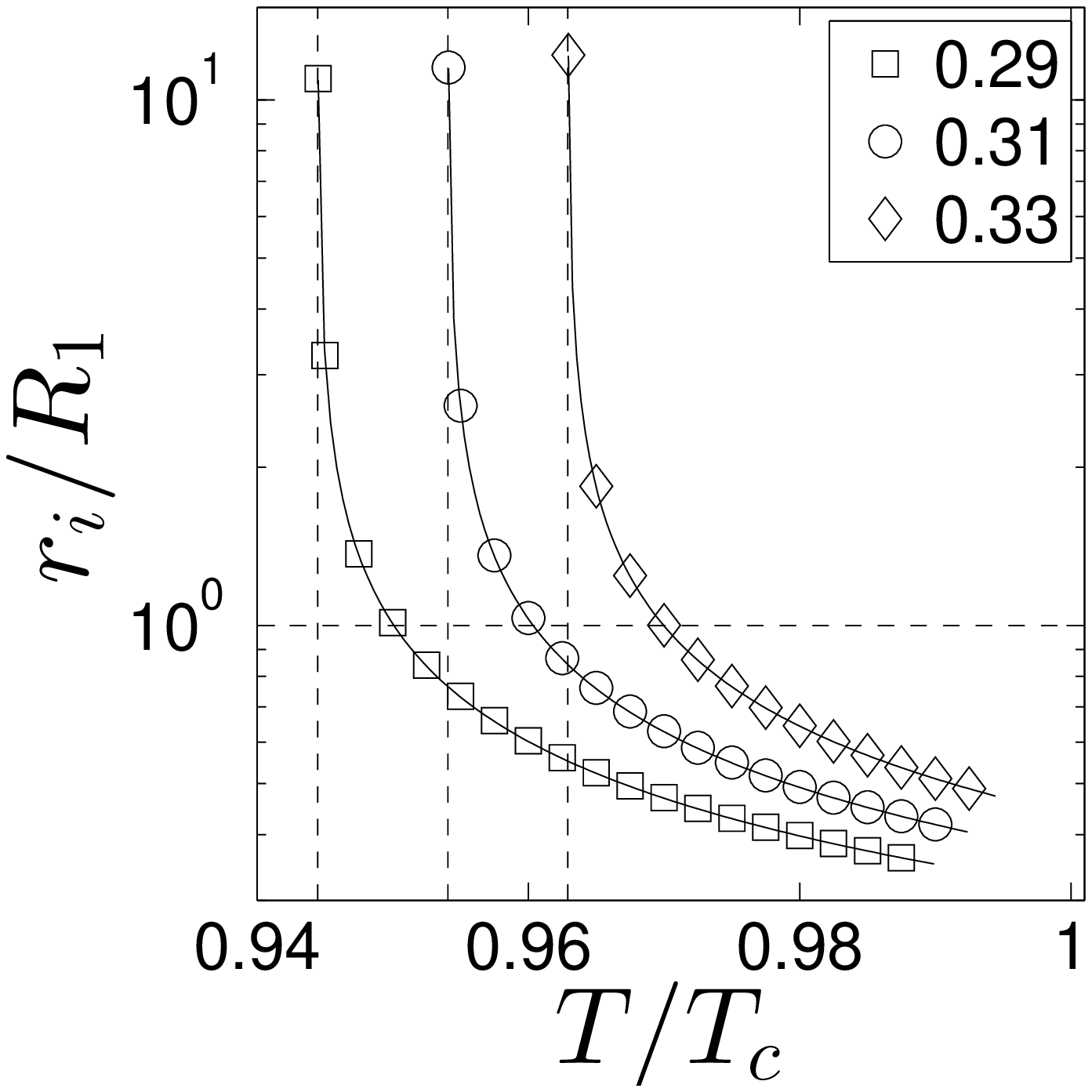}}
\subfigure[\label{fig_ri_collapse}]{\includegraphics[keepaspectratio=true,width=0.232\textwidth]{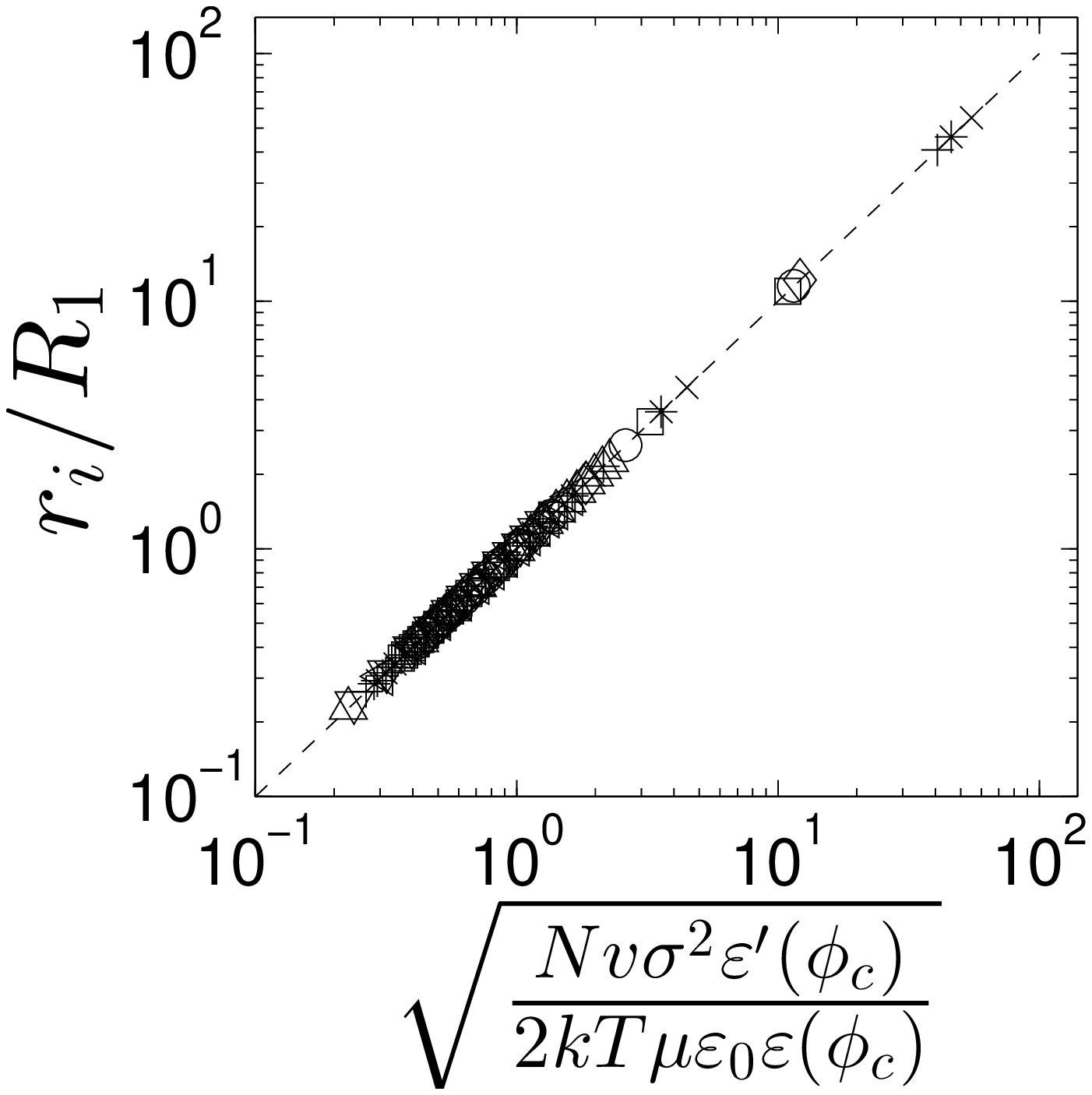}}
\caption{Controlling the location of the interface $r_i$ for an open cylinder system. 
(a) Normalized $r_i$ versus $\sigma$ [$\mathrm{C/m^2}$] for various values of $\phi_0$ where $T/T_c
= 0.975$. (b) Normalized $r_i$ versus $\phi_0$ for various values of $T/T_c$ where $\sigma
\approx 5.4\times 10^{-4}\,\mathrm{C/m^{2}}$. (c) Normalized $r_i$ versus $T/T_c$ for
various $\phi_0$ where $\sigma \approx 5.4\times 10^{-4}\,\mathrm{C/m^{2}}$. (d) Collapse
of all data from (a-c) when $r_i$ is plotted against eq.~\ref{eq_interface}. In (b,c),
vertical dashed lines mark the binodal for the given values of $\phi_0$ and $T$.
}
\label{fig_ri}%
\end{center}
\end{figure}

%========================================
% CONTROLLING THE INTERFACE
%========================================
The location of the interface, once it exists, is controlled by 
$\phi_0$, $T$, $\sigma$ and $R_2$. In general, $r_i$ increases with increasing $\sigma$
[Fig.~\ref{fig_phi_vs_r_varS} and~\ref{fig_sigma_vs_ri}], increasing $\phi_0$
[Fig.~\ref{fig_phi_vs_r_varP} and~\ref{fig_phi0_vs_ri}], decreasing $T$
[Fig.~\ref{fig_phi_vs_r_varT} and~\ref{fig_T_vs_ri}], and increasing $R_2$ [discussed in
Sect.~\ref{Section_closed}, Fig.~\ref{fig_phi_vs_r_varR2_closed}).
%%%%%%%%%%%%%%%%%%%%%%%%%%%%%%%%%%%%%%%%%%%%%%%%%%%%%%%%%%%%%%%%%%
%%%%%%%%%%%%%%%%%%%%%%%%%%%%%%%%%%%%%%%%%%%%%%%%%%%%%%%%%%%%%%%
Besides solving the full $\phi(r)$ profile, a quicker method for determining the location
of the interface $r_i$ consists of solving these three
equations~\cite{Samin2009,Samin2011}
\bea
f^{\prime}_{\mathrm{m}}(\phi_{iH}) - f^{\prime}_{\mathrm{e}}(\phi_{iH},r_i) - \mu = 0\nonumber\\
f^{\prime}_{\mathrm{m}}(\phi_{iL}) - f^{\prime}_{\mathrm{e}}(\phi_{iL},r_i) - \mu = 0\nonumber\\
f_{\mathrm{m}}(\phi_{iH}) \pm f_{\mathrm{e}}(\phi_{iH},r_i) - \phi_{iH}\mu = \nonumber\\
 f_{\mathrm{m}}(\phi_{iL}) \pm f_{\mathrm{e}}(\phi_{iL},r_i) - \phi_{iL}\mu
\label{eq_interface3}
\eea
for three unknowns: $r_i$ and the high and low concentrations $\phi_{iH}$ and $\phi_{iL}$,
respectively, at $r_i$.
The plus (minus) sign in the third equation is for constant charge
(potential) boundary conditions. The first two equations find extrema points and are
simply eq.~\ref{eq_EL_cylSph} or~\ref{eq_EL_wedge}, depending on system geometry. The
third equation ensures that the free energy for the high concentration $\phi_{iH}$ is as
favorable as the low concentration $\phi_{iL}$. 

An even simpler method for finding $r_i$ consists in recalling that there exists a third
solution to $f^{\prime}$---the local maxima $\phi_u$, Fig.~\ref{fig_df}. For cylindrical
($n=1$) and spherical ($n=2$) geometries, the explicit equation for $f'=0$ is 
\bea
0 &=& 4\left(1-\frac{T_c}{T}\right)\left(\phi_u-\phi_c\right) +
\frac{16}{3}\left(\phi_u-\phi_c\right)^3 \nonumber\\&& - \frac{Nv}{2kT\VE_0}
\left(\frac{\sigma R_1^n}{r_i^n} \right)^2 \frac{\VE^{\prime}(\phi_u)}{\VE(\phi_u)^2} -
\mu \label{eq_interface_long}
\eea
If $\phi_u$ is known, then $r_i$ can, in principle, be deduced from experimental
parameters ($\phi_0$, $T$, etc.). For now, we will borrow ideas from the binodal curve and
make the assumption $\phi_u=\phi_c=0.5$, but will see later that $\phi_u$ indeed
approximately equals $\phi_c$ under many conditions. Rearranging
eq.~\ref{eq_interface_long}, we now have the useful relation
\be
\left(\frac{\sigma R_1^n}{r_i^n} \right)^2 =   - \mu \frac{2kT\VE_0}{Nv} 
\frac{\VE(\phi_c)^2}{\VE^{\prime}(\phi_c)}
\label{eq_interface}
\ee
In open systems, this equation is further simplified by substituting $\mu = \mu_0(\phi_0)
= f^{\prime}_{\mathrm{m}}(\phi_0,T)$. 
The lines in Figs.~\ref{fig_sigma_vs_ri},~\ref{fig_phi0_vs_ri}, and~\ref{fig_T_vs_ri} use
eq.~\ref{eq_interface} to solve $r_i$ in an open cylinder system, and reveal an excellent
agreement to the solutions from eqs.~\ref{eq_interface3} (symbols).
Figure~\ref{fig_ri_collapse} combines all data from
Figs.~\ref{fig_sigma_vs_ri},~\ref{fig_phi0_vs_ri}, and~\ref{fig_T_vs_ri}, revealing that
the agreement spans many orders of magnitude.

The analogous equation for finding $r_i$ in a wedge geometry is
\be
\left(\frac{V}{\theta r_i} \right)^2 = - \frac{2kT\mu}{Nv\VE_0 \VE^{\prime}(\phi_c)}
\label{eq_interface_wedge}
\ee

%========================================
% STABILITY DIAGRAMS and ELECTROSTATIC BINODAL
%========================================
\section{Stability Diagram and Electrostatic Binodal}\label{Sect_eleBin}

\begin{figure}[!th]%
\begin{center}
\subfigure[\label{fig_stabilityCurve}]{\includegraphics[keepaspectratio=true,width=0.232\textwidth]{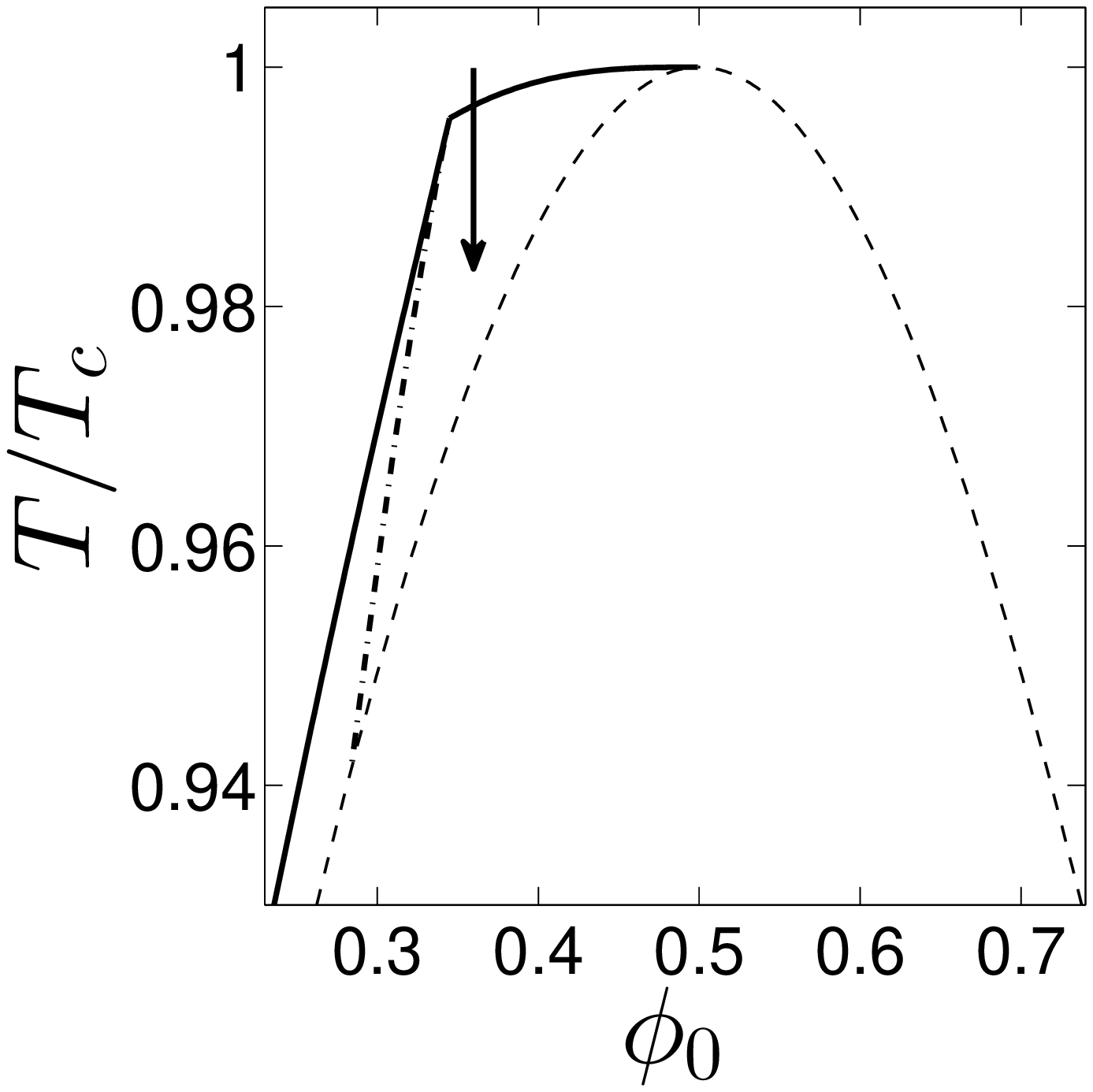}}
\subfigure[\label{fig_eleBinodal}]{\includegraphics[keepaspectratio=true,width=0.232\textwidth]{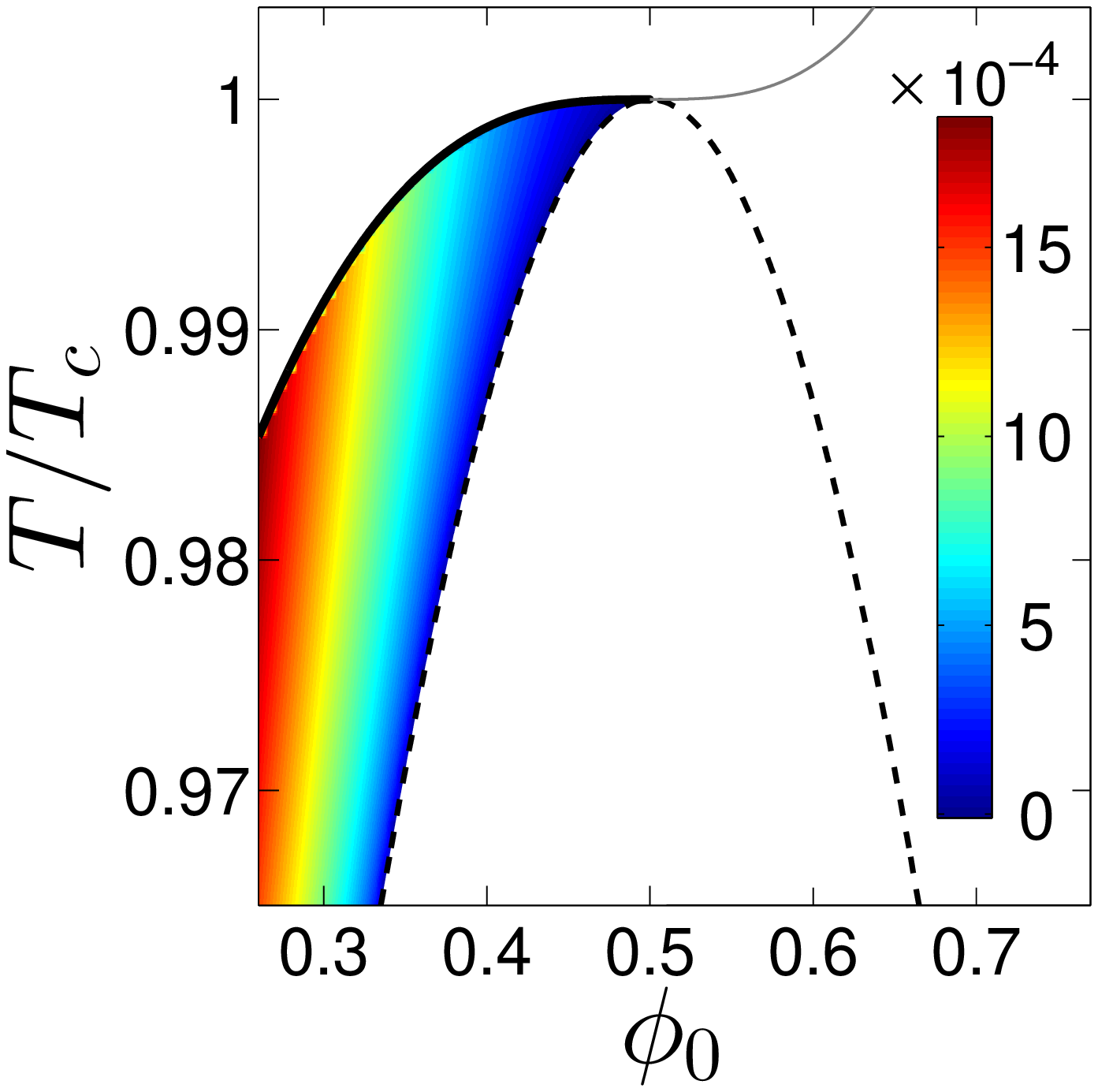}}\\
\subfigure[\label{fig_phiu}]{\includegraphics[keepaspectratio=true,width=0.232\textwidth]{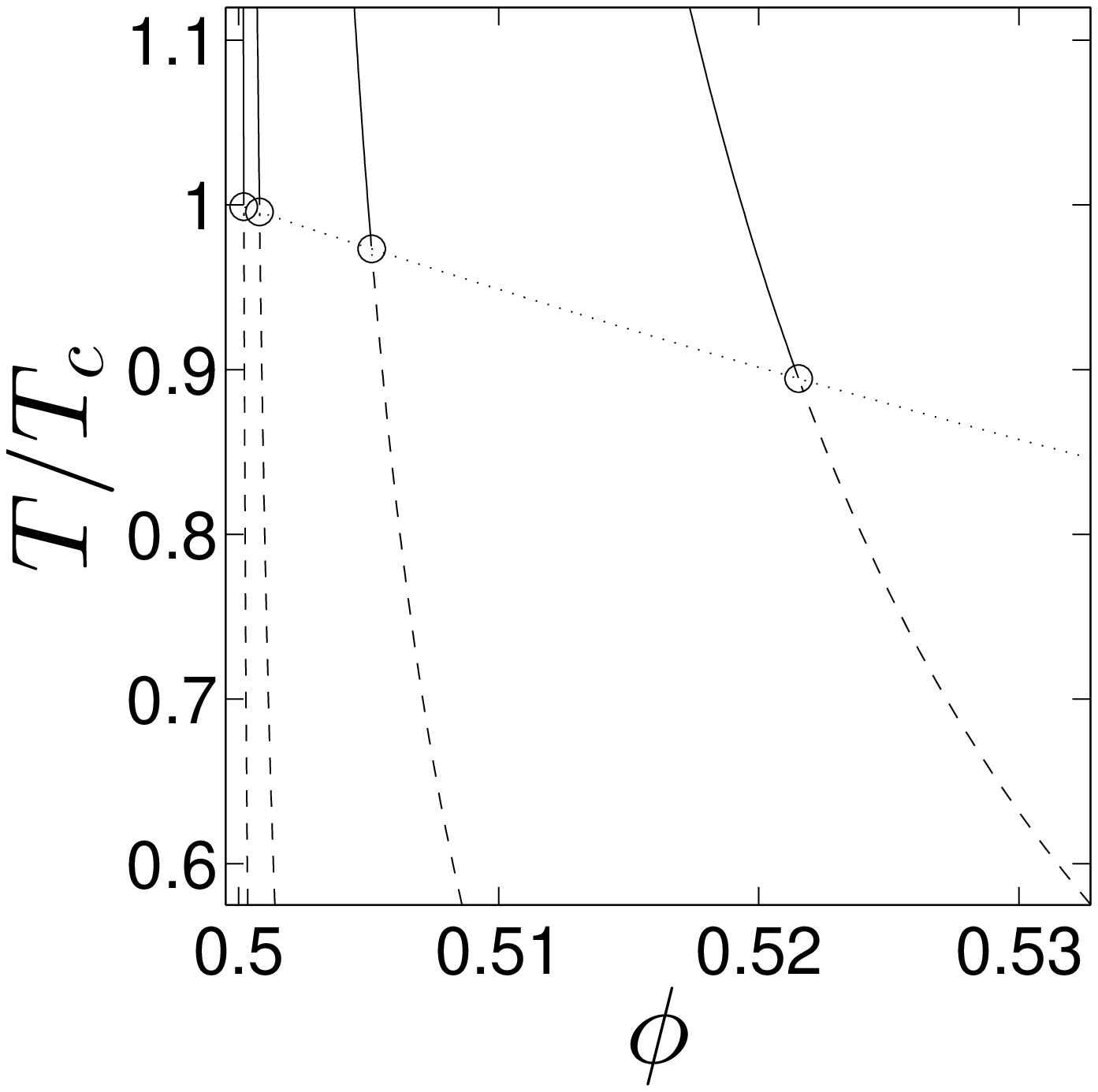}}
\subfigure[\label{fig_Scrit}]{\includegraphics[keepaspectratio=true,width=0.232\textwidth]{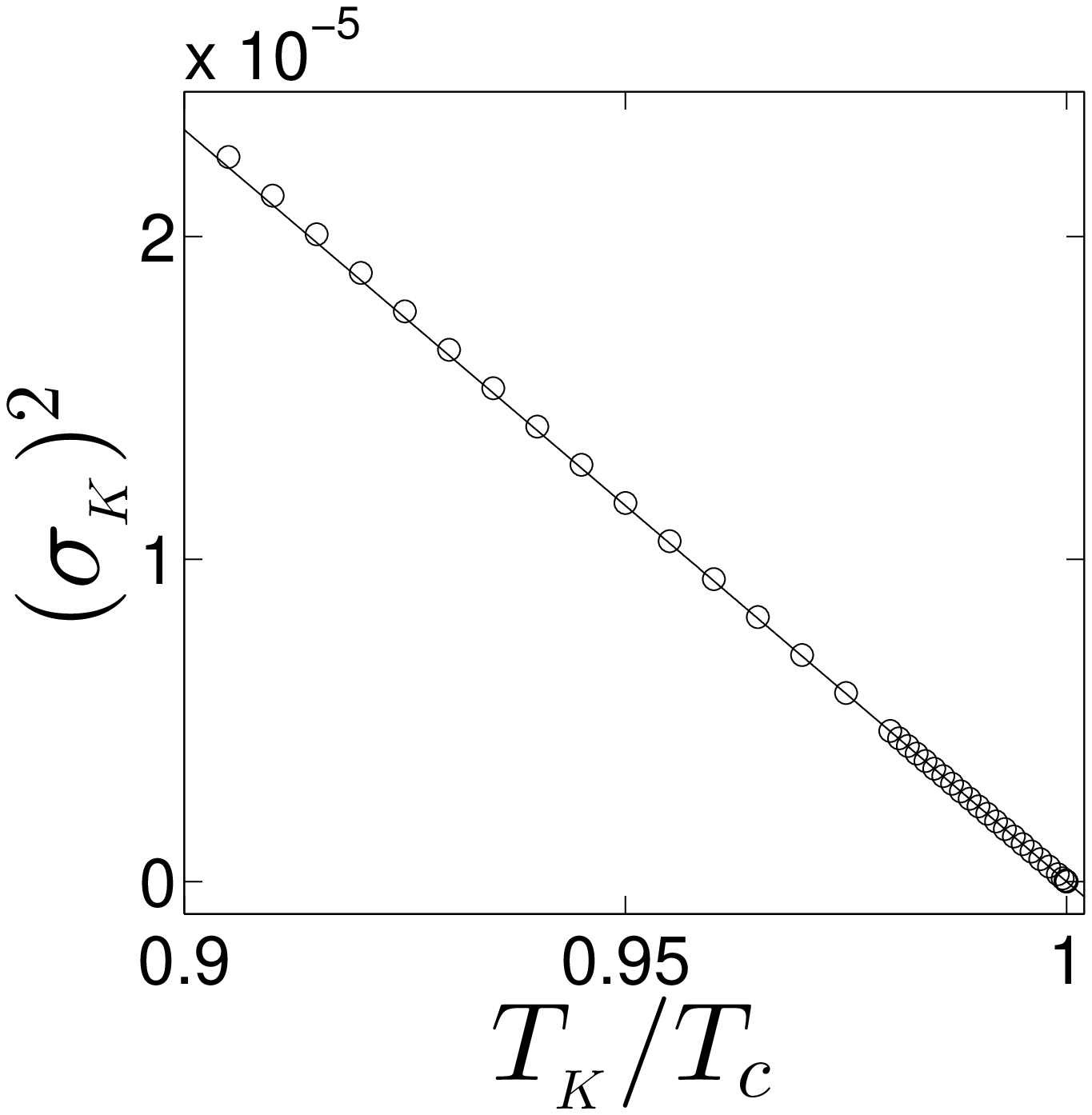}}\\
\subfigure[\label{fig_epsilon}]{\includegraphics[keepaspectratio=true,width=0.232\textwidth]{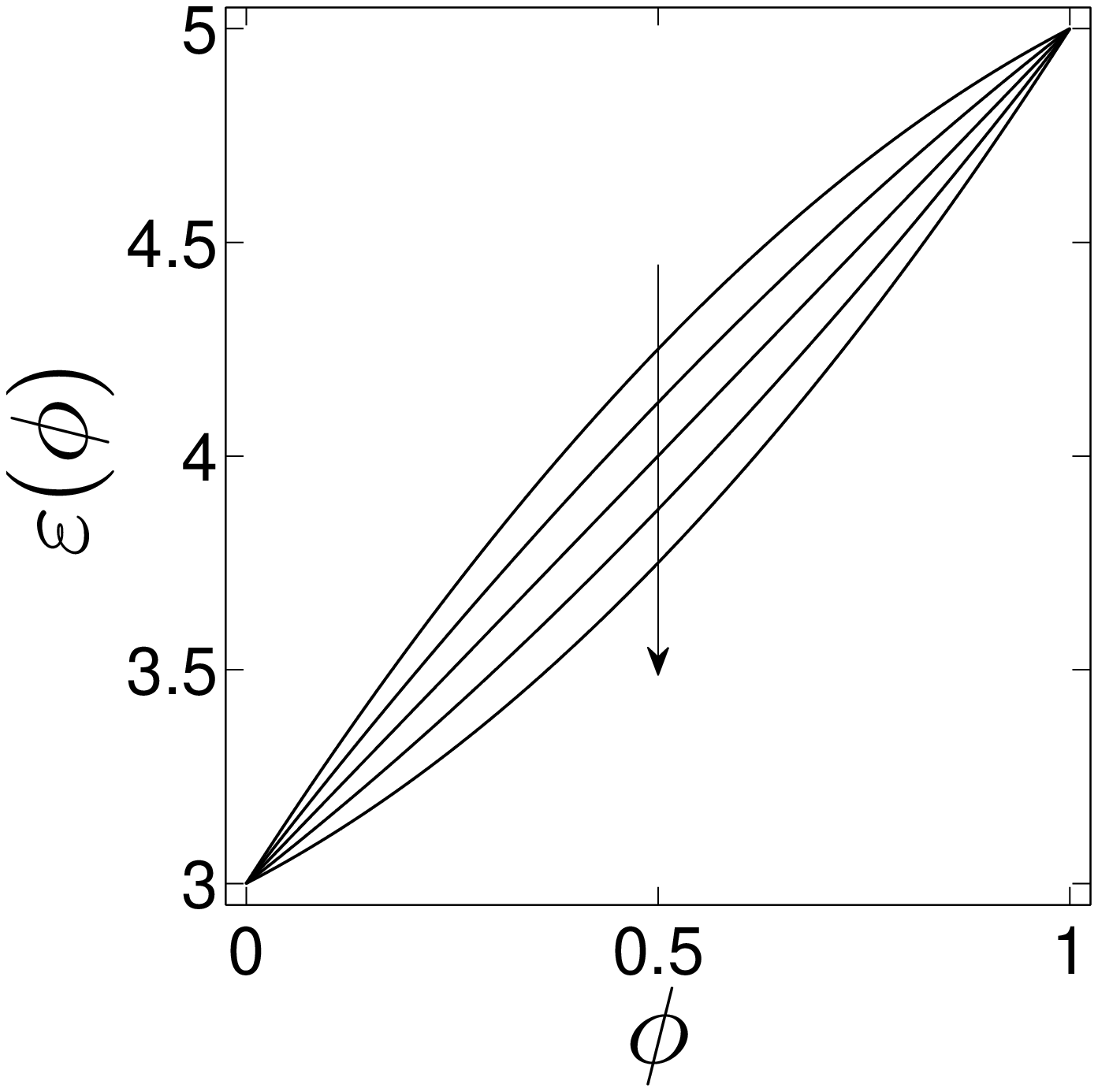}}
\subfigure[\label{fig_eleBinodal_wedge}]{\includegraphics[keepaspectratio=true,width=0.232\textwidth]{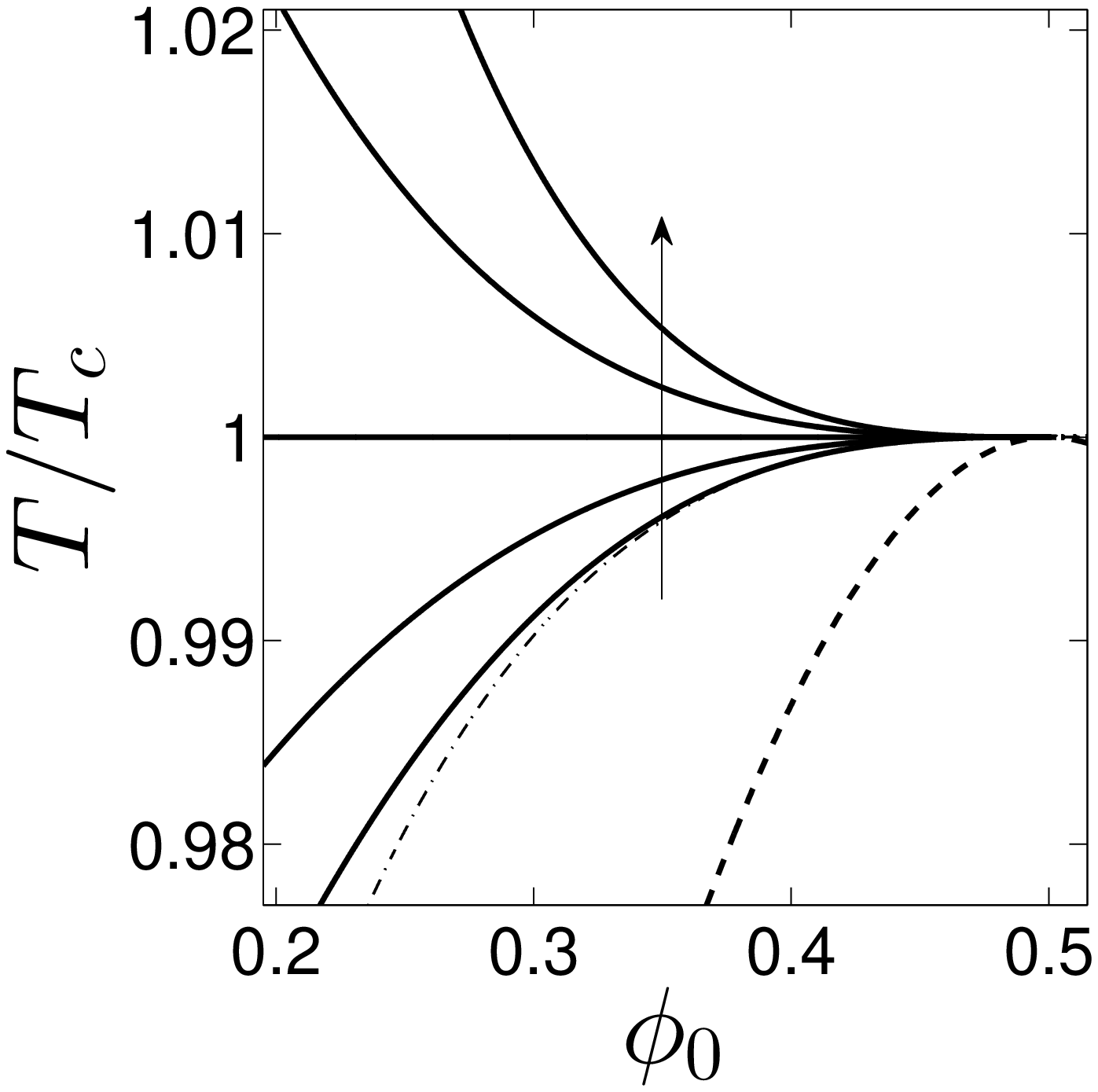}}
\caption{Electrostatic binodal in open systems. (a) Stability diagram (solid line) 
and electrostatic spinodal (dash-dotted line) in $\phi_0-T$ plane for $\sigma =
1\times 10^{-3}\,\mathrm{C/m^{2}}$. Dashed line is binodal curve. Path of arrow marks
location of data in Fig.~\ref{fig_phi_vs_r_varT}. (b) Overlay of many stability diagrams,
where color indicates transition $\sigma_t$ [$\mathrm{C/m^2}$]. Solid line is 
electrostatic binodal from eq.~\ref{eq_eleBinodal}, where thick and thin marks show where
values of $\sigma_t = \sigma_K$ are real and imaginary, respectively. (c) Solutions
$\phi_i$ (solid lines), $\phi_{iC}$ (symbols), and $\phi_{u}$ (dashed lines) to
$f^{(3)}=0$ versus $T$ where $\sigma = 0.5\times 10^{-3}$, $1\times 10^{-3}$,
$2.5\times 10^{-3}$, and $5\times 10^{-3}\,\mathrm{C/m^{2}}$ (lines, left to right).
Dotted curve shows $\phi_{iC}$ for all $\sigma$. (d) Critical $\sigma_K$ [$\mathrm{C/m^2}$] along 
electrostatic binodal versus $T_K$. (e)
Quadratic forms of $\VE(\phi)$ versus 
$\phi$, where the second derivative $\VE^{(2)}(\phi)=-2$, $-1$, $0$, $1$, and $2$ (arrow).
(f) Solid lines show electrostatic binodal for an open wedge system using
$\VE^{(2)}(\phi)$ from (e) (arrow). Dash-dotted line shows electrostatic binodal using
Flory-Huggins theory for $f_{\mathrm{m}}$ and $\VE^{(2)}(\phi)=-2$. Dashed line is 
binodal curve.} 
\label{F_Dphi_Dt}%
\end{center}
\end{figure}

If an electric field can cause phase separation in a region of $\phi_0-T$ space
\emph{above} the binodal curve, a natural question arises: what is the new stability
diagram for a particular value of surface charge density $\sigma$? 
This can be constructed by holding $\sigma$ constant
and probing $\phi_0-T$ space for liquid-liquid demixing. Since the electric field breaks
the symmetry of the free energy with respect to composition ($\phi_0\to 1-\phi_0$), the
stability diagram is asymmetric with respect to $\phi_0-\phi_c$.
Figure~\ref{fig_stabilityCurve} compares a typical stability curve for an open cylindrical
system, solid line, to the binodal curve, dashed line. Clearly, nonuniform fields can
produce large changes to the phase diagram.
%%%%%%%%%%%%%%%%%%%%%%%%%%%%%%%%%%%%%%%%%%%%%%%%%%%%%%%%%%%%%%%%
%%%%%%%%%%%%%%%%%%%%%%%%%%%%%%%%%%%%%%%%%%%%%%%%%%%%%%%%%%%%%%%%

Figure~\ref{fig_eleBinodal} shows the superposition of stability diagrams from a wide
range of $\sigma$ in an open cylindrical system, where the color indicates the transition
$\sigma_t$ for each point $(\phi_0,T)$. (Points beneath the binodal curve are omitted since
phase separation occurs there without an electric field.) Figure~\ref{fig_eleBinodal}
clearly illustrates two distinct regions in the $\phi_0-T$ plane. In the ``demixed''
region, there exists a $\sigma_t$ for each ($\phi_0$,$T$) such that any
$\sigma\ge\sigma_t$ results in liquid demixing. In the ``mixed'' region, there does not
exist \emph{any} $\sigma$ that results in liquid demixing. Notice how the mixed region
extends well below $T_c$, indicating that simply setting $T<T_c$ is not sufficient for
producing a phase transition with an electric field. We will call the curve that divides
these two regions the ``electrostatic binodal''.

To derive the electrostatic binodal, we draw inspiration from the ``regular'' binodal
curve. The convergence of the interface concentrations $\phi_{iH}$ and $\phi_{iL}$ in
Figs.~\ref{fig_phi_vs_r_varP} and~\ref{fig_phi_vs_r_varT} suggest the existence of a
critical point at $r_i$. If there exists a critical point at $r_i$, then the two minima
$\phi_{iH}$ and $\phi_{iL}$, the local maximum $\phi_u$, and the two inflection points
$\phi_{sH}$ and $\phi_{sL}$ converge to a single point $\phi_{iC}$, resulting in
$f^{\prime}(r_i)=f^{(2)}(r_i)=f^{(3)}(r_i)=0$ and $f^{(4)}(r_i)>0$. We call the
coordinates in the $\phi_0-T$ plane that produce a critical point at $r_i$ the critical
$\phi_K$ and critical $T_K$.

We now show one method for finding $\phi_K$ and $T_K$, using an open cylinder system as an
example and begin with $f^{(2)} = 0$:
\bea
2\left(1-\frac{T_c}{T_K}\right) +  8\left(\phi_{iC}-\phi_c\right)^2 + \nonumber\\
\frac{Nv}{2kT_K\VE_0}\frac{[\VE^{\prime}(\phi_{iC})]^2}{\VE(\phi_{iC})^3}\left(\frac{\sigma
R_1}{r_i} \right)^2  &=& 0\label{equ_d2fdp2}
\eea
The derivation of eq.~\ref{eq_interface} depends on finding $\phi$ that satisfies
$f^{\prime}=0$, but does not specify $\phi$ as a local maximum or minimum. In particular,
$\phi_{iC}$ also satisfies eq.~\ref{eq_interface}. We therefore substitute
eq.~\ref{eq_interface} for $(\sigma R_1/r_i)^2$ into eq.~\ref{equ_d2fdp2}, use $\mu =
\mu_0(\phi_K) = f^{\prime}_{\mathrm{m}}(\phi_K,T_K)$ for an open system, and rearrange to
obtain
\bea
\frac{T_K}{T_c} & = & \left[ \frac{ 4\left(\phi_{iC}-\phi_c\right)^2 -
\frac{8}{3}\left(\phi_K-\phi_c\right)^3
\frac{\VE^{\prime}(\phi_{iC})}{\VE(\phi_{iC})}}{1-2\left(\phi_K-\phi_c\right)\frac{\VE^{
\prime}(\phi_{iC})}{\VE(\phi_{iC})} } +1 \right]^{-1} \nonumber\\&&{\rm open~cylinder}
\label{eq_eleBinodal_full}
\eea
Notice that as $\VE^{\prime}\to 0$, we recover the solution to $f_{\mathrm{m}}^{(2)}=0$,
and $T_K\to T_c$ when $\phi_i$ equals the critical composition $\phi_c$.

Proceeding, $\phi_{iC}$ must also satisfy $f^{(3)} = 0$ at $r_i$. Figure~\ref{fig_phiu}
shows the solutions to $f^{(3)} = 0$ 
for a wide range of $T$, where the curves from left to right are low to high $\sigma$. The
values of $\phi_i$ (solid lines), $\phi_{iC}$ (symbols), and $\phi_u$ (dashed lines) form
a continuous variation with $T$, Fig.~\ref{fig_phiu}, analogous to  $\phi_i$, $\phi_{c}$,
and $\phi_u$ with $f_{\mathrm{m}}$, Fig.~\ref{fig_Fm_binodal}. 
Since $\phi_{iC}\approx \phi_c$, Fig.~\ref{fig_phiu}, we can simplify
eq.~\ref{eq_eleBinodal_full} to obtain the expression for the electrostatic binodal in an
open cylinder system
\be
\frac{T_K}{T_c}\approx \left[ \frac{ - \frac{8}{3}\left(\phi_K-\phi_c\right)^3
\frac{\VE^{\prime}(\phi_{c})}
{\VE(\phi_{c})}}{1-2\left(\phi_K-\phi_c\right)\frac{\VE^{\prime}(\phi_{c})}{\VE(\phi_{c})}
} +1\right]^{-1}~~~~{\rm open~cylinder}
\label{eq_eleBinodal}
\ee
Interestingly, this equation only depends on $\phi_K$ and the functional form of
$\VE(\phi)$, but is independent of $\sigma$ and $r_i$. This finding is a consequence of
the self-similarity of solutions in open systems for a constant $\phi_0$ and $T$,
described in Sect.~\ref{Sect_Interface} and shown in
Fig.~\ref{fig_phi_vs_r_varS_collapse}. Moreover, the geometry difference between cylinders
and spheres does not influence the electrostatic binodal. Equation~\ref{eq_eleBinodal} is,
in fact, the same equation for the electrostatic binodal in an open sphere system, using
the same assumptions.

The thick solid line in Fig.~\ref{fig_eleBinodal} shows the results from
eq.~\ref{eq_eleBinodal}, as it accurately divides the $\phi_0-T$ plane into mixed and
demixed regions. With each point $(\phi_K,T_K)$, there is an associated critical
$\sigma_K$: the $\sigma$ that places $r_i$ exactly at $R_1$. It's important to recognize
that $\sigma_K$ is not constant along the electrostatic binodal---$\sigma_K$ is $0$ at
$T_K=T_c$ and increases as $T_K$, Fig.~\ref{fig_Scrit}, and/or $\phi_K$ decrease.
Figure~\ref{fig_Scrit} compares $\sigma_K$ from calculations (symbols) versus $\sigma_K$
derived from eq.~\ref{eq_interface} using $\phi_K$, $T_K$, and $r_i=R_1$ (line).
Equation~\ref{eq_eleBinodal} predicts that the electrostatic binodal also exists for
$\phi_0>\phi_c$, thin solid line in Fig.~\ref{fig_eleBinodal}; however, the associated
values of $\sigma_K$ are imaginary and not possible in real physical systems.

The electrostatic binodal is a line of critical points, or simply a ``critical line''.
This finding explains some curious observations found previously~\cite{Samin2009}: If
$\phi_0$ and/or $T$ is changed such that the stability diagram for a constant $\sigma$ is
crossed on the boundary between the kink and $(\phi_c,T_c)$, for example the arrow in
Fig.~\ref{fig_stabilityCurve}, then $r_i$ emerges at some distance \emph{greater} than
$R_1$, Fig.~\ref{fig_phi_vs_r_varT}. The kink marks $(\phi_K,T_K,\sigma=\sigma_K)$. The
boundary of the stability diagram to the right of the kink is exactly the electrostatic
binodal. On this boundary, $\sigma$ is now \emph{larger} than $\sigma_K$. In other words,
$\sigma$ is no longer the minimum surface charge that induces the transition; therefore,
$r_i$ necessarily emerges at some distance greater than $R_1$.

The open wedge system produces analogous results; however, we will use the simplicity of
the equations in this geometry to demonstrate the effects of quadratic $\VE(\phi)$
relations, Fig.~\ref{fig_epsilon}. Following the same reasoning as for an open cylinder
system, we find the electrostatic binodal for an open wedge
\bea
\frac{T_K}{T_c} &=& \left[ \frac{ \frac{4}{3}\left(\phi_K-\phi_c\right)^3
\frac{\VE^{(2)}(\phi_{c})}{\VE^{\prime}(\phi_{c})}}{ 1 +
\left(\phi_K-\phi_c\right)\frac{\VE^{(2)}(\phi_{c})}{\VE^{\prime}(\phi_{c})}}+1 \right]^{-1}\nonumber \\ &&
{\rm open~wedge}
\label{eq_eleBinodal_wedge}
\eea
We add that $\phi_i$, $\phi_{iC}$, and $\phi_u$ exactly equal $\phi_c$ if $\VE^{(2)}(\phi)$ 
and higher derivatives vanish. 
Notice the similarity between
eqs.~\ref{eq_eleBinodal} and~\ref{eq_eleBinodal_wedge}, where the main difference is that
higher derivatives of $\VE(\phi)$ control the electrostatic binodal in the wedge geometry.
Figure~\ref{fig_eleBinodal_wedge} shows how the electrostatic binodal for the wedge curves
downwards to upwards as $\VE^{(2)}(\phi_{c})$ changes from negative to positive. And if
$\VE^{(2)}(\phi_{c}) = 0$, then $T_K$ for the electrostatic binodal simply equals $T_c$
for all $\phi_K$. By comparing Fig.~\ref{fig_epsilon} to the results in
Fig.~\ref{fig_eleBinodal_wedge}, it is evident that small amounts of curvature in
$\VE(\phi)$ can create large changes in the electrostatic binodal, in agreement with
previous findings~\cite{Samin2009}.

We briefly discuss an alternate derivation presented in Ref.~\cite{Samin2009} to emphasize
that we have not exhausted all possible relations between parameters. Beginning with
$f^{(2)}=f^{(3)}=0$ for $r_i=R_1$ and $\VE^{(3)}(\phi)=0$ in the wedge geometry, we obtain
\be
\frac{T_K}{T_c}=1+ \frac{Nv\VE_0\VE^{(2)}(\phi)}{8kT_c}\left(\frac{V_K}{\theta R_1}
\right)^2
\label{eq_Samin}
\ee
Interestingly, using the Flory-Huggins approximation for $f_{\mathrm{m}}$ results in
exactly the same relation, eq.~\ref{eq_Samin}, as the Landau approximation. The
differences between the two approximations
instead arise when determining $\phi_K$, where the biggest discrepancies occur, as
expected, for values of $\phi_K$ that are far from $\phi_c$,
Fig.~\ref{fig_eleBinodal_wedge}.

%========================================
% ELECTROSTATIC SPINODAL
%========================================
\section{Electrostatic Spinodal}\label{Sec_spin}
\begin{figure}[!tb]%
\begin{center}
\subfigure[\label{fig_phii_vs_T}]{\includegraphics[keepaspectratio=true,width=0.232\textwidth]{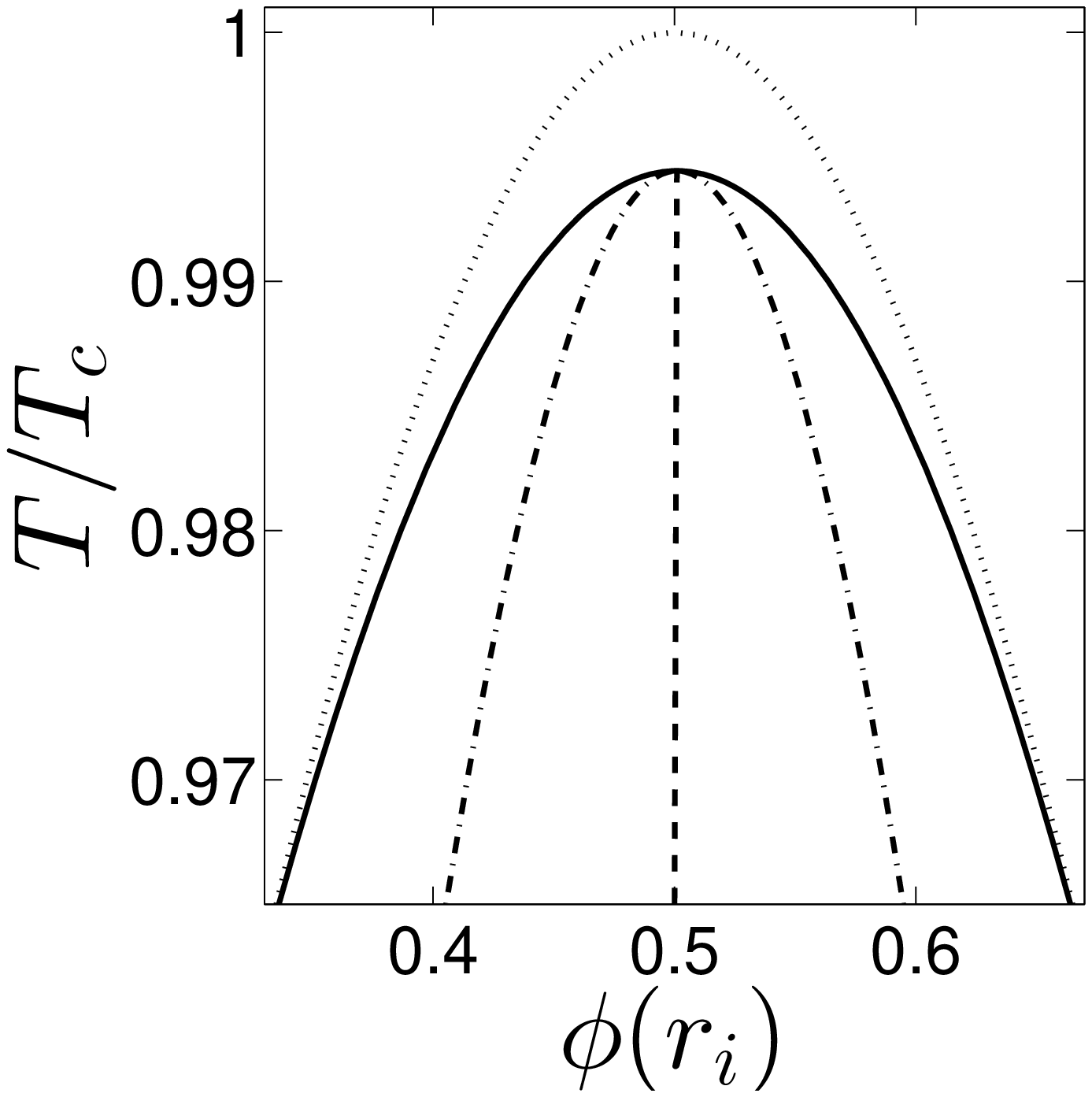}}
\subfigure[\label{fig_phi_vs_r_3solns}]{\includegraphics[keepaspectratio=true,width=0.232\textwidth]{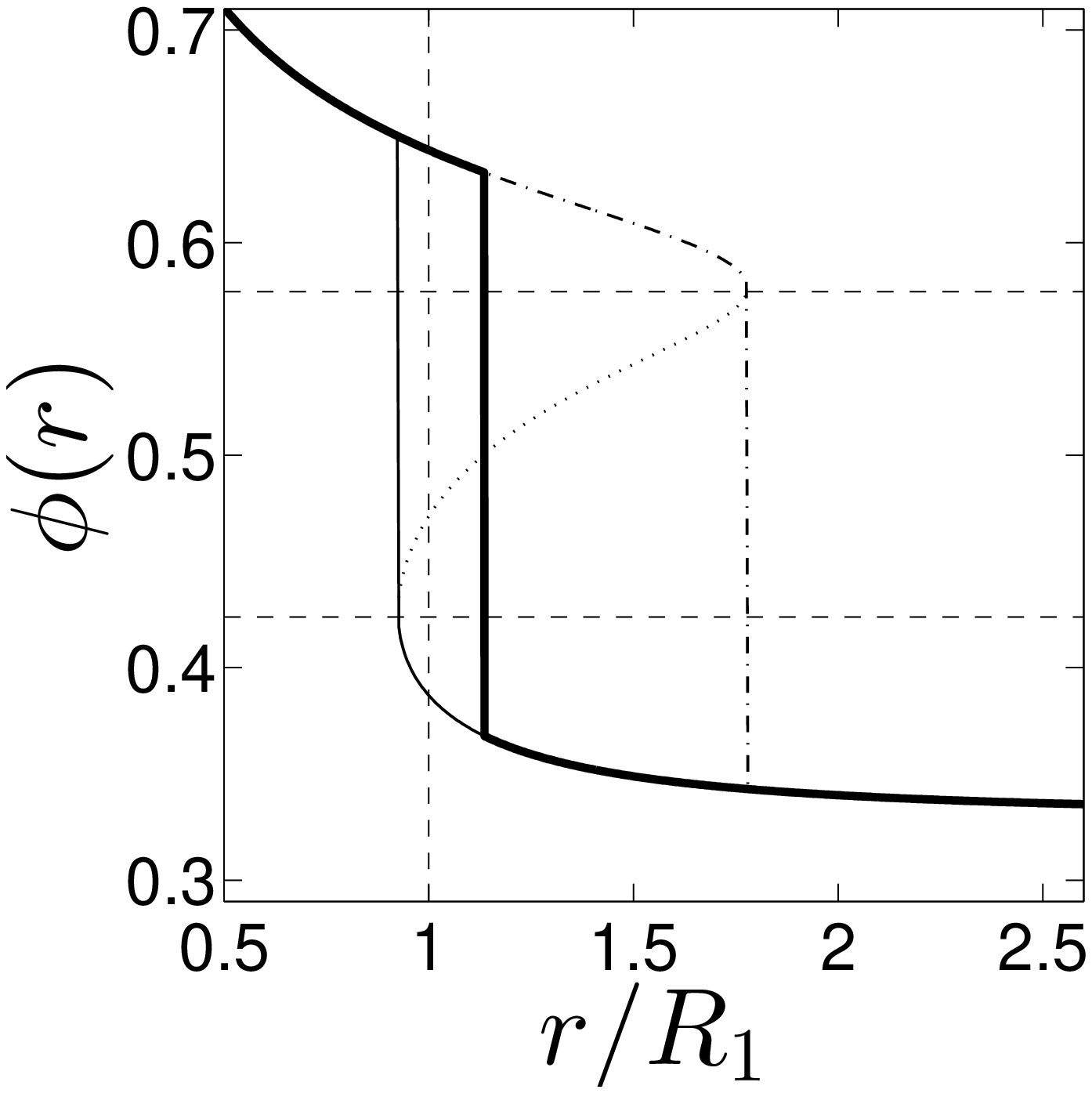}}\\
\subfigure[\label{fig_ri_vs_phi0_dyn}]{\includegraphics[keepaspectratio=true,width=0.232\textwidth]{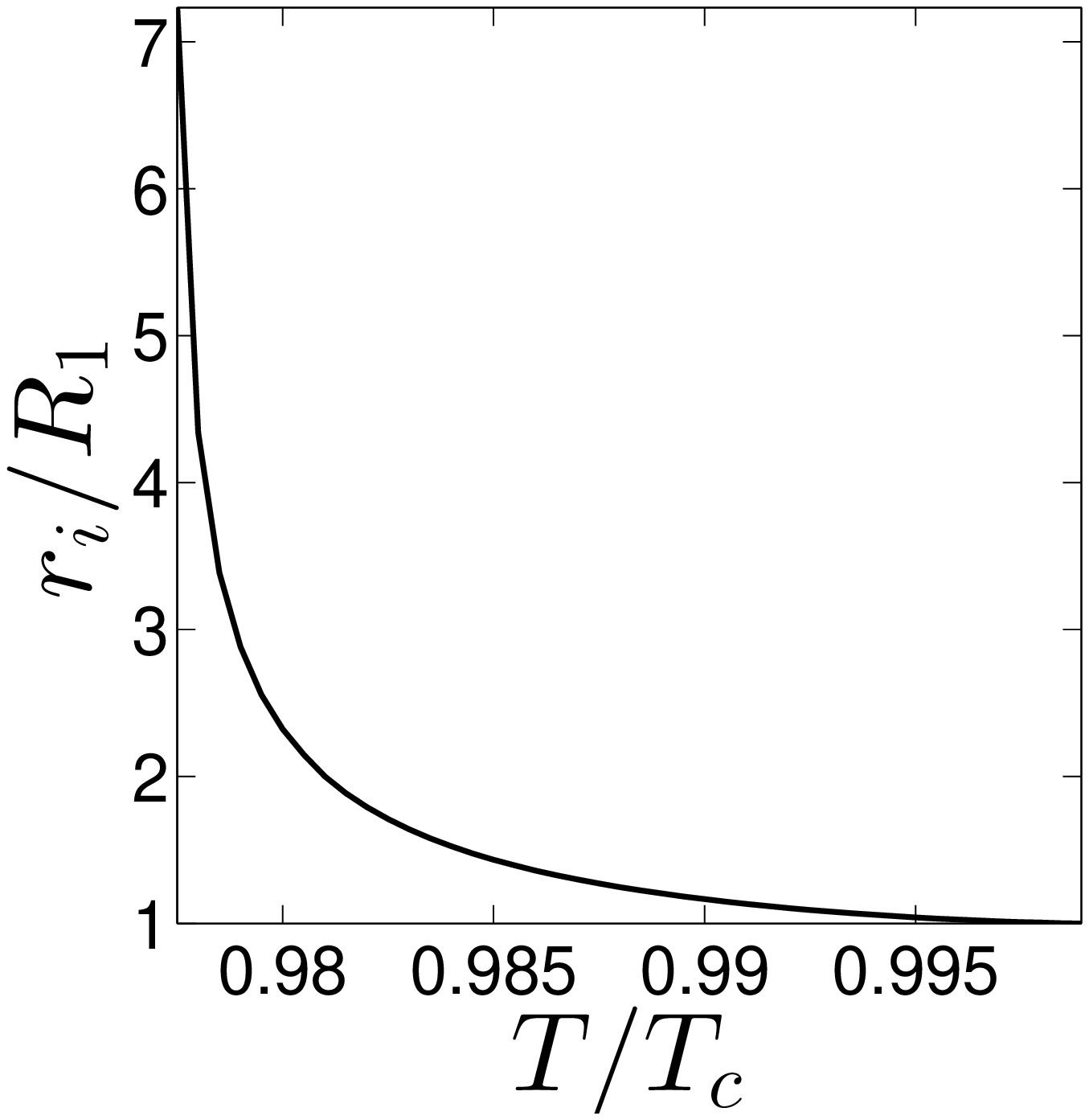}}
\subfigure[\label{fig_eleSpinodal}]{\includegraphics[keepaspectratio=true,width=0.232\textwidth]{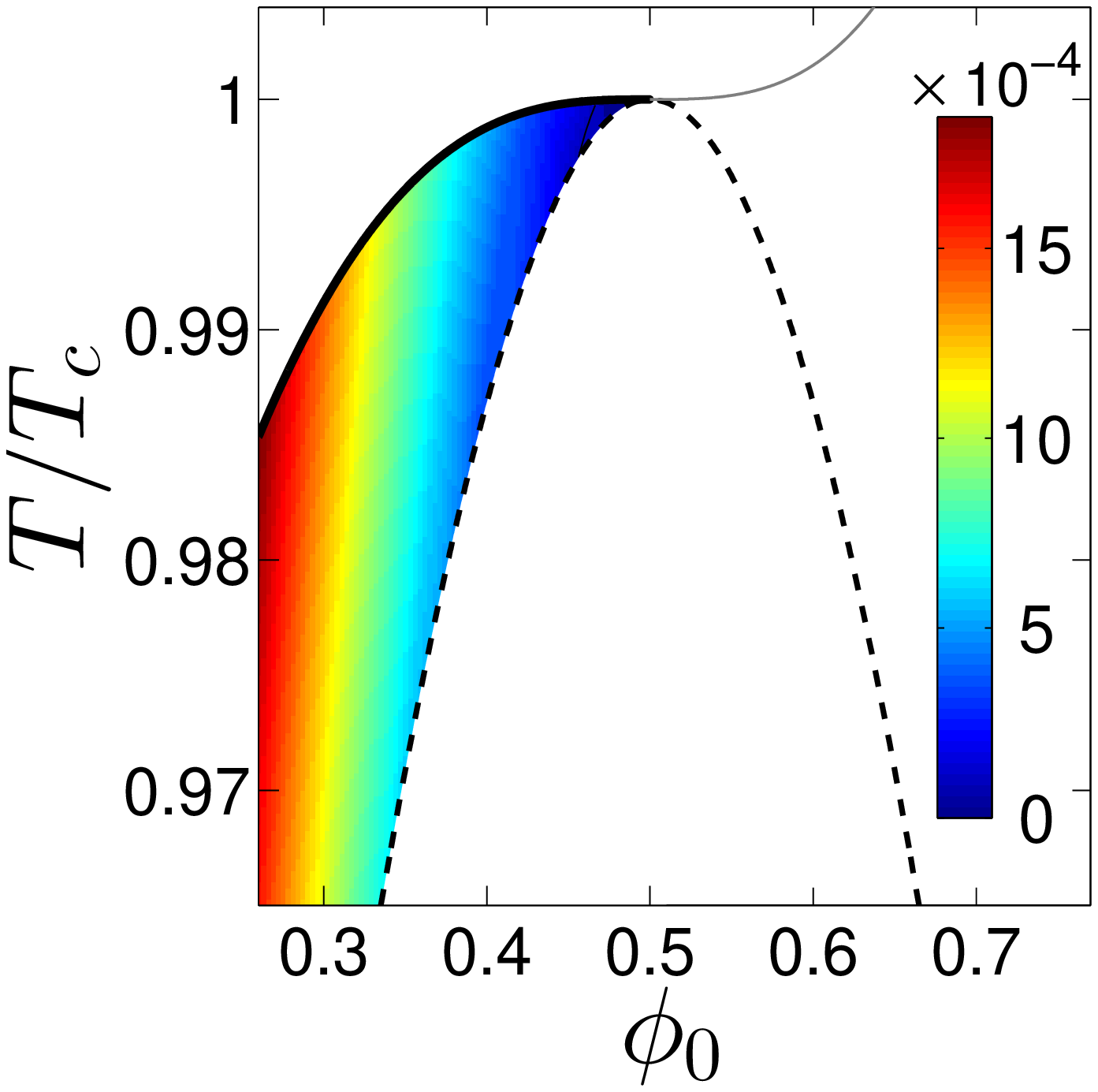}}
\caption{Electrostatic spinodal for an open cylinder system. (a) Behavior of $\phi(r_i)$ with $T$, 
showing $\phi_{iH}$, $\phi_{iL}$ (solid line), $\phi_{sH}$, $\phi_{sL}$
(dash-dotted line), and $\phi_u$ (dashed line). Lines converge at the critical point
$\phi_{iC}$. In all data, $\phi_0 = 0.33$ and $\sigma\approx 2.09\times 10^{-3}\,\mathrm{C/m^{2}}$. Dotted
lines show $\phi_b$. (b) All $\phi(r)$ solutions to $f'=0$
versus normalized $r$.
Thin solid, dash-dotted, and dotted lines show lower, upper, and unstable solutions,
respectively, for $\phi_0=0.33$, $T/T_c = 0.975$, and $\sigma=8\times
10^{-4}\,\mathrm{C/m^{2}}$. Thick line shows the solution that minimizes $f$. Horizontal
dashed lines show $\phi_{sL}$ and $\phi_{sH}$. 
(c) Location of the interface $r_i$ versus
normalized $T$ along the spinodal line in Fig.~\ref{fig_stabilityCurve} for $\sigma = 1\times 10^{-3}\,\mathrm{C/m^2}$. 
(d) Overlay of electrostatic spinodals for many
$\sigma$ (color, [$\mathrm{C/m^2}$]).}
\end{center}
\end{figure}

We now turn the discussion to possible metastable states, recalling
the meaning of the spinodal curve in the mean-field theory ~\cite{Binder2009}.
Earlier in the manuscript, we rationalized the existence of inflection points $\phi_s$ at
$r_i$ through the presence of a maximum $\phi_u$ and minima $\phi_{iH}$, $\phi_{iL}$. Both
high and low values $\phi_{sH}$, $\phi_{sL}$ satisfy $f^{(2)}=0$ and exist at all
interfaces. Figure~\ref{fig_phii_vs_T}, for example, explicitly shows the
mathematical features [$\phi_{iH}$, $\phi_{iL}$ (solid line), $\phi_{sH}$, $\phi_{sL}$
(dash-dotted line), $\phi_u$ (dashed line), and critical point] occurring at $r_i$ with
changing $T$ in an open cylinder system. For comparison, the
dotted lines display the behavior of the binodal points $\phi_b$ with $T$.
%%%%%%%%%%%%%%%%%%%%%%%%%%%%%%%%%%%%%%%%%%%%%%%%%%%%%%%%%%%%%%%%
%%%%%%%%%%%%%%%%%%%%%%%%%%%%%%%%%%%%%%%%%%%%%%%%%%%%%%%%%%%%%%%

Despite the ubiquitous presence of $\phi_{sH}$ and $\phi_{sL}$, only $\phi_{sL}$ carries
physical meaning in open systems, and only in a limited region of the stability diagram.
To see how this occurs, we return to the solutions of $f'=0$. Thus far,
we focused
on $r_i$, the location of the interface for the minimized $f$; however, there can be many
$r$ that posses the same mathematical features. Figure~\ref{fig_phi_vs_r_3solns} shows all
possible solutions to $f'=0$, where the thin solid, dash-dotted,
and dotted lines
are the ``lower'', ``upper'', and ``unstable'' solutions, respectively. The heavy solid
line depicts the solution that actually minimizes $f$, and the two dashed lines denote $\phi_{sH}$
and $\phi_{sL}$ found at $r_i$.

We start from a homogeneous mixture at composition $\phi_0$ and perform the thought experiment of turning
on an electric field. Considering diffusive liquid movement in the absence of other
factors (ex. liquid convection, noise), this experimental setup implies that the
profile $\phi(r)$ initially develops along the free energy ``well'' created by the lower
solution. If the electric field can sufficiently ``pull'' the higher dielectric material
such that there is at least one distance $r$ where $\phi(r)\ge\phi_{sL}$, then the liquid
can escape the metastable (mixed) state at the local free energy minimum to find the
global minimum (demixed). We call $r_s$ the distance where
$\phi(r_s)=\phi_{sL}$ and find $r_s$ by solving $f^{\prime}=0$ at $\phi_{sL}$. For a
cylindrical geometry we have
\be
\frac{r_s}{R_1} = \sqrt{ \frac{Nv}{2kT\VE_0} \left[\frac{\sigma^2}{
f^{\prime}_m(\phi_{sL})-\mu}\right] \frac{\VE^{\prime}(\phi_{sL})}{\VE(\phi_{sL})^2} }
\ee

Knowing that the highest value of $\phi(r)$ occurs closest to the electrode at $R_1$, we
seek the conditions where $r_s=R_1$. These conditions, therefore, mark the electrostatic
spinodal: If $r_s\ge R_1$ at a particular $(\phi_0,T,\sigma)$, then demixing occurs
spontaneously. If $r_s< R_1$ at a particular $(\phi_0,T,\sigma)$, for example
Fig.~\ref{fig_phi_vs_r_3solns}, then the liquids can be metastabaly mixed. The long time
solution for dynamics in these cases therefore resides along the thin solid curve,
Fig.~\ref{fig_phi_vs_r_3solns}.

Figure~\ref{fig_stabilityCurve} shows the location of the electrostatic spinodal for a
particular value of $\sigma$. The curve begins at the critical point $(\phi_K,T_K)$ and
travels
down, on the right side of the stability diagram boundary. Similar to the ``regular''
spinodal curve, demixing occurs spontaneously (non-spontaneously) for $(\phi_0,T)$ to the
right (left) of the electrostatic spinodal. Since the electrostatic spinodal cuts inside
the stability diagram, the location of the interface $r_i$ emerges at distances greater
than $R_1$, with $r_i=R_1$ only at $(\phi_K,T_K)$. Figure~\ref{fig_ri_vs_phi0_dyn}
displays the behavior of $r_i$ along the spinodal in 
Fig.~\ref{fig_stabilityCurve}.
Finally, the electrostatic spinodal exists for all $\sigma$. Figure~\ref{fig_eleSpinodal}
shows the superposition of the electrostatic spinodal curves, where the color indicates
the associated $\sigma$.

%========================================
% CLOSED SYSTEMS
%========================================
\section{Closed Systems}\label{Section_closed}
Up until now, we focused on liquid behavior in open systems, where we considered the
location of the second boundary as $R_2\to\infty$. A closed system with a finite $R_2$
markedly alters the phase diagram~\cite{Samin2009}; however, we will show that these
alterations naturally arise from the solutions of open systems.

\begin{figure}[!t]%
\begin{center}
\subfigure[\label{fig_phi_vs_r_varS_closed}]{\includegraphics[keepaspectratio=true,width=0.232\textwidth]{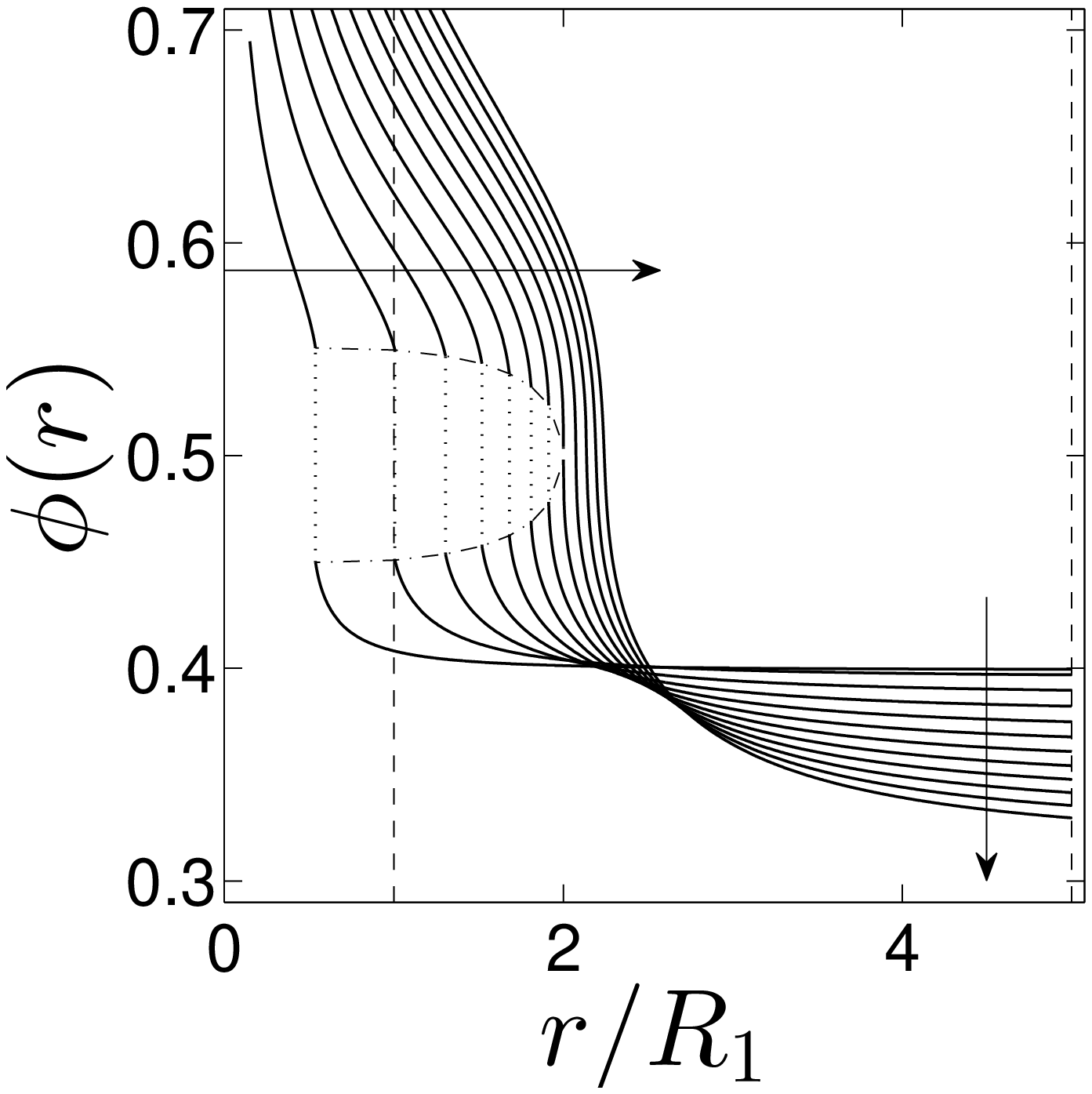}}
\subfigure[\label{fig_phi_vs_r_varR2_closed}]{\includegraphics[keepaspectratio=true,width=0.232\textwidth]{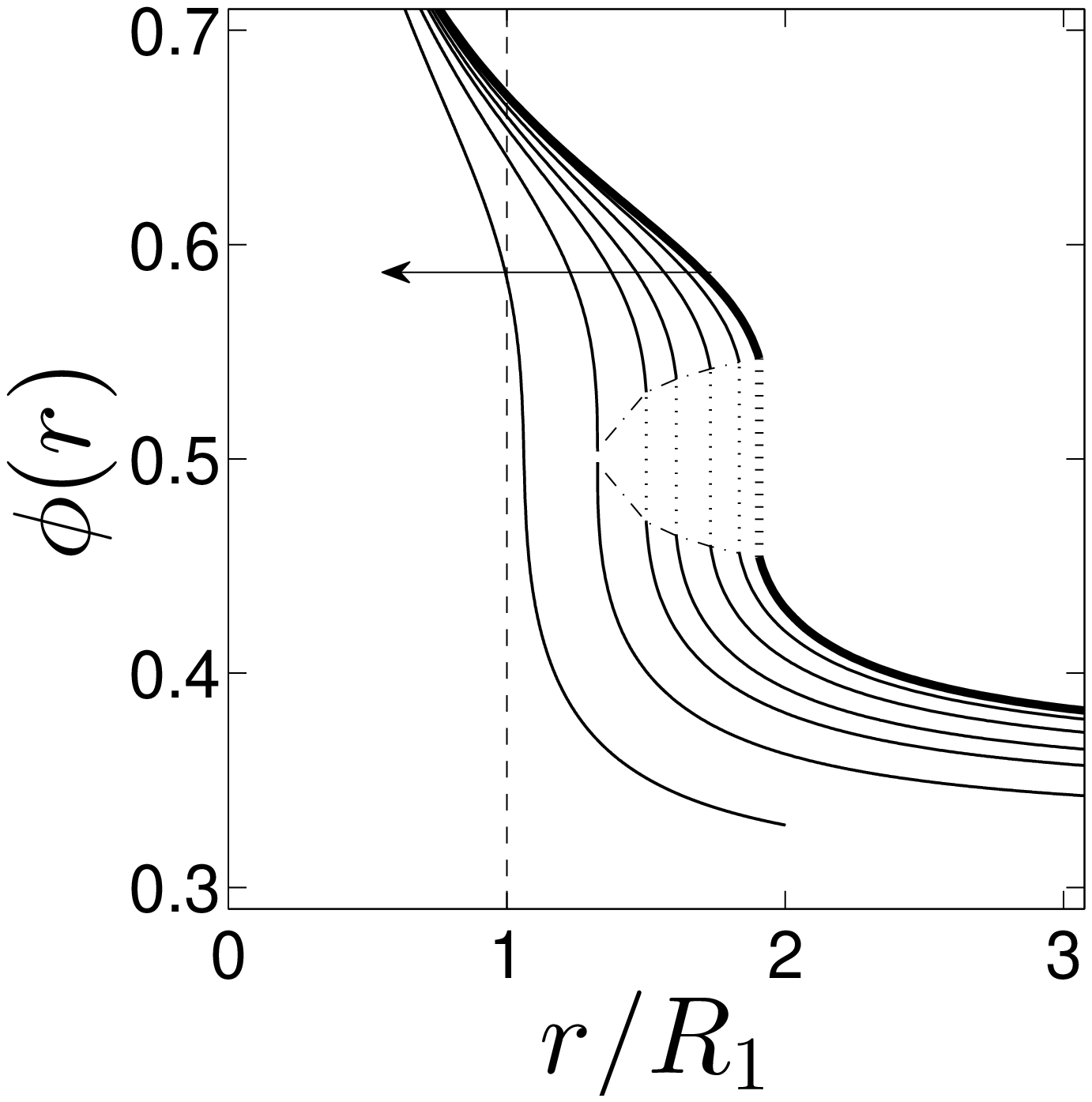}}\\
\subfigure[\label{fig_st1_st2}]{\includegraphics[keepaspectratio=true,width=0.232\textwidth]{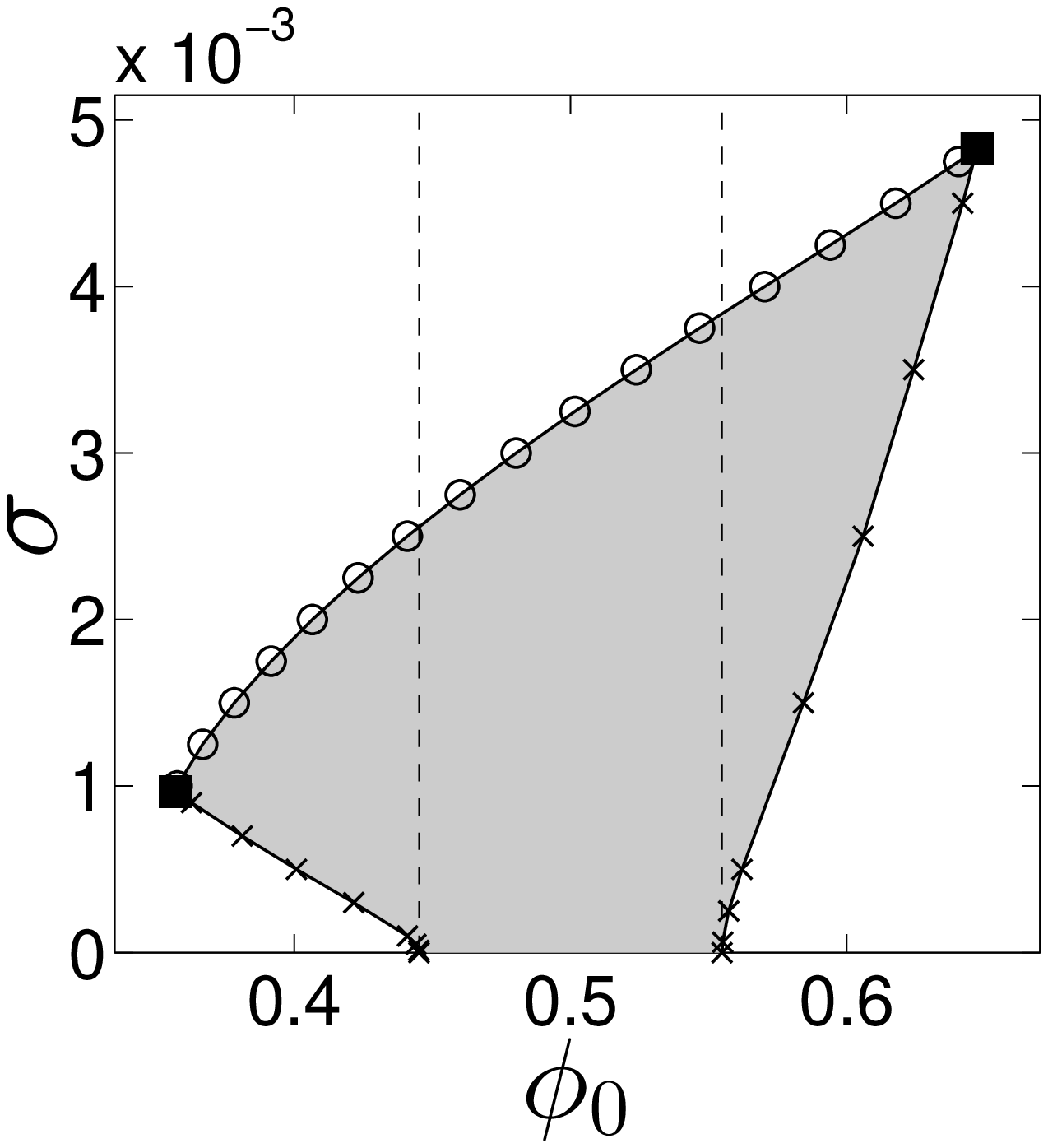}}
\caption{Variation of concentration profiles $\phi(r)$ in a closed cylinder system. (a)
$\phi(r)$ versus a normalized distance $r$ for a constant $\phi_0=0.4$, $T/T_c\approx
0.996$, $R_2/R_1=5$ and varying $\sigma = 0.25\times 10^{-3}$ to $3\times
10^{-3}\,\mathrm{C/m^{2}}$ in $0.25\times 10^{-3}$ increments (arrows). (b) $\phi(r)$
versus a normalized distance $r$ for a constant $\phi_0=0.36$, $T/T_c\approx 0.995$,
$\sigma =1.5\times 10^{-3}\,\mathrm{C/m^{2}}$ for an open system (thick line) and closed
systems (thin lines) with decreasing $R_2/R_1 = 20$, $12$, $8$, $6$, $4$ to $2$ (arrow). 
(c) $\sigma_{t1}$ ($\times$), $\sigma_{t2}$ ($\circ$), and $\sigma_K$ (filled squares) versus $\phi_0$
for $T/T_c=0.996$ and $R_2/R_1=5$. Dashed lines show binodal $\phi_b$ at same $T$. Phase separation occurs in shaded region.}
\label{F_bin}%
\end{center}
\end{figure}

We begin as previously, with variations in the concentration profiles $\phi(r)$, and
identify interesting changes with two parameters, $\sigma$ and $R_2$. Both
Figs.~\ref{fig_phi_vs_r_varS_closed} and~\ref{fig_phi_vs_r_varR2_closed} clearly reveal
that the discontinuity at the interface decreases and vanishes with increasing $\sigma$
and decreasing $R_2$, respectively, in closed cylinder systems. Intriguingly, the profiles
in Fig.~\ref{fig_phi_vs_r_varS_closed} stand in sharp contrast to the self-similar
solutions found in open systems, Figs.~\ref{fig_phi_vs_r_varS}
and~\ref{fig_phi_vs_r_varS_collapse}. Closer inspection of
Fig.~\ref{fig_phi_vs_r_varS_closed} also reveals that the parabolic-like shape in the
discontinuity with various $\sigma$ opens to the left, rather than to the right as in
Figs.~\ref{fig_phi_vs_r_varP},~\ref{fig_phi_vs_r_varT},
and~\ref{fig_phi_vs_r_varR2_closed}. An important consequence is that for closed systems
there are \emph{two} transition surface charge densities 
$\sigma_t$: the first $\sigma_{t1}$ is the $\sigma$ that places $r_i$ 
exactly at $R_1$, while the second $\sigma_{t2}$ is the $\sigma$ where the interface
discontinuity vanishes. 
Therefore, the interface between the liquids in closed systems only exists when $\sigma$
satisfies $\sigma_{t1}\le\sigma\le\sigma_{t2}$, shaded region in Fig.~\ref{fig_st1_st2}.
%%%%%%%%%%%%%%%%%%%%%%%%%%%%%%%%%%%%%%%%%%%%%%%%%%%%%%%%
%%%%%%%%%%%%%%%%%%%%%%%%%%%%%%%%%%%%%%%%%%%%%%%%%%%%%%%%%%%%

Material conservation drives all differences between closed and open systems, thus, the
key to understanding these differences resides in understanding $\mu$. Recall that $\mu =
\mu_0=f_{\mathrm{m}}^{\prime}(\phi_0,T)$ in open systems, while $\mu$ is adjusted to
account for material conservation in closed systems. Mathematically, the adjusted $\mu$
for a closed system at $(\phi_0,T)$ exactly matches the $\mu_0$ for an open system with a
different ``effective'' concentration $\phi_E$ in the bath. Consequently, the $\phi(r)$
profile between $R_1$ and $R_2$ at $(\phi_0,T)$ in a closed system exactly matches the
$\phi(r)$ profile at $(\phi_E,T)$ in an open system. In other words, the behavior of a
closed system maps onto that of an open system via $\phi_E$.

We can explain the variation of $\phi(r)$ with $\sigma$ in closed systems using this
construct. Intuitively, the higher dielectric material is pulled closer to the electrode
as the value of $\sigma$ increases. In order to conserve material in a closed system,
$\phi(r)$ necessarily decreases near $R_2$, Fig.~\ref{fig_phi_vs_r_varS_closed}. This
shift in liquid concentration translates as a decrease in $\phi_E$, hence increasing
$\sigma$ in a closed system maps as increasing $\sigma$ \emph{and} decreasing $\phi_E$ in
an open system. Recall that the interface discontinuity becomes smaller with lower
$\phi_E$ in an open cylinder system, Fig~\ref{fig_phi_vs_r_varP}, and eventually vanishes
when $\phi_E$ crosses the electrostatic binodal. The same principles apply to closed
systems, where the second transition $\sigma_{t2}$ marks this crossing.

\begin{figure}[!tb]%
\begin{center}
\subfigure[\label{fig_stabilityCurve_closed}]{\includegraphics[keepaspectratio=true,width=0.22\textwidth]{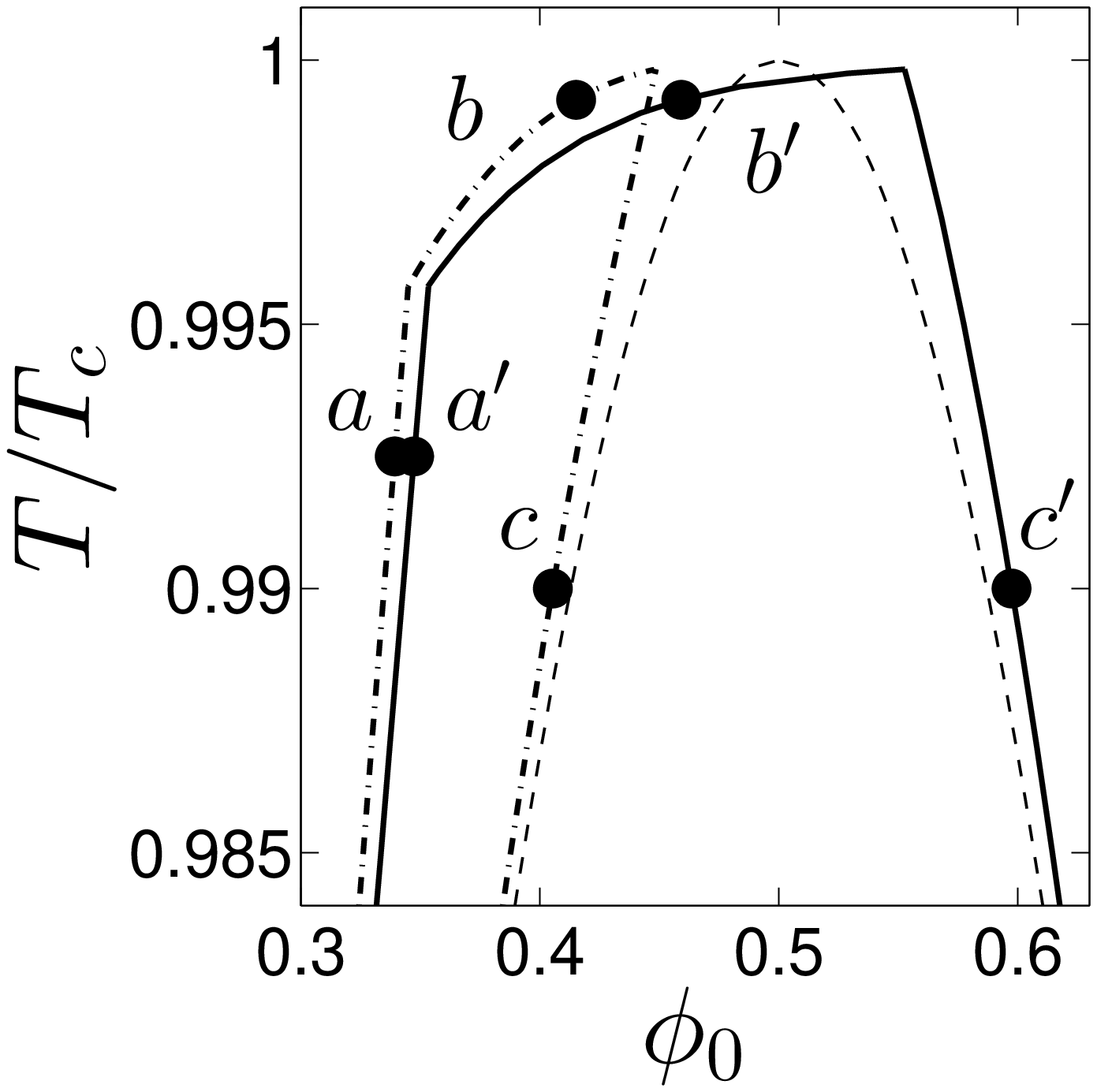}}
\subfigure[\label{fig_phi_r_3regions}]{\includegraphics[keepaspectratio=true,width=0.22\textwidth]{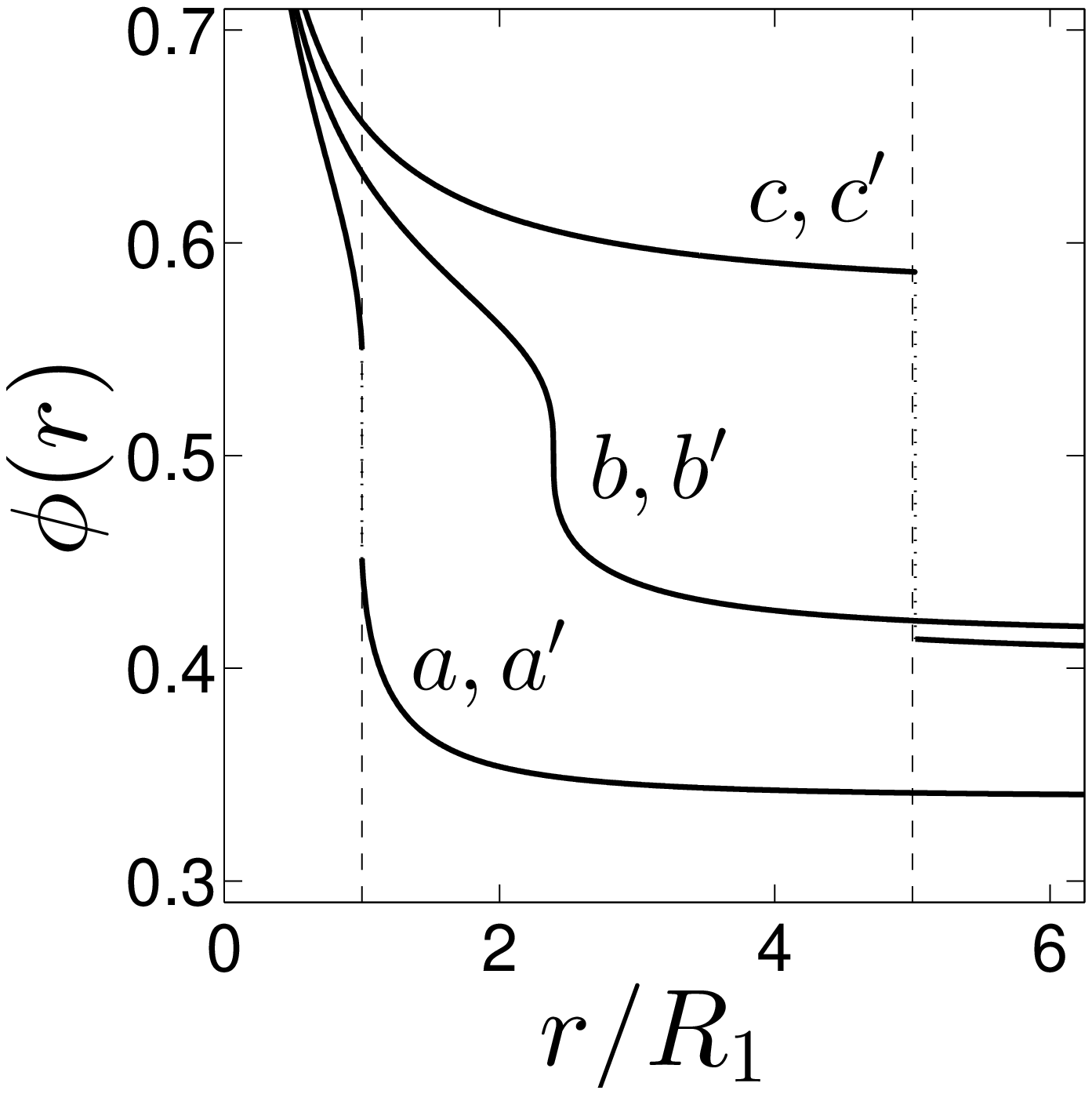}}\\
\subfigure[\label{fig_eleBinodal_closed_S1}]{\includegraphics[keepaspectratio=true,width=0.22\textwidth]{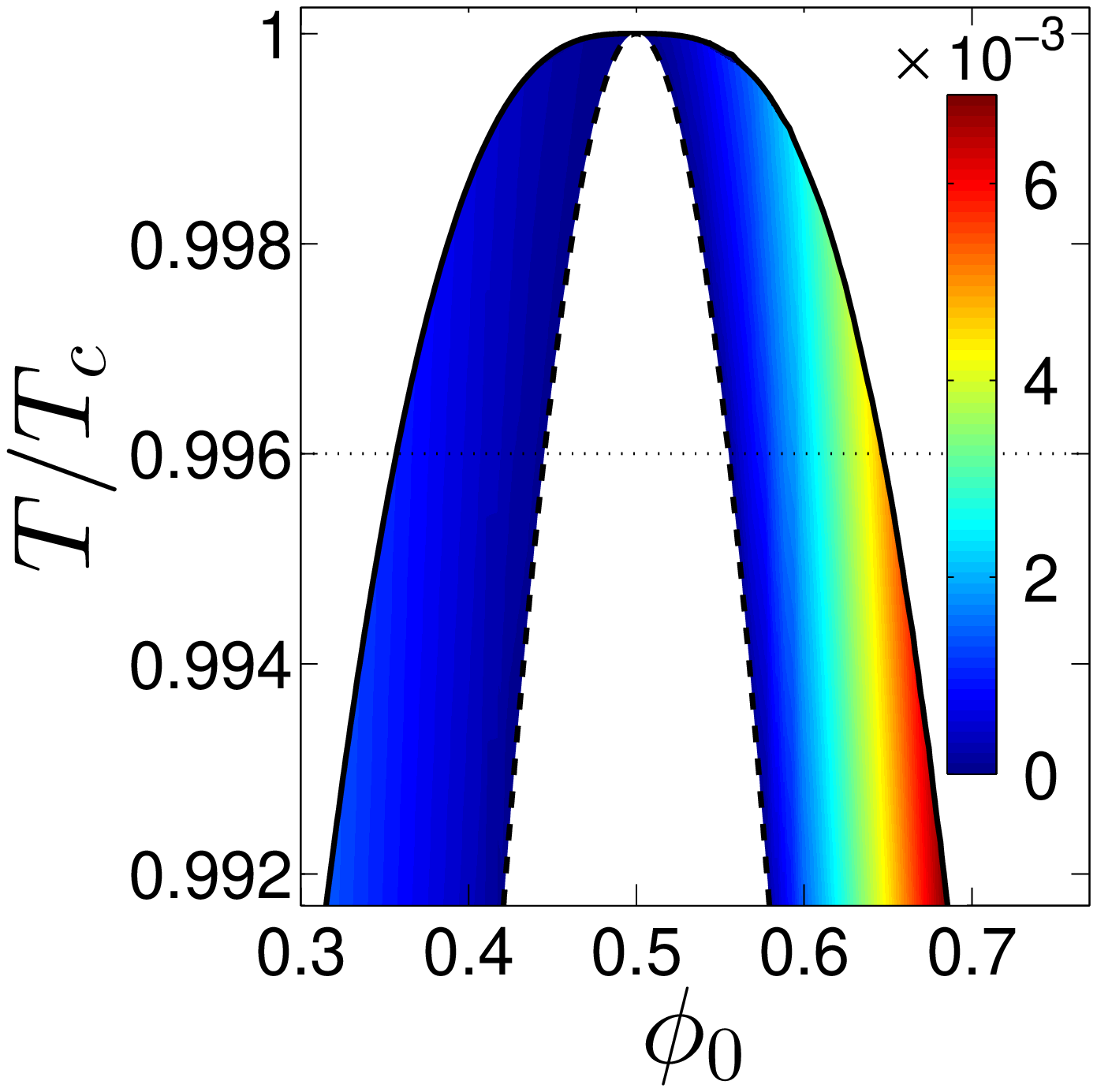}}
\subfigure[\label{fig_eleBinodal_closed_S2}]{\includegraphics[keepaspectratio=true,width=0.22\textwidth]{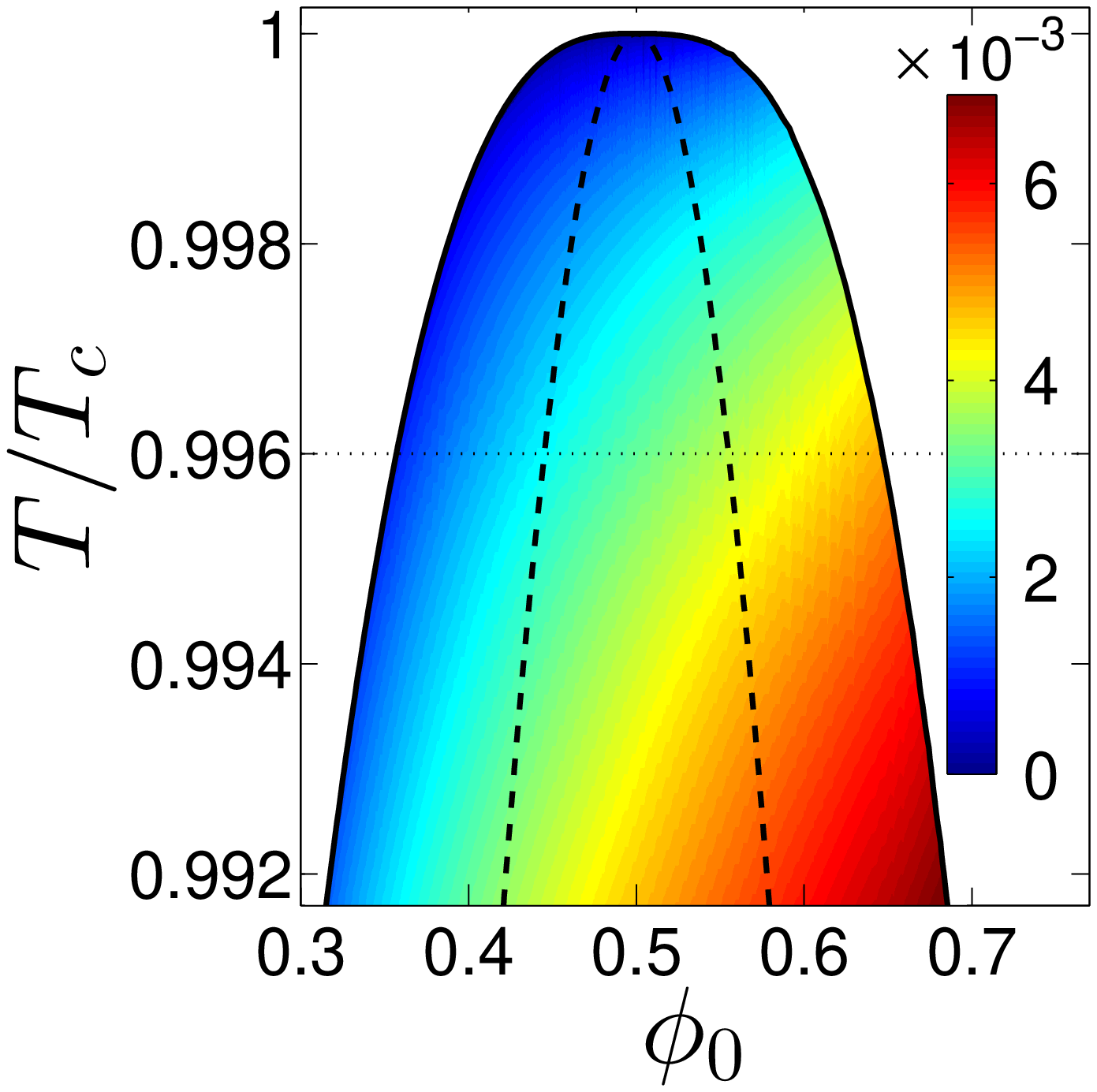}}\\
\subfigure[\label{fig_eleBinodal_closed_S_T}]{\includegraphics[keepaspectratio=true,width=0.22\textwidth]{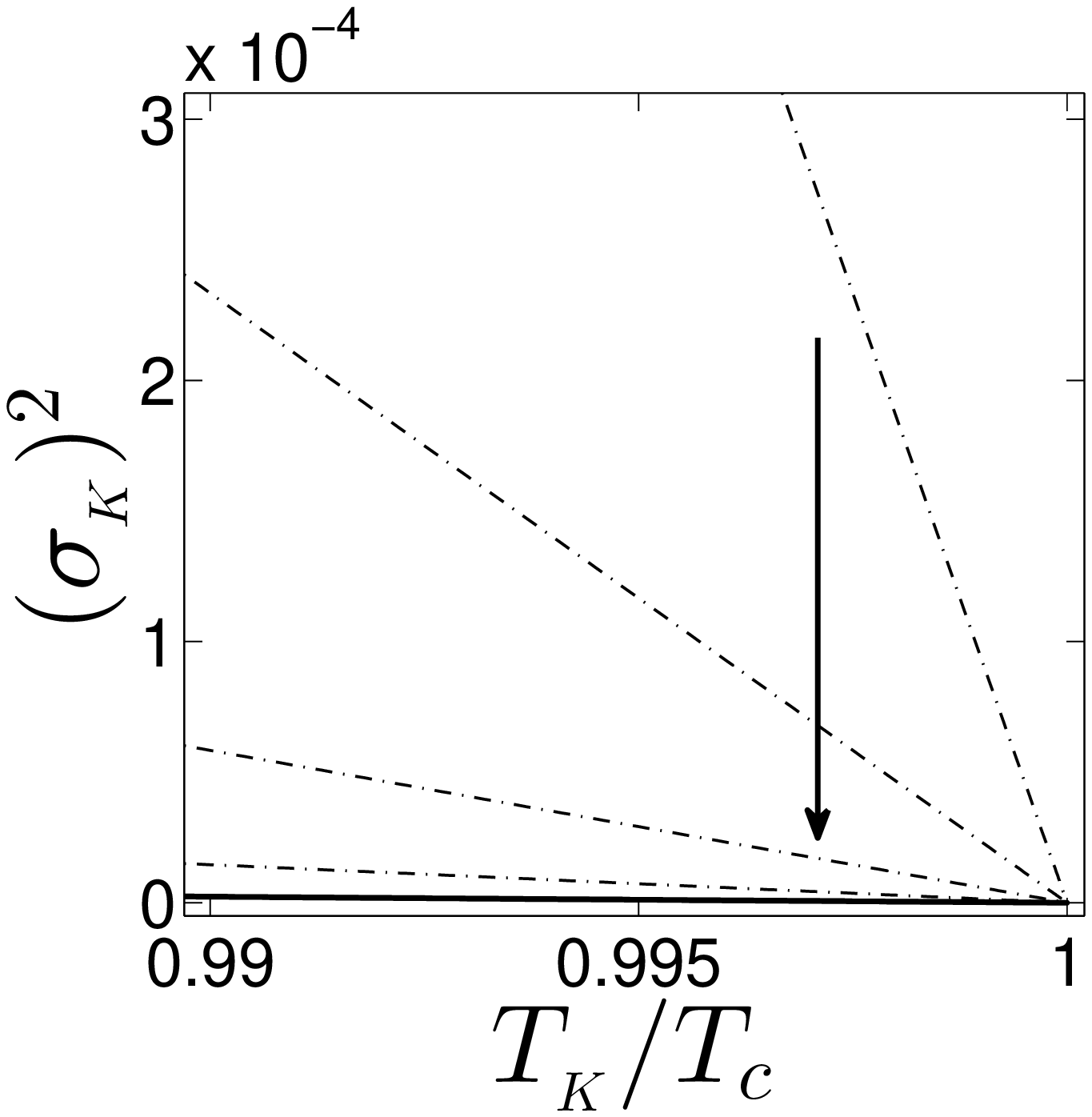}}
\subfigure[\label{fig_eleBinodal_closed_R2}]{\includegraphics[keepaspectratio=true,width=0.22\textwidth]{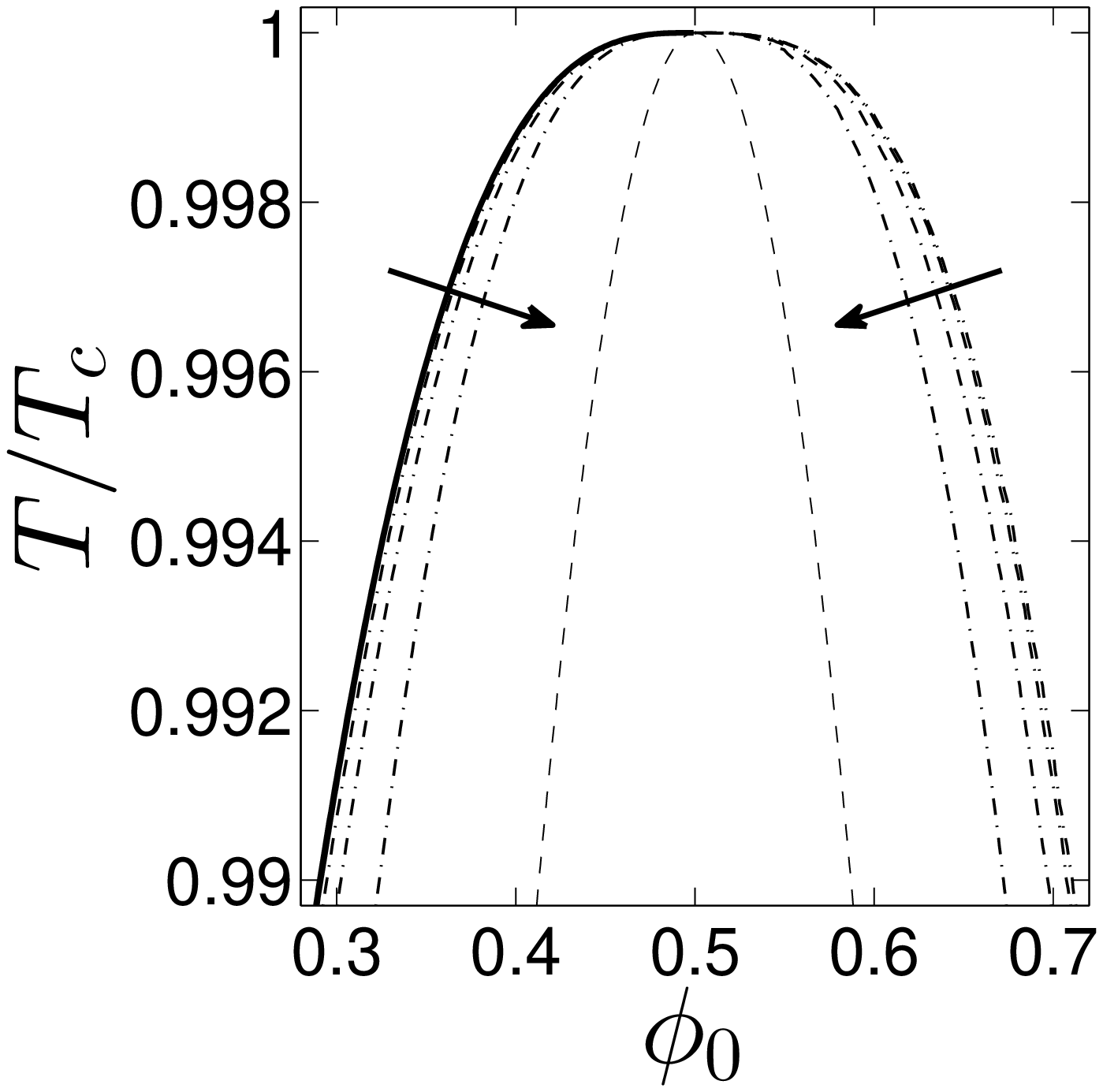}}
  \caption{Electrostatic binodal in a closed cylinder system. (a) Stability curve (solid
line) in $\phi_0-T$ space for a constant $\sigma = 1\times 10^{-3}\,\mathrm{C/m^{2}}$ and
$R_2/R_1=5$. Dash-dotted line shows the mapping of the stability diagram boundaries to an open system via $\phi_E$ (see text). Dashed line shows binodal
curve. (b) $\phi(r)$ versus normalized $r$ for points marked by symbols in (a). Dashed
lines mark the boundaries of the container. (c,d) Overlay of many stability diagrams,
where color indicates transition $\sigma_{t1}$ and $\sigma_{t2}$ [$\mathrm{C/m^2}$], respectively. 
Dotted lines show location of data in Fig.~\ref{fig_st1_st2}. 
(e) Critical surface charge density $\sigma_K$ on the electrostatic binodal for $\phi_K\le
\phi_c$
(solid line) and $\phi_K\ge \phi_c$ (dash-dotted lines) for $R_2 = 20$, $10$, $5$, and
$2.5\,\mu\mathrm{m}$ (arrow). (f) Electrostatic binodal for open (solid line) and closed
(dash-dotted lines) systems for the same $R_2$ as (e). Curves are {\it not} symmetric around
$\phi_0=0.5$. Dashed line shows binodal curve.}
\end{center}
\end{figure}

Now that we understand how experimental parameters change $\phi(r)$, we focus on how these
changes affect the stability diagram. Figure~\ref{fig_stabilityCurve_closed} shows a
typical stability diagram for a constant $\sigma = 1\times 10^{-3}\,\mathrm{C/m^{2}}$ and
$R_2/R_1=5$ in a closed cylinder system. One striking difference between open,
Fig.~\ref{fig_stabilityCurve}, and closed, Fig.~\ref{fig_stabilityCurve_closed}, systems
is that liquid separation can now occur for $\phi_0>\phi_c$. Experimentally, this
manifests as an interface emerging close to $R_2$, rather than $R_1$. A second more subtle
difference is that the stability diagram for closed systems occupies a slightly smaller
region of $\phi_0-T$ space for $\phi_0\le \phi_c$ compared to open systems with the same
$\sigma$. Finally, the upper boundary for the closed system stability curve travels
\emph{below} $T_c$ to \emph{exclude} a portion of the binodal curve. Closed systems,
therefore, provide the interesting possibility of an electric field {\it mixing}
liquids that normally demix. 

%%%%%%%%%%%%%%%%%%%%%%%%%%%%%%%%%%%%%%%%%%%%%%%%%%%%%%%%%%%%%%%%%%
%%%%%%%%%%%%%%%%%%%%%%%%%%%%%%%%%%%%%%%%%%%%%%%%%%%%%%%%%%%%

We can use the mapping construct to not only comprehend these changes but also to produce
the stability diagram of closed system. Open systems link to closed systems
via integration. Specifically,
integrating $\phi(r)$ between $R_1$ and $R_2$ in an open system at $(\phi_E,T)$ gives the
corresponding $(\phi_0,T)$ for the closed system. We begin with the left boundary of the
stability diagram for an open system, label $a$ in Fig.~\ref{fig_stabilityCurve_closed},
and integrate $\phi(r)$ between $R_1$ and $R_2$ to determine the location of the left
boundary in a closed system, label $a^{\prime}$ in Fig.~\ref{fig_stabilityCurve_closed}.
The difference between $\phi_E$ and $\phi_0$ along this boundary is small. If we look at
an example $\phi(r)$ profile, Fig.~\ref{fig_phi_r_3regions}, we see that the interface
location $r_i$ equals $R_1$ and that the electric field for $r>R_1$ produces only small
variations in $\phi(r)$. Truncating the integration at $R_2$, 
therefore, only minimally alters the liquid concentration.

Next, we consider the upper boundary of the open system stability diagram, label $b$ in
Fig.~\ref{fig_stabilityCurve_closed}, and integrate from $R_1$ to $R_2$ to obtain the
upper boundary for the closed system stability diagram, label $b^{\prime}$ in
Fig.~\ref{fig_stabilityCurve_closed}. Here, large differences between $\phi_E$ and
$\phi_0$ can occur. This boundary for open systems is the electrostatic binodal. As
previously described in Sect.~\ref{Sect_eleBin}, $\sigma>\sigma_K$, which causes the
location of the interface $r_i$ to emerge at distances greater than $R_1$. The inclusion
of high dielectric material from $R_1$ to $r_i$ can substantially increase $\phi_0$ when
integration stops at $R_2$.

The upper boundary of the stability diagram for a closed system ends when $r_i=R_2$. And
to form the right boundary in a closed system, we must find the conditions where $\sigma$
places $r_i$ at $R_2$ in an open system. There are two methods by which to proceed. First,
we present the simple straightforward approach. We use eq.~\ref{eq_interface} with
$\sigma$, $r_i=R_2$, and various $T$ to determine the appropriate $\phi_E$, label $c$ in
Fig.~\ref{fig_stabilityCurve_closed}, and then integrate $\phi(r)$ profiles from $R_1$ to
$R_2$ to create the right boundary for the closed system, label $c^{\prime}$ in
Fig.~\ref{fig_stabilityCurve_closed}. The second method relies on the self-similarity of
the solutions in open systems. We recognize that the line labeled $c$ in
Fig.~\ref{fig_stabilityCurve_closed} is the stability line (where $r_i=R_1$) for a
rescaled surface charge, namely $\sigma R_1/R_2$ for cylindrical geometry. The ability to
shift the interface and rescale the solution with a modified $\sigma$ will prove useful in
creating the closed system electrostatic binodal.

The superposition of the stability diagrams from many $\sigma$ produces
Figs.~\ref{fig_eleBinodal_closed_S1} and~\ref{fig_eleBinodal_closed_S2}, where color
indicates $\sigma_{t1}$ and $\sigma_{t2}$, respectively, for $R_2/R_1=5$.
These figures reveal striking asymmetry with respect to $\phi_c$ in the values of both
$\sigma_{t1}$ and $\sigma_{t2}$. Notably, higher $\sigma$ are necessary to both create
($\sigma_{t1}$) and eventually destroy ($\sigma_{t2}$) the interface when $\phi_0>\phi_c$.
The outer bounding line in 
Figs.~\ref{fig_eleBinodal_closed_S1} and~\ref{fig_eleBinodal_closed_S2} represents the
electrostatic binodal for a closed system. This line is also asymmetric with respect to
$\phi_c$. And due to the structure of the stability diagram in closed systems, $\sigma_K$
is both $\sigma_{t1}$ \emph{and} $\sigma_{t2}$ for all $(\phi_K,T_K)$, see Fig.~\ref{fig_st1_st2}.

In order to find this electrostatic binodal, we follow the same methods we used for
finding the stability diagram of the closed system. 
We begin with the open system solutions at $(\phi_E=\phi_K,T=T_K,\sigma=\sigma_K)$, and
integrate $\phi(r)$ between $R_1$ and $R_2$ to determine $\phi_0$ (the $\phi_K$ for the
closed system). Notice that this procedure accounts for interfaces emerging at $R_1$;
however, closed systems can also have interfaces emerging from $R_2$. Therefore, we
rescale the open system solutions by increasing $\sigma$ so that $r_i=R_2$ [precisely
$(\phi_E=\phi_K,T=T_K,\sigma=\sigma_K R_2/R_1)$ for cylindrical geometry], and integrate
$\phi(r)$. This rescaling links $\sigma$ and $R_2$, as evident in
Fig.~\ref{fig_eleBinodal_closed_S_T}. Consequently, phase separation for concentrations
greater than $\phi_c$ technically exist for open systems, and requires infinitely large
$\sigma$ to produce an interface at $R_2\to\infty$. Practically speaking however, even
closed systems with a ``large enough'' $R_2$ would need unreasonably high values of
$\sigma$ to induce a transition in this region of $\phi_0-T$ space. Under these
conditions, other events, such as heating, liquid ionization, bubble formation, and 
electrical breakdown of the liquids would need to be
considered~\cite{Coelho1971,Schmidt1984,Denat1988}.

Figure~\ref{fig_eleBinodal_closed_R2} shows how the electrostatic binodal changes with
$R_2$, where the curve surrounds a smaller region of $\phi_0-T$ space as $R_2$ decreases.
This change, however, is relatively minor, unless $R_2/R_1$ becomes sufficiently
``small''.

%%%%%%%%%%%%%%%%%%%%%%%%%%%%%%%%%%%%%%%%%%%%%%%%%%%%%%%%%%%%%%%%%%
\begin{figure}[!tb]%
\begin{center}
\subfigure[\label{fig_phi_vs_r_spinodal1_closed}]{\includegraphics[keepaspectratio=true,width=0.232\textwidth]{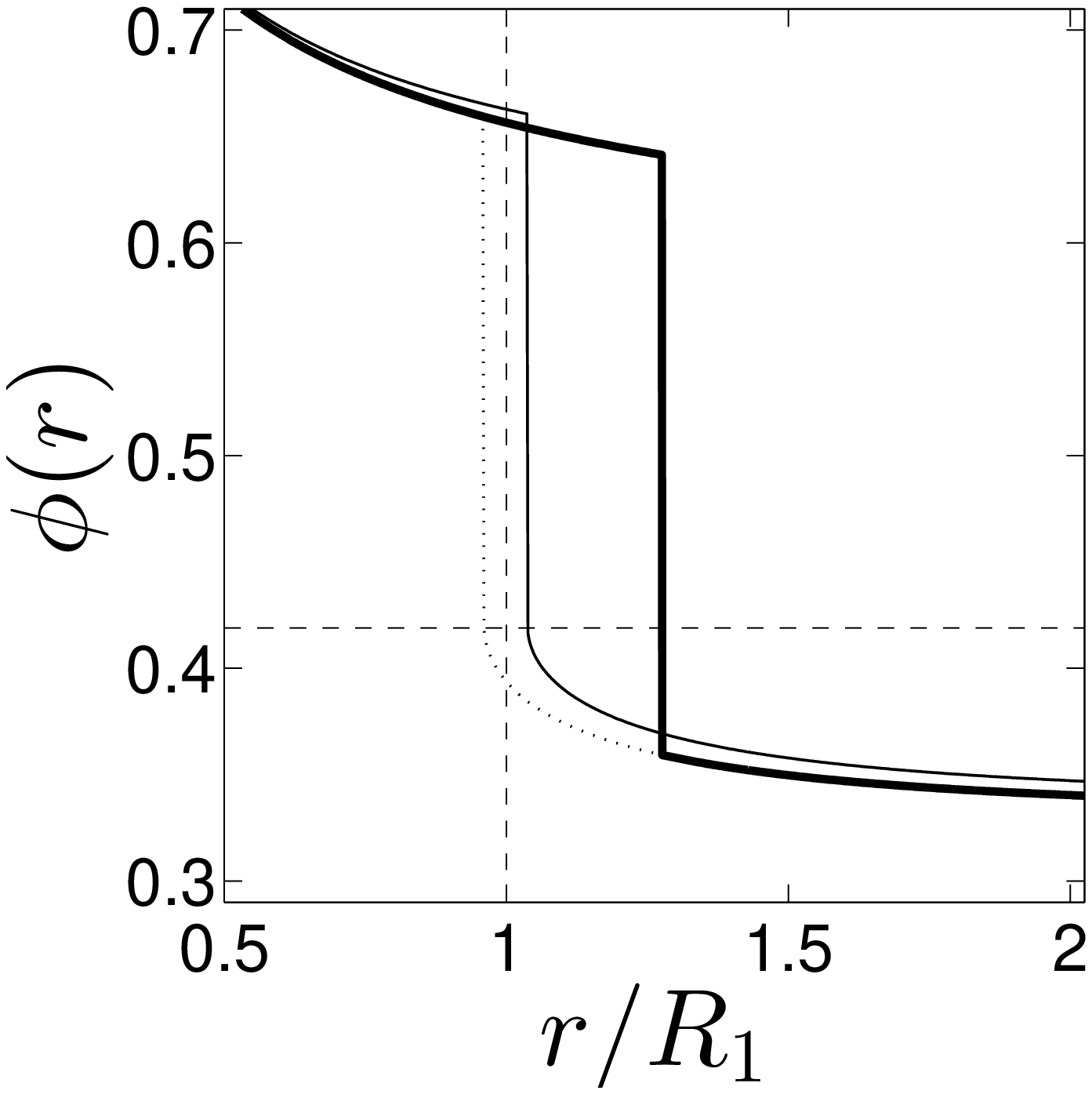}}
\subfigure[\label{fig_phi_vs_r_spinodal2_closed}]{\includegraphics[keepaspectratio=true,width=0.232\textwidth]{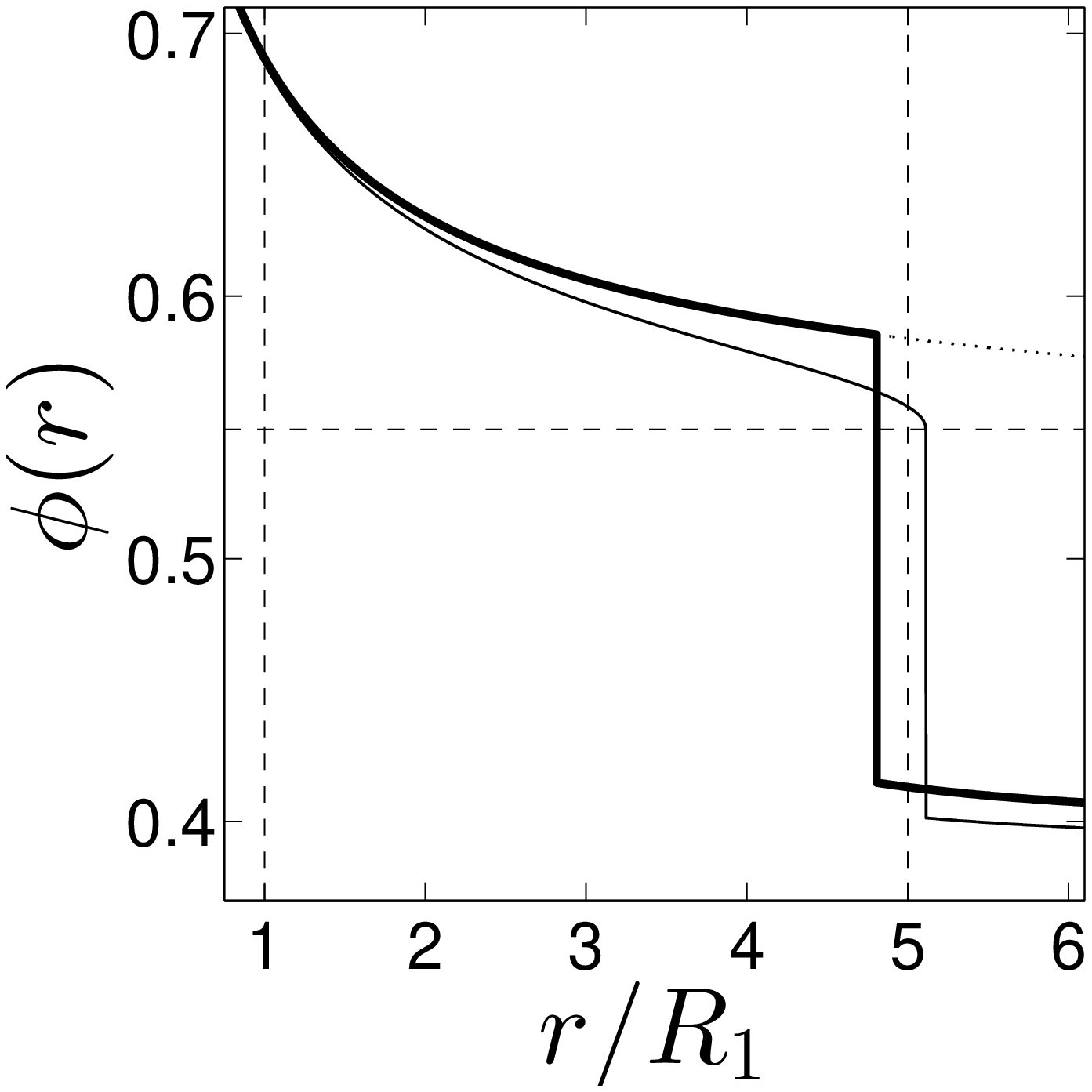}}\\
\subfigure[\label{fig_spinodal_closed}]{\includegraphics[keepaspectratio=true,width=0.232\textwidth]{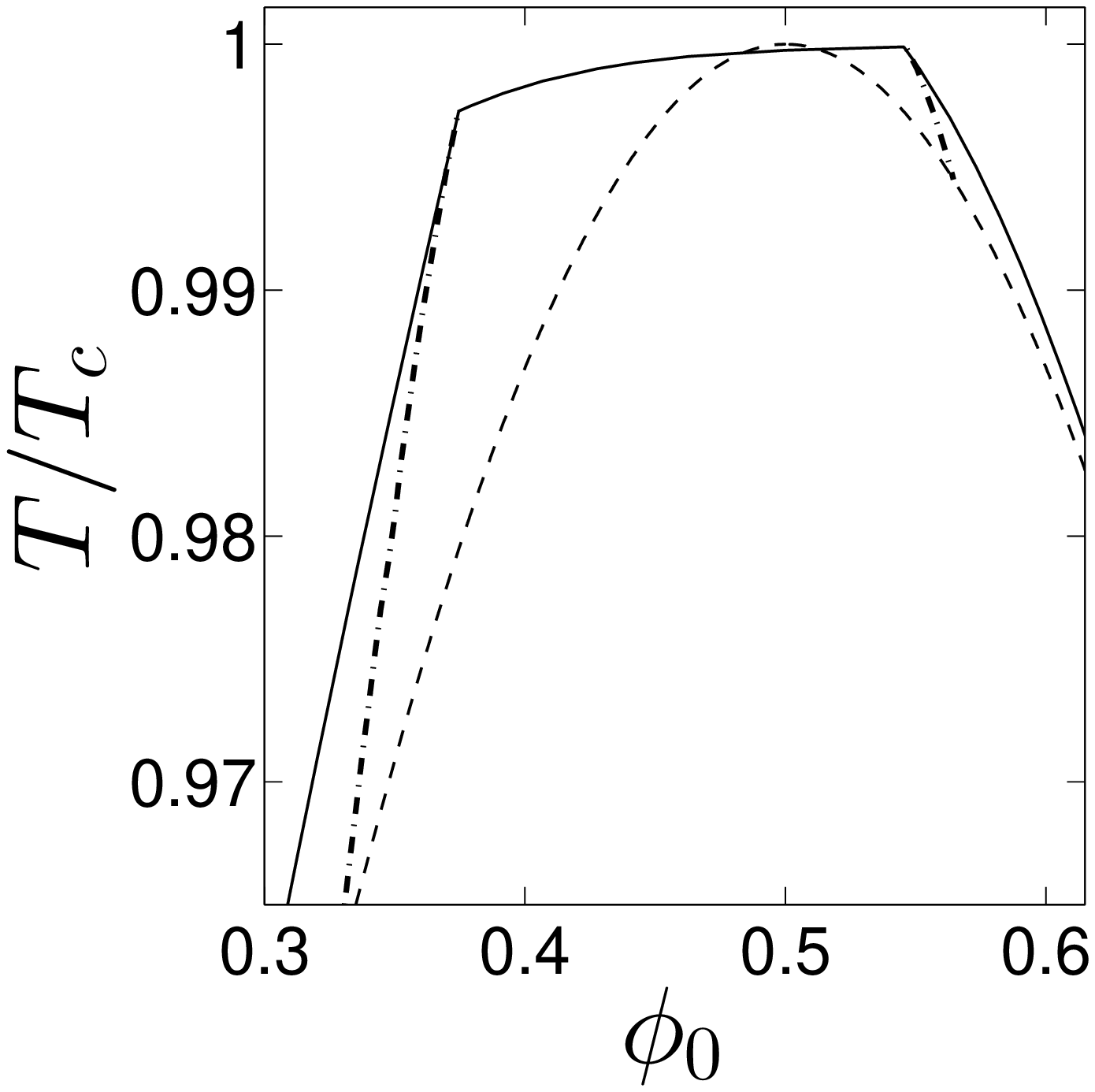}}
\caption{Electrostatic spinodal in closed systems. (a) $\phi(r)$ versus normalized $r$
for $\phi_0\approx 0.34$, $T/T_c\approx 0.973$, $\sigma = 8\times 10^{-4}\,\mathrm{C/m^{2}}$,
and $R_2/R_1=5$ for the minimized solution to $f$ (thick solid line) and lower metastable
solution (thin solid line). Horizontal dashed line shows $\phi_{sL}$. Thick solid line
also shows the minimized open system solution for $\phi_E = 0.33$, while dotted line shows
the corresponding lower solution. (b) $\phi(r)$ versus normalized $r$ for $\phi_0\approx
0.59$, $T/T_c=0.99$, $\sigma = 1.46\times 10^{-3}\,\mathrm{C/m^{2}}$, and $R_2/R_1=5$ for
the minimized solution to $f$ (thick solid line) and upper metastable solution (thin solid
line). Horizontal dashed line shows $\phi_{sH}$. Thick solid line also shows the minimized
open system solution for $\phi_E \approx 0.40$, while dotted line shows the corresponding
upper solution. (c) Stability diagram (solid curve) with spinodal lines (dash-dotted
curves) for $\sigma = 8\times 10^{-4}\,\mathrm{C/m^2}$ and $R_2/R_1=5$. Dashed line is
the binodal curve.}
\end{center}
\end{figure}
%%%%%%%%%%%%%%%%%%%%%%%%%%%%%%%%%%%%%%%%%%%%%%%%%%%%%%%%%%%%

Material conservation produces two spinodal lines in a closed system---one line
associated with each boundary. Finding the spinodal line associated with $R_1$ consists of
finding $\phi(r_s)=\phi_{sL}$ on the lower solution of $f$ and ensuring $r_s=R_1$, similar
to open systems. However, the lower solution from $\phi_E$ in an open system does not
fulfill the material conservation requirement. Instead, the lower solution from yet
another open system concentration $\phi_E$ must be used.
Figure~\ref{fig_phi_vs_r_spinodal1_closed} shows example $\phi(r)$ profiles associated
with $R_1$. The heavy line corresponds to the profile $\phi(r)$ that satisfies the free 
energy minimum of $f$, the dotted line is the lower solution for the open system, and the
thin line is the lower solution for the closed system with $R_2/R_1=5$. In
Fig.~\ref{fig_phi_vs_r_spinodal1_closed}, the open system could be in a metastable state
(compare thick solid and dotted lines), while the closed system would not be metastable
(compare thick and thin solid lines). Similar behavior applies for the location of the
spinodal line at $R_2$; however, this line consists of finding $\phi(r_s)=\phi_{sH}$ on
the \emph{upper} solution of $f$. The line styles in
Fig.~\ref{fig_phi_vs_r_spinodal2_closed} are as those in
Fig.~\ref{fig_phi_vs_r_spinodal1_closed}. In Fig.~\ref{fig_phi_vs_r_spinodal2_closed}, the
closed system could be metastable, while the open system would not be metastable (recall
that the upper solutions have no meaning in open systems).

Finally, Fig~\ref{fig_spinodal_closed} shows the location of
the electrostatic spinodal lines in a closed system for particular values of $\sigma$ and
$R_2/R_1$. Each line begins at the critical points $(\phi_K,T_K)$ on either side of
$\phi_c$ and travel down ``inside'' the stability diagram.

%========================================
% CONCLUSION
%========================================
\section{Conclusion}

In summary, we describe the mixing-demixing phase diagram for two
dielectric liquids in an electric field. 
By focusing on the liquid-liquid interface and adapting standard methods for determining 
phase diagrams, we found the
electrostatic-equivalent of binodal lines, spinodal lines, and critical points. 
Given this new perspective, the dynamics of phase 
separation with non-uniform electric fields requires reinvestigation, 
with an emphases on validating predicted metastable 
states and uncovering critical dynamic behavior. 
Perhaps similar adaptations of existing theory for dynamics will uncover new features in the 
electric-field modified liquid-liquid phase diagram. 

In addition, we restricted our analysis to solutions with radial symmetry, enforcing one dimensional 
solutions that only depend on the distance $r$. This
constraint, however, might not satisfactorily 
apply to all experimental conditions, and allowing for full two- or three-dimensional 
theoretical 
investigations could uncover non-radially symmetric solutions. 
For example, interfacial energies, both liquid-liquid and liquid-surface energies,
dominate the liquid patterning for phase separation beneath the regular binodal curve in
the absence of a field. And in the case where both liquids have an equal preference for
the surface, liquid-liquid interfaces emerge normal to a surface. This configuration, 
however, can be electrostatically unfavorable since the low dielectric material is
adjacent to the charge. It will be interesting to 
determine if, when, and
how instabilities in the interface develop and if these instabilities modify the phase
diagram. 

Also, highly confined cylindrical geometries do not show a true liquid-liquid phase
transition~\cite{Winkler2010}. Here, the system can be approximated as one dimensional,
with the expectation that correlations diverge as the length of the cylinder goes to
infinity. It is unknown if the addition of a non-uniform electric field is
sufficient to induce a true transition in this case. 
An appropriate investigation on this topic would, of course, require theories 
that go beyond the mean-field approach.

Finally, we have not considered the fluid wetting behavior on the electrode surfaces.
In the wedge geometry, for example, these phenomena include wedge filling, where a liquid transitions between partial and complete filling~\cite{Parry1999,Parry2000,Bruschi2002}. This transition can be either first or second order and depends on factors like the wedge opening angle, liquid contact angle, and temperature. 
Since our results show that the interface location directly ties with the electric field, 
it currently remains unclear if the electric field enhances or diminishes the effects of wetting, or possibly both (depending on experimental conditions).

%========================================
% ACKNOWLEDGEMENTS
%========================================
\subsection*{Acknowledgements}
This work was supported by the Israel Science Foundation under grant No. 11/10,
the COST European program MP1106 ``Smart and green interfaces - from single bubbles and drops to industrial, environmental and biomedical applications'',  and the
European Research Council ``Starting Grant'' No. 259205.

%===========================================================

\end{document}